\documentclass[acmsmall]{acmart}
\AtBeginDocument{%
  }

\usepackage{xcolor}
\usepackage{array}
\usepackage{subfigure}
\usepackage{graphicx}
\usepackage{multirow}
\usepackage{color}
\usepackage{url}
\usepackage[lined,linesnumbered,ruled]{algorithm2e}
\usepackage{todonotes}
\setuptodonotes{inline}

\setcopyright{acmcopyright}
\copyrightyear{2023}
\acmYear{2023}
\acmDOI{XXXXXXX.XXXXXXX}

\acmJournal{TOSN}
\acmVolume{37}
\acmNumber{4}
\acmArticle{111}
\acmMonth{2}

\begin{document}

\title{VibHead: An Authentication Scheme for Smart Headsets through Vibration}

\author{Feng Li}
\email{fli@sdu.edu.cn}
\affiliation{%
  \institution{Shandong University}
  \streetaddress{72 Binhai Road}
  \city{Qingdao}
  \state{Shandong}
  \country{China}
  \postcode{266237}
}

\author{Jiayi Zhao}
\email{202135184@mail.sdu.edu.cn}
\affiliation{%
  \institution{Shandong University}
  \streetaddress{72 Binhai Road}
  \city{Qingdao}
  \state{Shandong}
  \country{China}
  \postcode{266237}
}

\author{Huan Yang}
\email{cathy\_huanyang@hotmail.com}
\affiliation{%
  \institution{Qingdao University}
  \streetaddress{308 Ningxia Road}
  \city{Qingdao}
  \state{Shandong}
  \country{China}
  \postcode{266071}
}

\author{Dongxiao Yu}
\email{dxyu@sdu.edu.cn}
\affiliation{%
  \institution{Shandong University}
  \streetaddress{72 Binhai Road}
  \city{Qingdao}
  \state{Shandong}
  \country{China}
  \postcode{266237}
}

\author{Yuanfeng Zhou}
\email{yfzhou@sdu.edu.cn}
\affiliation{%
  \institution{Shandong University}
  \streetaddress{1500 Shunhua Road}
  \city{Jinan}
  \state{Shandong}
  \country{China}
  \postcode{250101}
}

\author{Yiran Shen}
\email{yiran.shen@sdu.edu.cn}
\affiliation{%
  \institution{Shandong University}
  \streetaddress{27 South Shanda Road}
  \city{Jinan}
  \state{Shandong}
  \country{China}
  \postcode{250100}
}

\renewcommand{\shortauthors}{Li et al.}

\begin{abstract}
  Recent years have witnessed the fast penetration of Virtual Reality (VR) and Augmented Reality (AR) systems into our daily life, the security and privacy issues of the VR/AR applications have been attracting considerable attention. Most VR/AR systems adopt head-mounted devices (i.e., smart headsets) to interact with users and the devices usually store the users' private data. Hence, authentication schemes are desired for the head-mounted devices. Traditional knowledge-based authentication schemes for general personal devices have been proved vulnerable to shoulder-surfing attacks, especially considering the headsets may block the sight of the users. Although the robustness of the knowledge-based authentication can be improved by designing complicated secret codes in virtual space, this approach induces a compromise of usability. Another choice is to leverage the users' biometrics; however, it either relies on highly advanced equipments which may not always be available in commercial headsets or introduce heavy cognitive load to users.

  In this paper, we propose a vibration-based authentication scheme, VibHead, for smart headsets. Since the propagation of vibration signals through human heads presents unique patterns for different individuals, VibHead employs a CNN-based model to classify registered legitimate users based the features extracted from the vibration signals. We also design a two-step authentication scheme where the above user classifiers are utilized to distinguish the legitimate user from illegitimate ones. We implement VibHead on a Microsoft HoloLens equipped with a linear motor and an IMU sensor which are commonly used in off-the-shelf personal smart devices. According to the results of our extensive experiments, with short vibration signals ($\leq 1s$), VibHead has an outstanding authentication accuracy; both FAR and FRR are around $5\%$.

\end{abstract}

\ccsdesc[500]{Security and privacy~Authentication}
\ccsdesc[500]{Human-centered computing~Ubiquitous and mobile computing systems and tools}
\keywords{User authentication, vibration signals, smart headsets}

\received{20 February 2007}
\received[revised]{12 March 2009}
\received[accepted]{5 June 2009}

\maketitle

\section{Introduction} \label{sec:intro}
  Recent years have witnessed the rapid development of Virtual Reality (VR) and Augmented Reality (AR) technologies, and the VR/AR devices have been widely used in a large variety of application scenarios, e.g., e-commerce~\cite{FlavianIO-JBR19,SCGC2020}, healthcare~\cite{FuHS-TCH22,WangBJM-NMI22}, social networking, industry~\cite{LiZLPL-RCIM22,KarnikBBKD-JIII22}, entertainment~\cite{TanprasertY-CHI22} and education~\cite{Babich2019,MarinV-TLT21}. According to \cite{IDC2022}, the worldwide spending on AR/VR is forecast to reach $50.9$ billion in 2026.  There have been a number of  VR/AR headsets available in market, e.g., Microsoft HoloLens, PlayStation VR, HTC VIVE, Oculus Quest, Samsung Gear VR, PICO etc, which provides convenient human-computer-interaction functionalities enabled by embedded sensors.

  Nevertheless, the rise of VR/AR applications induce substantial risks in security and privacy, especially considering VR/AR devices (e.g., smart headsets) may store users' personal information, e.g., emails, photos, videos, payment accounts, health data etc. Since tradition authentication techniques, e.g., password, digital PIN and patter drawing, are vulnerable to shoulder-surfing attacks, many recent studies propose to guarantee the robustness of authentication by introducing more complicated secret codes in virtual space, e.g., \cite{YuLFM-APCCAS19,FunkMMKMM-CHI19,MathisWVK-CHI20}. Nevertheless, users have to remember their secret codes and spend quite a while on entering the secret codes through complicated operations in virtual space and inconvenient interaction with VR/AR devices. Another choice is to utilize implicit biometrics to defend against shoulder-surfing attacks. For example, head and/or body movements are utilized in \cite{LiAZXLG-PerCom16,MustafaMSM-IWSPA18,ShenWLXZHR-TDSC19,WangZ-CHI21,PfeufferGPMBA-CHI19} to recognize different users, however, users have to spend several seconds (even minutes) to elaborately perform specific gestures. Gaze biometrics also have been explored for user authentication~\cite{SluganovicRRM-CCS16, SluganovicRRM-CCS16, ZhuJXML-IMWUT20, LiebersHBGS-VRST21, KumarLCSHPTH-MM22}; however, it relies on highly accurate eye movement tracking and thus entails special equipments which may not always be available on personal headset devices.

  It is well known that human heads can be described by many unique physiological features, e.g., size, bone shape, facial structure, ratio of fat, etc. Therefore, a possible authentication solution is to fully explore these physiological features with embedded sensors on headsets. Since vibration signals propagation can be well characterized in short range, they have been applied by many recent proposals to design authentication systems for hand-held devices (e.g., mobile phones, smart watches etc. which are operated by hand surfaces or fingers)~\cite{LiuWCS-CCS17,LiFK-CCS19,XuYCHZCL-MobiCom20,LeeCL-CCS21,CaoJLX-ICDCS21,HusaT-CNS21}. However, how to utilize vibration signals to authenticate users of smart headsets is still an open problem.

  In this paper,  we first investigate the feasibility of utilizing vibration signals to authenticate smart headset users and find that the propagation of vibration signals through human heads creates unique patterns for different individuals. Motivated by the observations, we propose a vibration-based authentication system, {\sl VibHead}, for smart headsets. The system utilizes a pair of linear motor and IMU sensors pervasively embedded on VR/AR devices. The motor generates vibration signals actively  on one side of VR/AR devices and the IMU sensors on the other side samples the propagated signals for authentication. Except for the primitive data features collected by the IMU sensors, we also extract \textit{Mel-frequency Cepstral Coefficient} (MFCC) features from the collected data samples. Both the raw readings of the IMU sensors and the MFCC coefficcents are fed to a CNN-based classifier for user classification. Furthermore, we design a two-step authentication scheme, by employing the user classifiers. We finally perform extensive experiments to verify the efficacy of our VibHead system, and the experiments results show that VibHead can accurately authenticate users in different typical gestures which are usually taken by the users in daily life.

  The reminder of this paper is organized as follows. Related literature is surveyed in Sec.~\ref{sec:relwk}. Before introducing our system architecture and threat model in Sec.~\ref{sec:overview}, we first report some preliminary observations to motivate the design of VibHead in Sec~\ref{sec:mot}. We then present how to collect vibration signals and extract features from the collected signals in Sec.~\ref{sec:vibfeature}. We report our user classification model and the two-step user authentication scheme in Sec.~\ref{sec:classification} and Sec.~\ref{sec:authen}, respectively. We finally evaluate our VibHead system in Sec.~\ref{sec:evaluation} and conclude this paper in Sec.~\ref{sec:conclusion}.

\section{Related Work} \label{sec:relwk}
  \subsection{Authentication for Smart Headsets} \label{ssec:auth4headset}
    To address the vulnerability of tradition authentication schemes to shoulder-surfing attacks, some studies propose to construct password systems in virtual space. For example, \cite{YuLFM-APCCAS19} proposes a 3D password system such that users enter their passwords in a virtual 3D space. In \cite{FunkMMKMM-CHI19}, an authentication system, called LookUnlock, is proposed, based on head-gaze tracking and spatial mapping. In LookUnlock, a password is a set of spatial and virtual objects which a user focuses on in the correct sequence. In \cite{MathisWVK-CHI20}, users select digits from a virtual 3D cube through a handheld controller. These methods improve the robustness of authentication by increasing the complexity of secret codes with a sacrifice in usability.

    Unlike the above knowledge-based authentication, which is based on ``what you know'', another choice is to leverage ``who you are'' by employing unique biometrics of different users for better robustness against guessing attacks and shoulder-surfing attacks. Some of recent studies extract distinctive biometric features from users' gestures. For sample, \cite{LiAZXLG-PerCom16} proposes a user authentication system where the unique patterns of the user's head movement in response to an external audio stimulus are utilized. In \cite{MustafaMSM-IWSPA18}, an authentication scheme for VR applications is designed. Specifically, the patterns of the head and body movement exhibited by a user traversing the VR space can be utilized to continuously authenticate the user. In \cite{ShenWLXZHR-TDSC19}, gait signatures are utilized for user authentication. An user is asked to walk a few step and the readings of the IMU sensor are analyzed to characterize the user's unique gait. \cite{WangZ-CHI21} develops an authentication scheme for AR/VR users such that they can unlock their profiles using simple head gestures (eg., nodding). \cite{PfeufferGPMBA-CHI19} utilizes combinations of gestures to authenticate users. In these methods based on gesture biometric, users are asked to spend several seconds (even minutes) to elaborately take specific gestures.

    Gaze biometrics also have been explored for user authentication recently. \cite{SluganovicRRM-CCS16} applies visual stimulus and extracts reliable biometric features from the resulting reflexive eye movements. \cite{SluganovicRRM-CCS16} also applies visual stimuli to an user, and utilizes eye movement for the purpose of continuous authentication. In \cite{ZhuJXML-IMWUT20}, users are authenticated according to the rhythms of blinking and the unique pupil size variation patterns. In \cite{LiebersHBGS-VRST21}, both gaze behaviors and head orientations are utilized for authentication in VR headsets. In \cite{KumarLCSHPTH-MM22}, an user’s gaze and lateral shifts (i.e., footsteps) are utilized simultaneously to drive the authentication. Utilizing gaze biometrics for authentication only work with precise eye movement tracking and thus relies on special equipments (e.g., gaze trackers); however, current add-on eye trackers on VR/AR headset may not always provide sufficient tracking accuracy.

  \subsection{Vibration-based Authentication} \label{ssec:relwk-vibauth}
    The propagation of Vibration signals can be well characterized in short range and thus has been widely used in various applications, e.g., gesture recognition~\cite{LaputXH-UIST16,WenRD-CHI16}, key-stroke detection~\cite{LiuCGW-SECON17,ChenGHWRHW-SECON18} and near-field communications~\cite{RoyGC-NSDI15,RoyC-NSDI16}.

    Recently, vibration signals are explored for user authentication. A vibration-based finger-input authentication system, VibWrite, is developed in \cite{LiuWCS-CCS17}. VibWrite can be implemented on any solid surface for smart access systems. It utilizes intrinsic human physical characteristics presenting at specific location/surface for authentication. \cite{LiFK-CCS19} exploits the uniqueness and nonlinearity of hand-surface vibration for user authentication. In \cite{XuYCHZCL-MobiCom20}, an on-touch user authentication system is designed for smartphones. In the system, active vibration signals are generated by a build-in motor. The physical characters of touching fingers are extracted from the vibration signals received by the IMU sensors on the the smartphone and then utilized for user authentication. An authentication scheme for smartwatches is designed in \cite{LeeCL-CCS21}. Since the propagations of vibration signals through different human bodies are distinct, \cite{LeeCL-CCS21} first generates a random sequence of vibration types and then utilize the vibration signals colleted by the IMU sensor for user authentication. \cite{CaoJLX-ICDCS21} studies continuous user authentication. The vibration responses of concealed hand biometrics, which are passively activated by the natural user-device interaction on the touchscreen, are utilized for authentication. \cite{HusaT-CNS21} proposes Vibe, which is an implicit two-factor authentication mechanism. Vibe uses a vibration communication channel to authenticate users. Although these methods pioneer in utilizing vibration for authentication, they are elaborately designed for either hand-made systems or hand-held \textit{Commercial Off The Shelf} (COTS) devices (e.g., mobile phones or smart watches), whereas designing vibration-based authentication for smart headsets is still an open problem.


\section{Motivational Examples}  \label{sec:mot}
\begin{figure*}[htb!]
  \begin{center}
    \parbox{.19\textwidth}{\center\includegraphics[width=.18\textwidth]{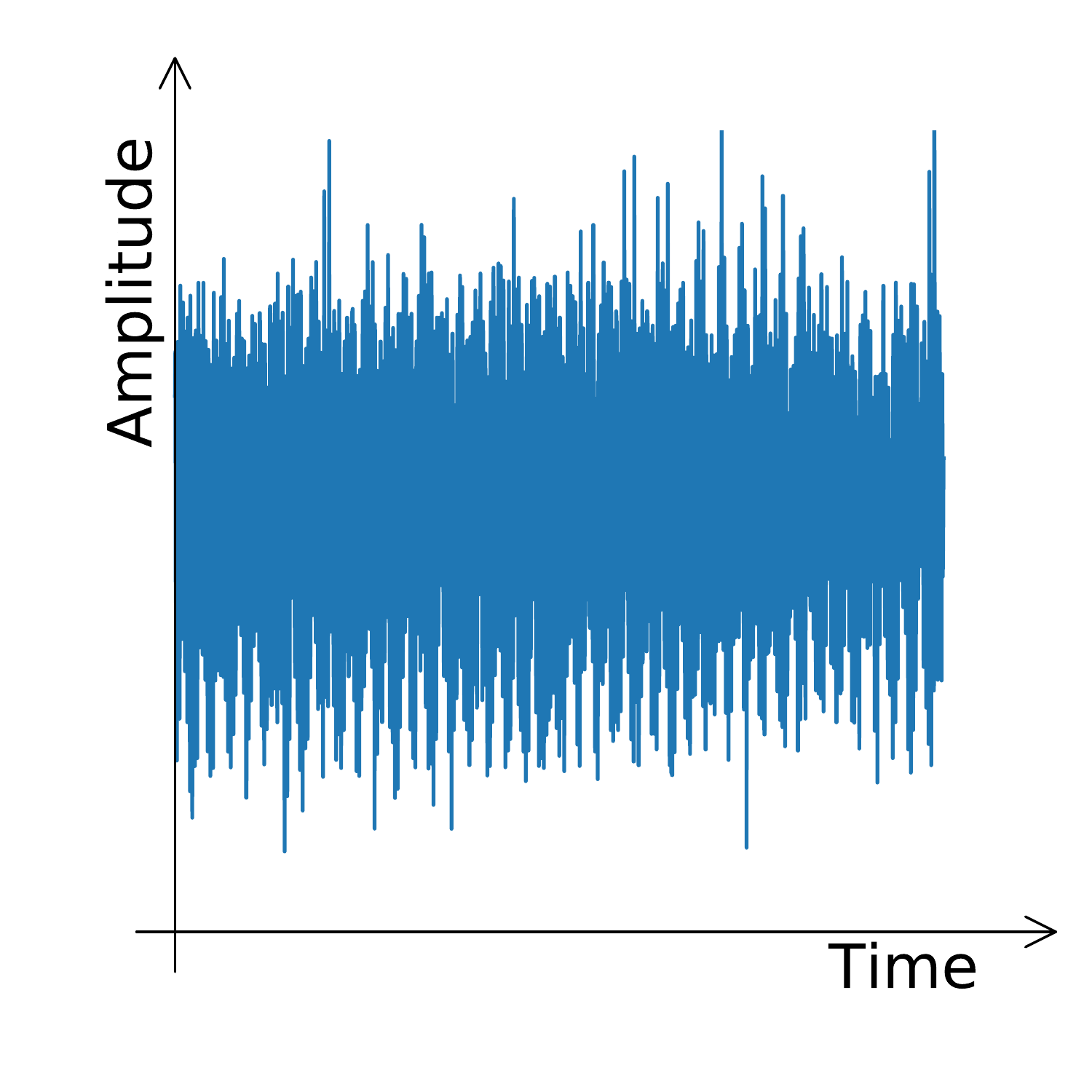}}
    \parbox{.19\textwidth}{\center\includegraphics[width=.18\textwidth]{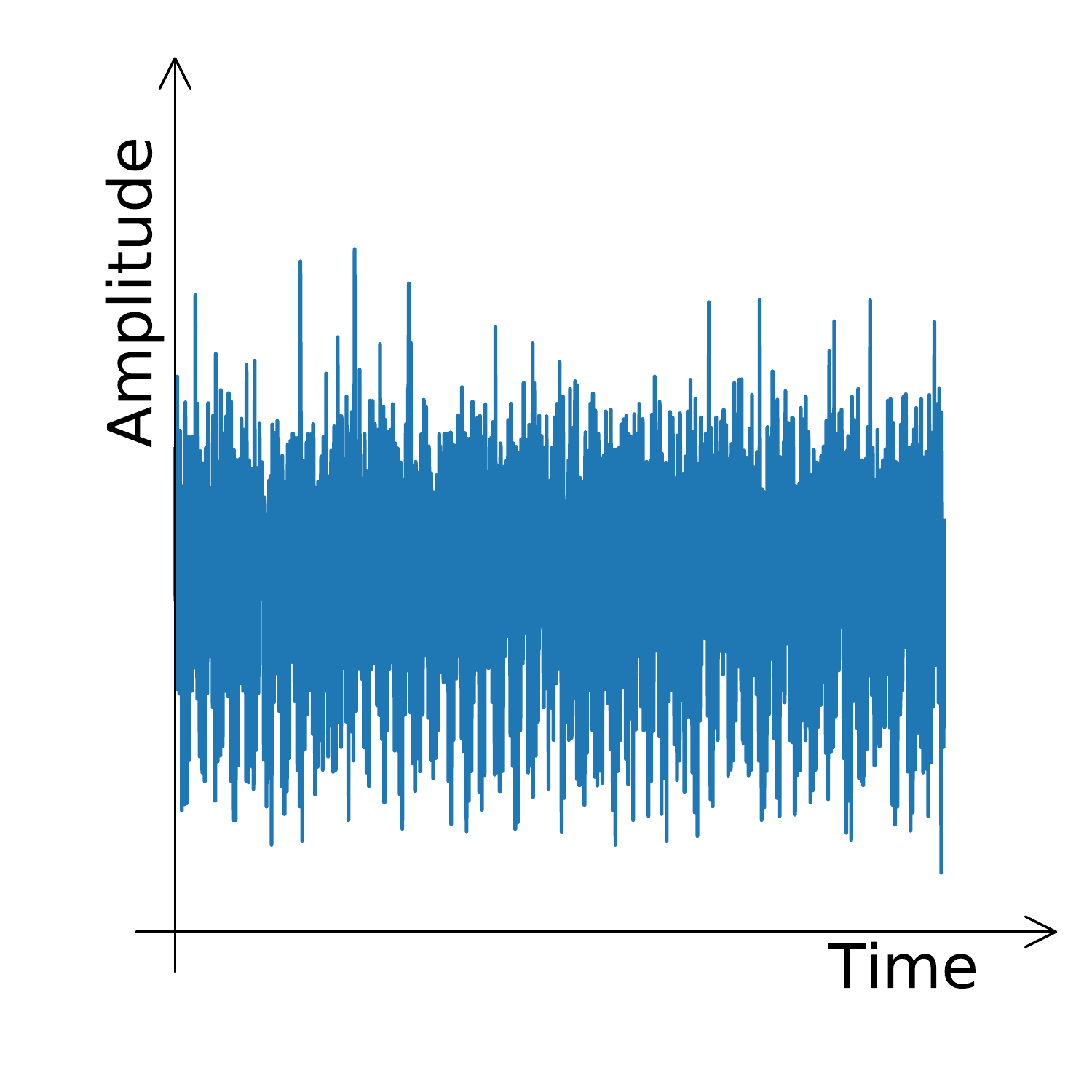}}
    \parbox{.19\textwidth}{\center\includegraphics[width=.18\textwidth]{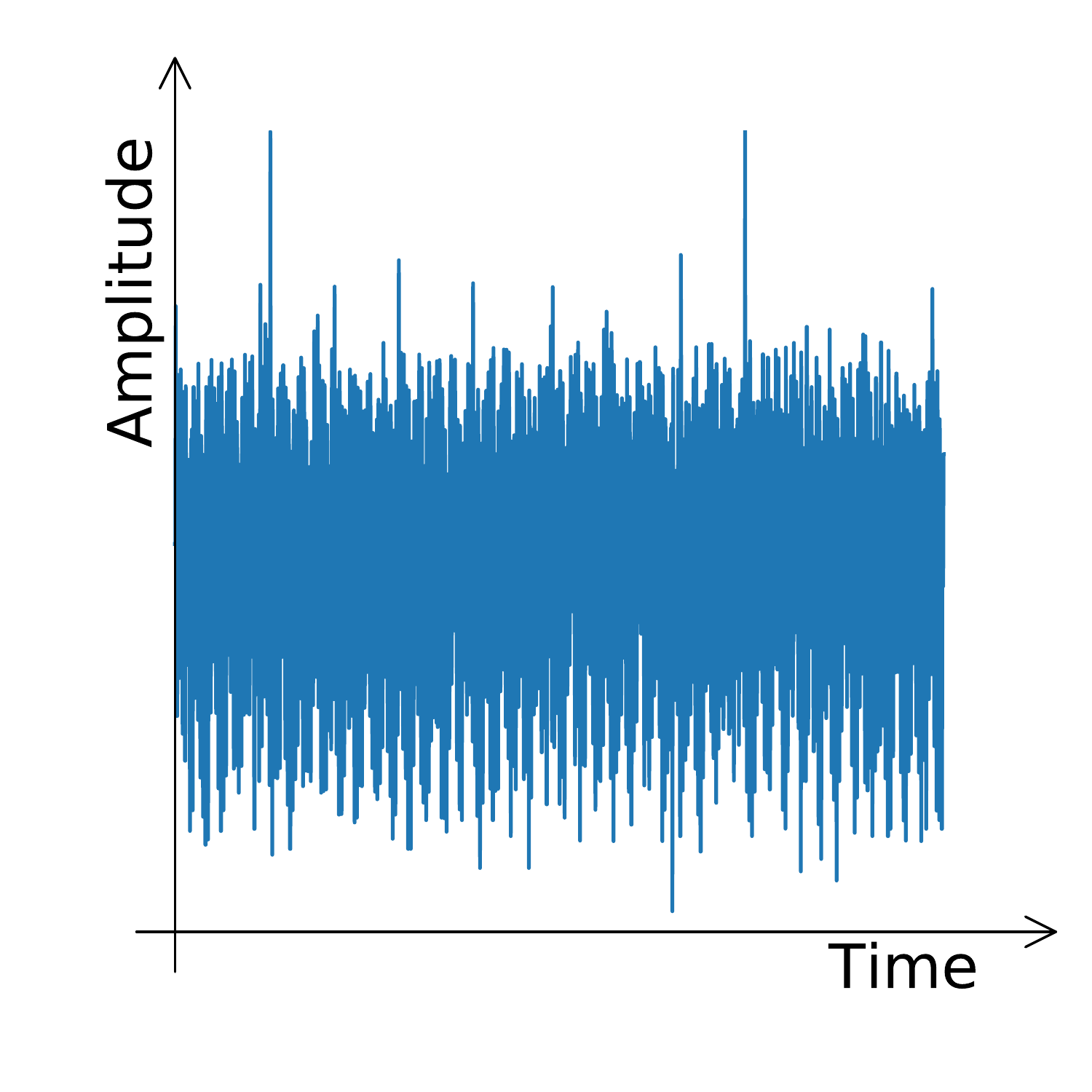}}
    \parbox{.19\textwidth}{\center\includegraphics[width=.18\textwidth]{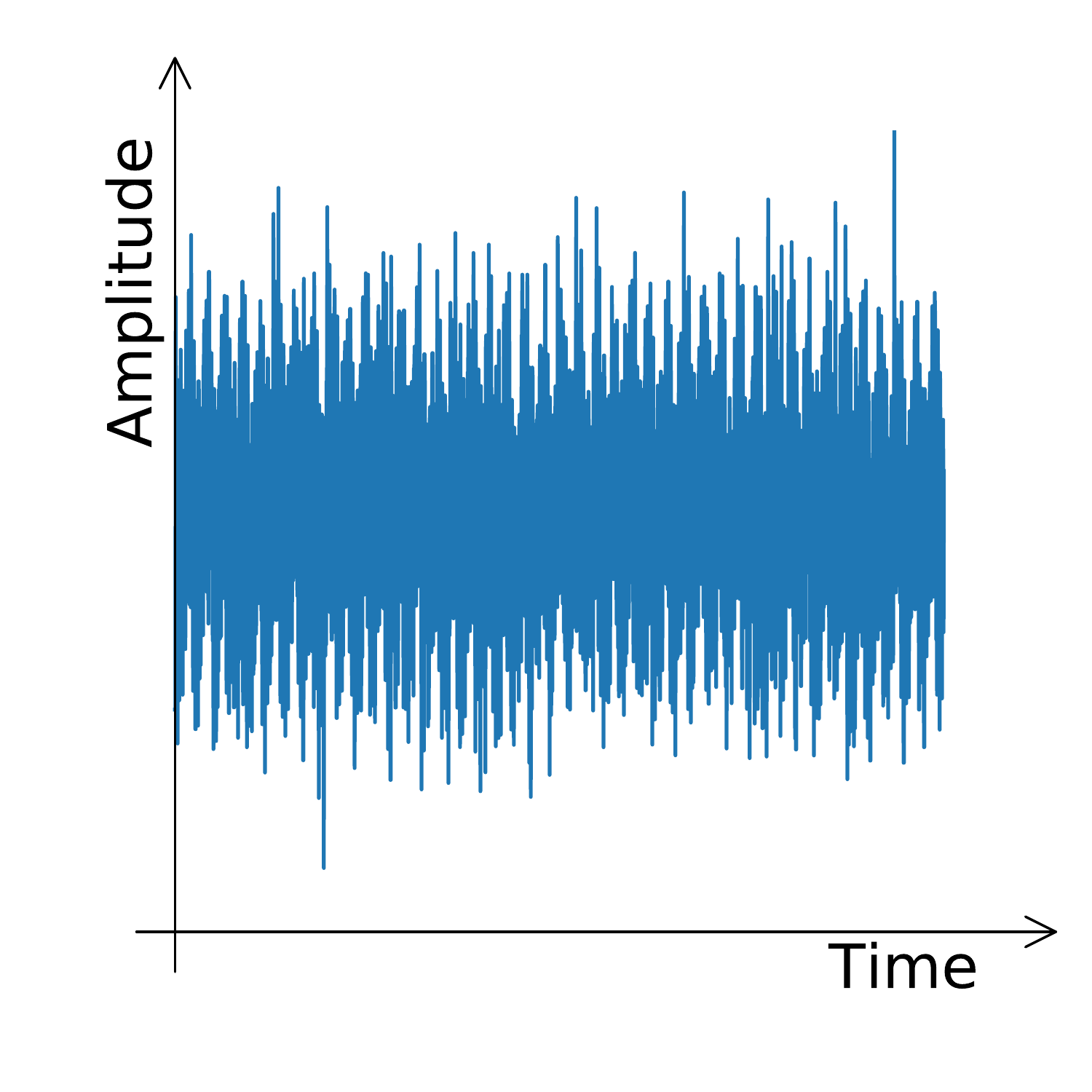}}
    \parbox{.19\textwidth}{\center\includegraphics[width=.18\textwidth]{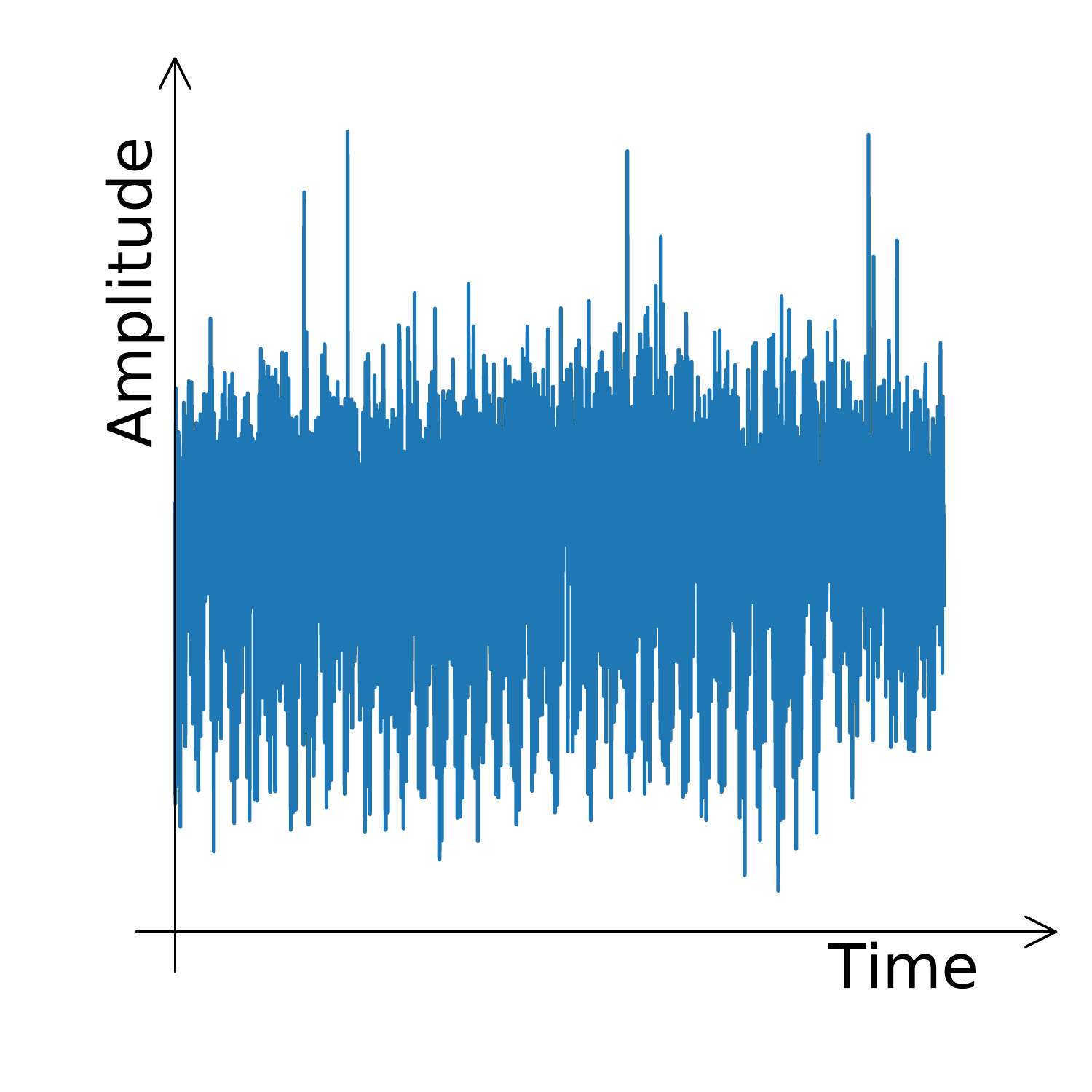}}
    \parbox{.19\textwidth}{\center\includegraphics[width=.18\textwidth]{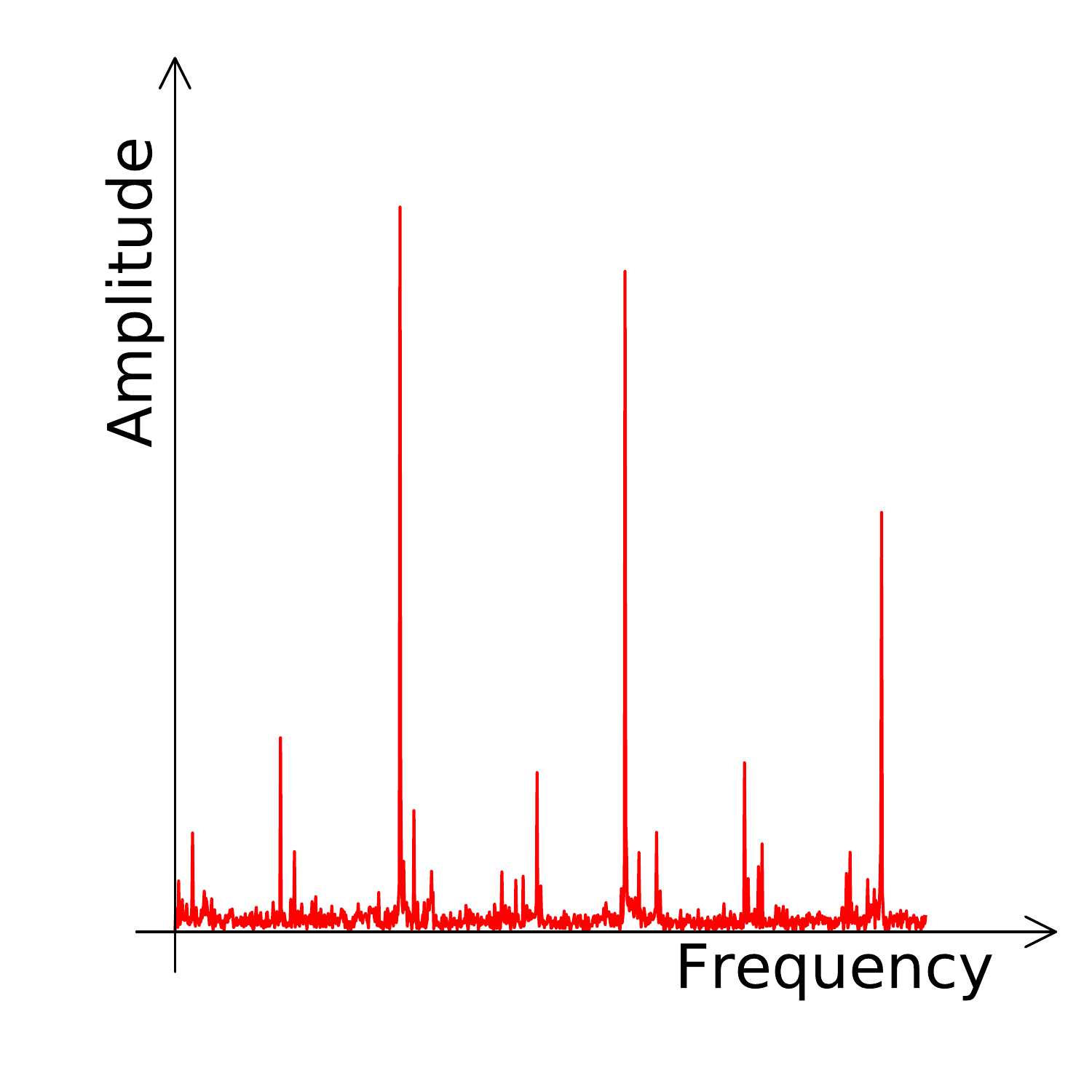}}
    \parbox{.19\textwidth}{\center\includegraphics[width=.18\textwidth]{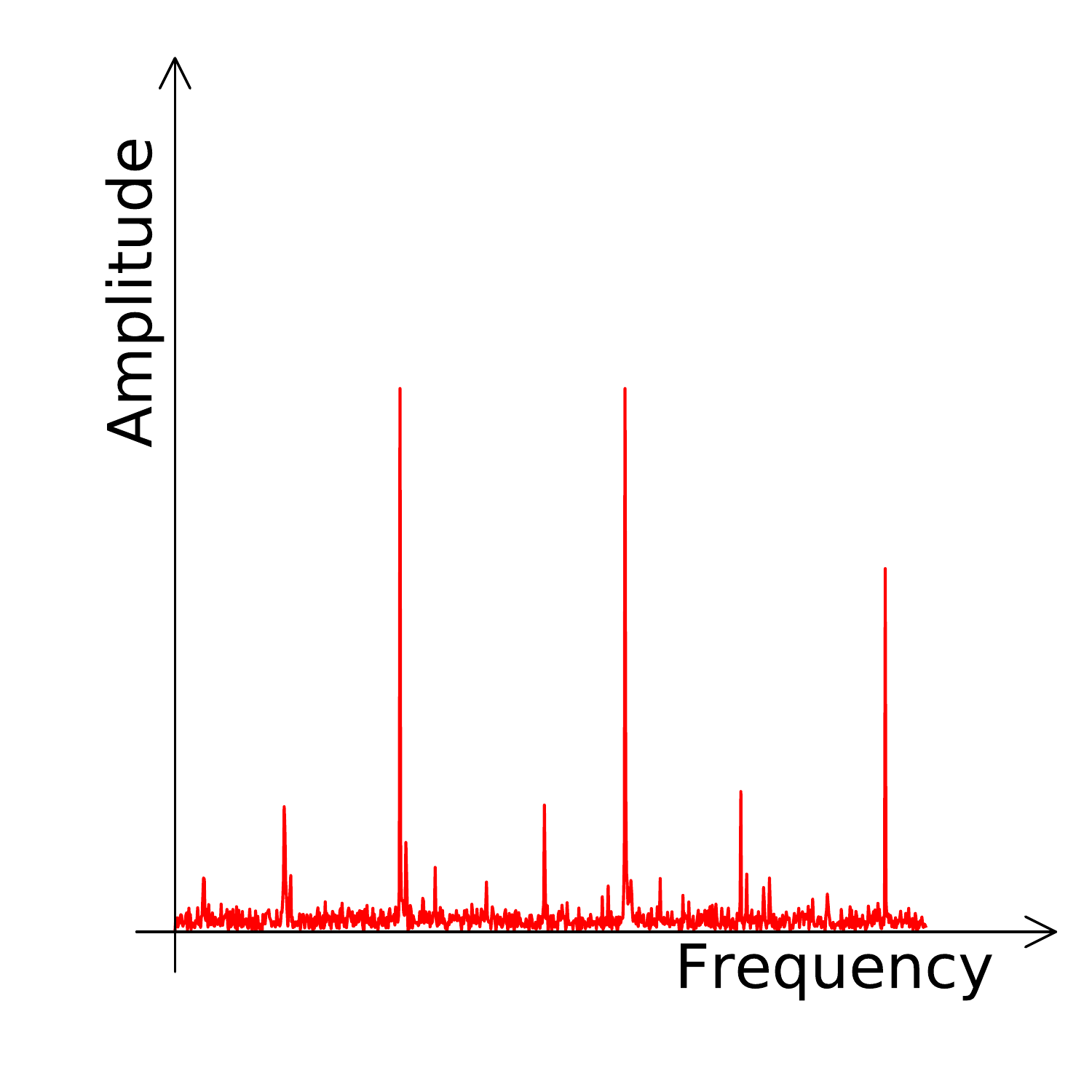}}
    \parbox{.19\textwidth}{\center\includegraphics[width=.18\textwidth]{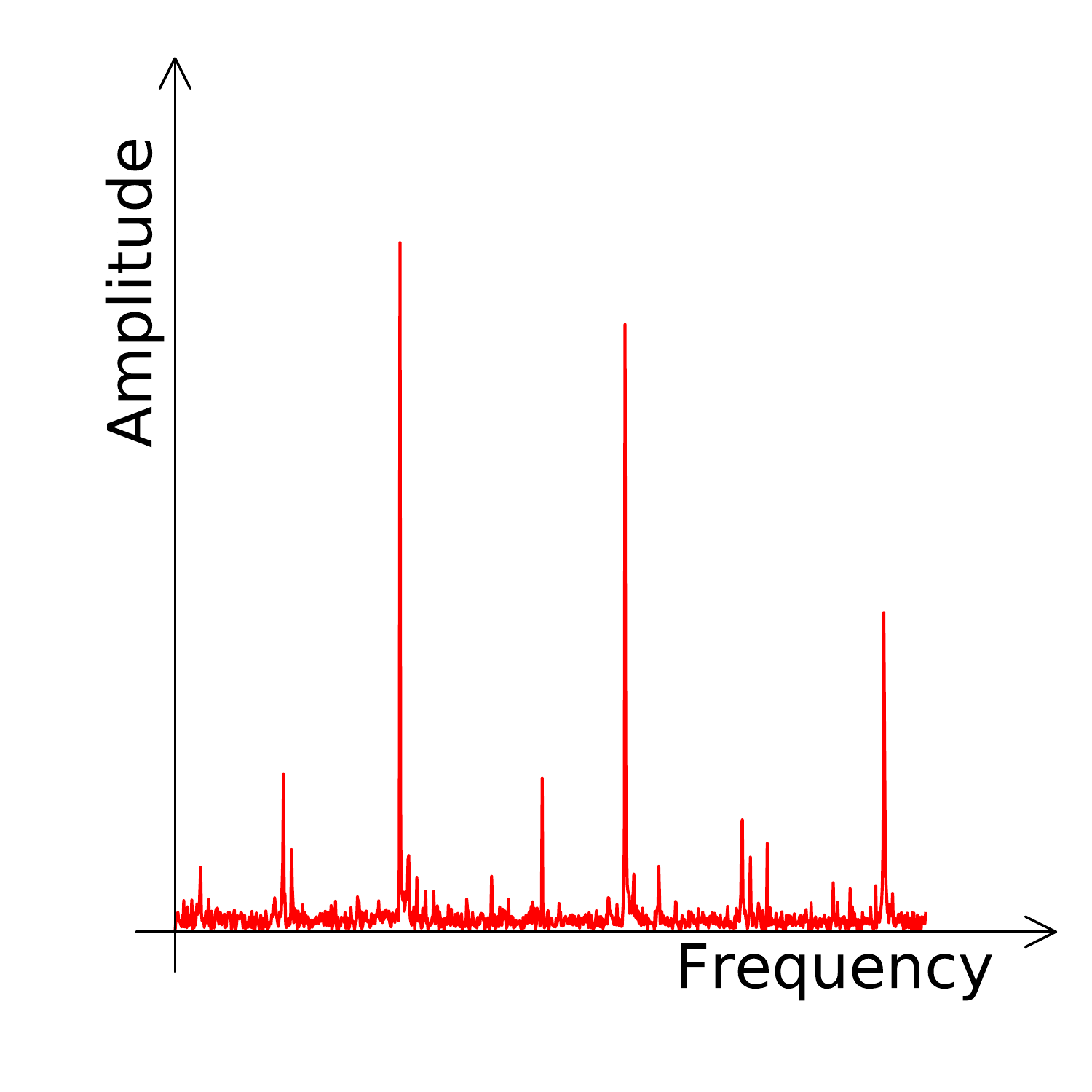}}
    \parbox{.19\textwidth}{\center\includegraphics[width=.18\textwidth]{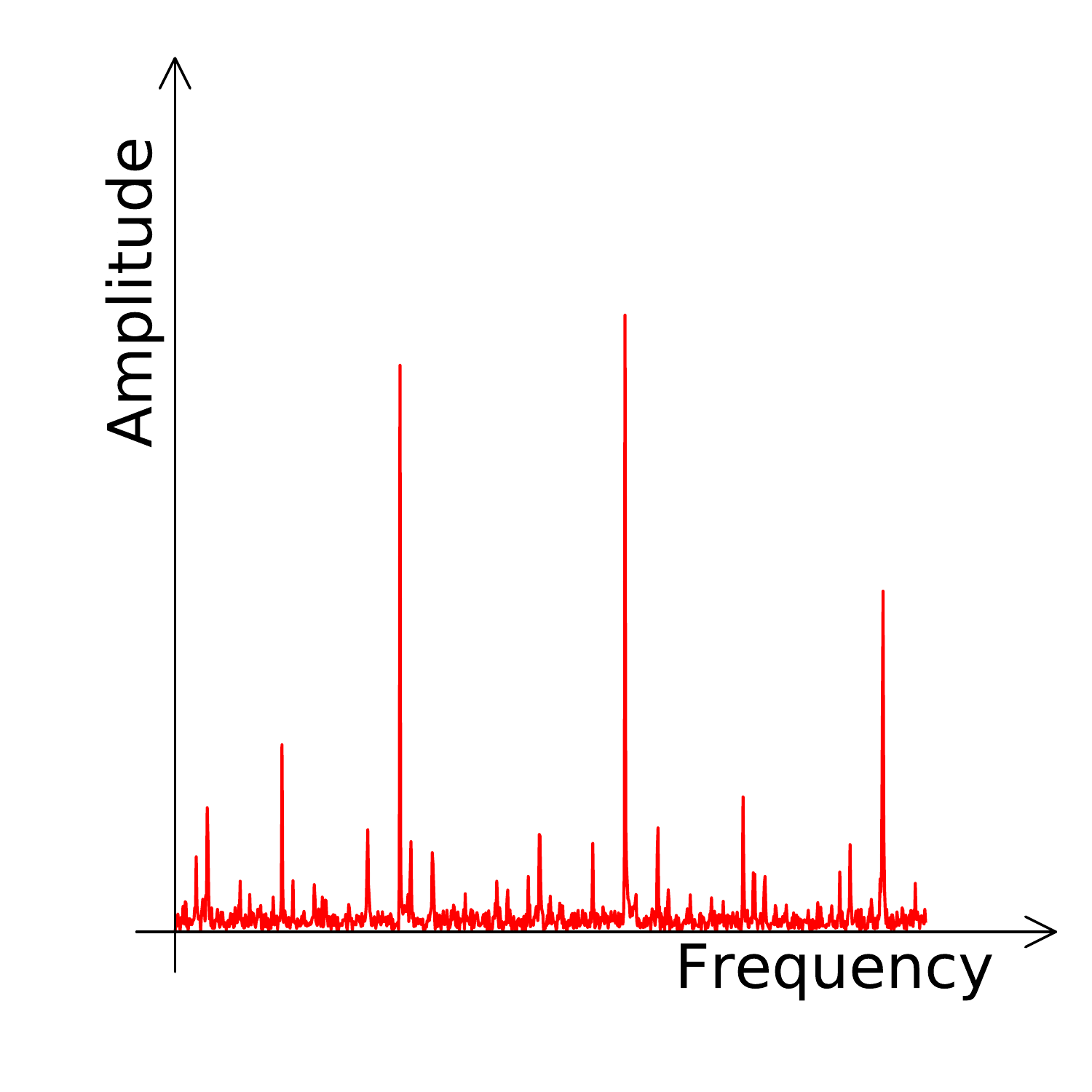}}
    \parbox{.19\textwidth}{\center\includegraphics[width=.18\textwidth]{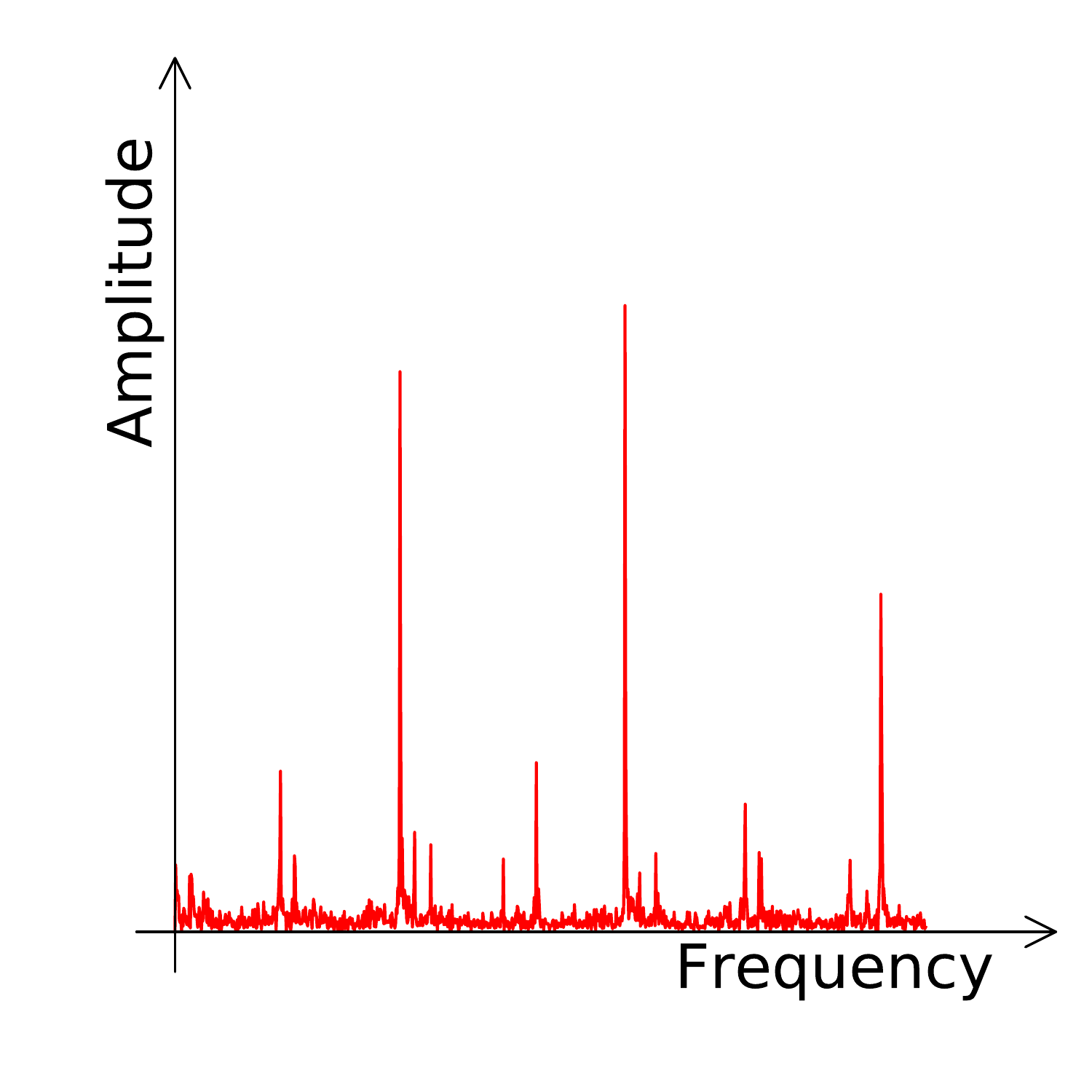}}
    \parbox{.19\columnwidth}{\center\scriptsize(a1)}
    \parbox{.19\columnwidth}{\center\scriptsize(a2)}
    \parbox{.19\columnwidth}{\center\scriptsize(a3)}
    \parbox{.19\columnwidth}{\center\scriptsize(a4)}
    \parbox{.19\columnwidth}{\center\scriptsize(a5)}
    \parbox{.19\textwidth}{\center\includegraphics[width=.18\textwidth]{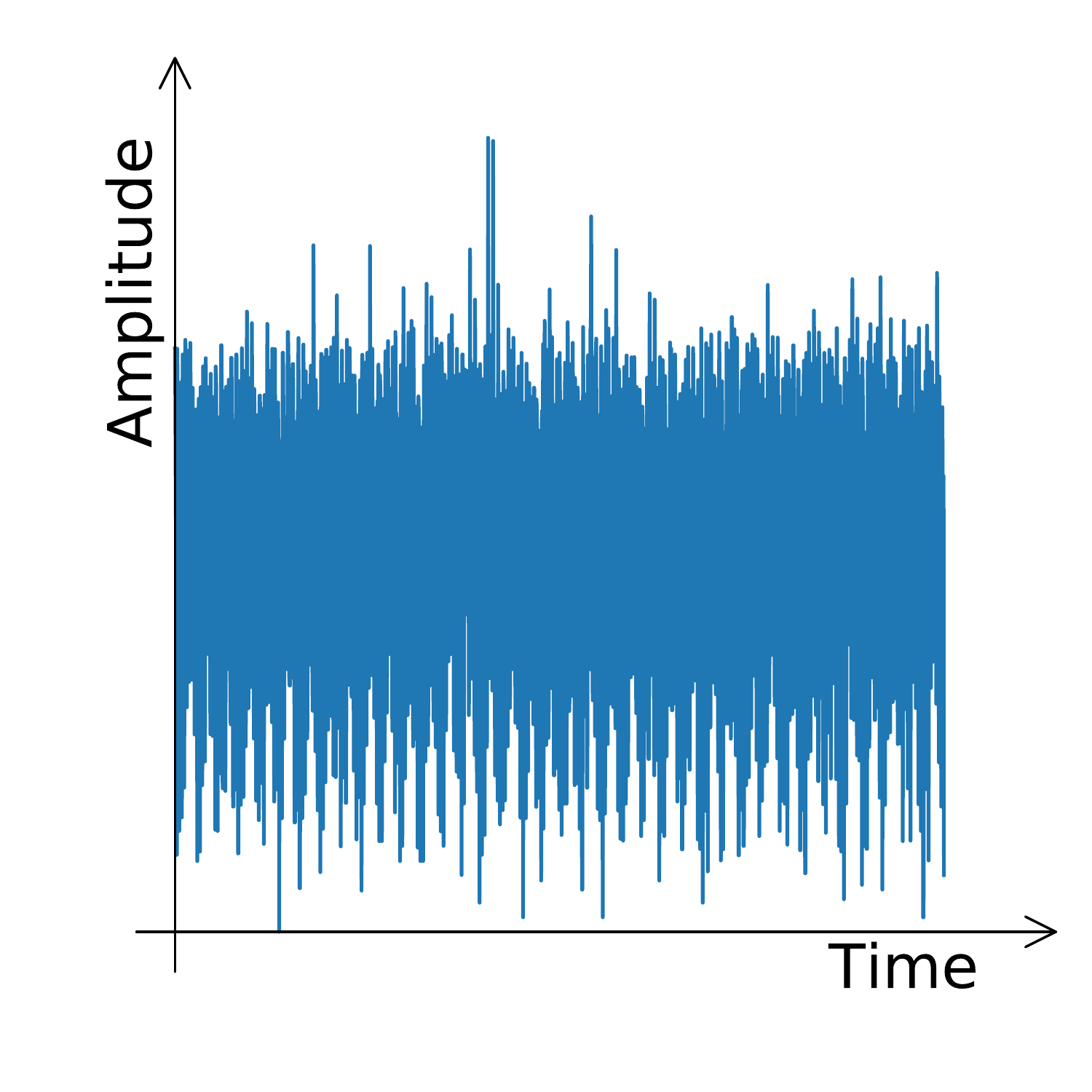}}
    \parbox{.19\textwidth}{\center\includegraphics[width=.18\textwidth]{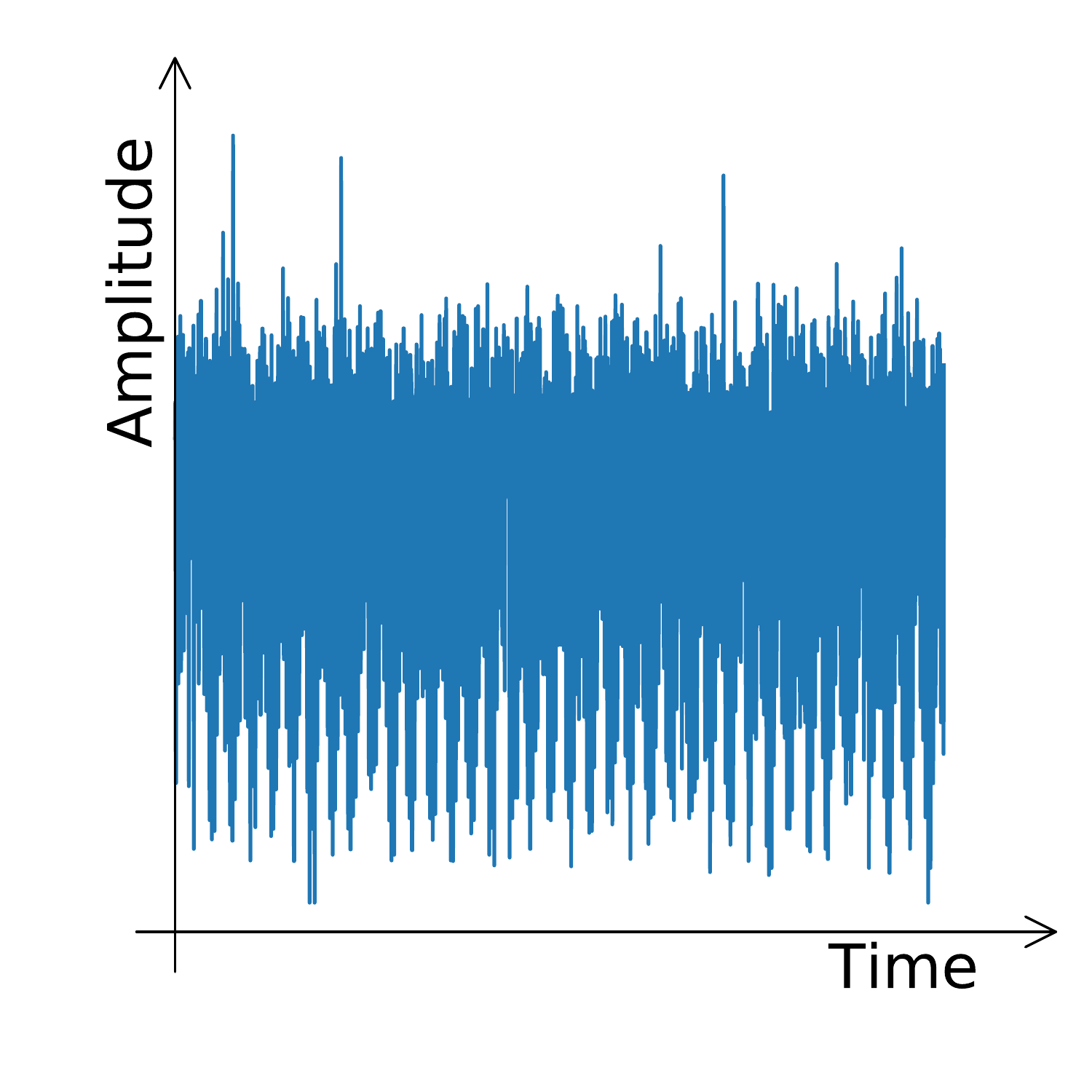}}
    \parbox{.19\textwidth}{\center\includegraphics[width=.18\textwidth]{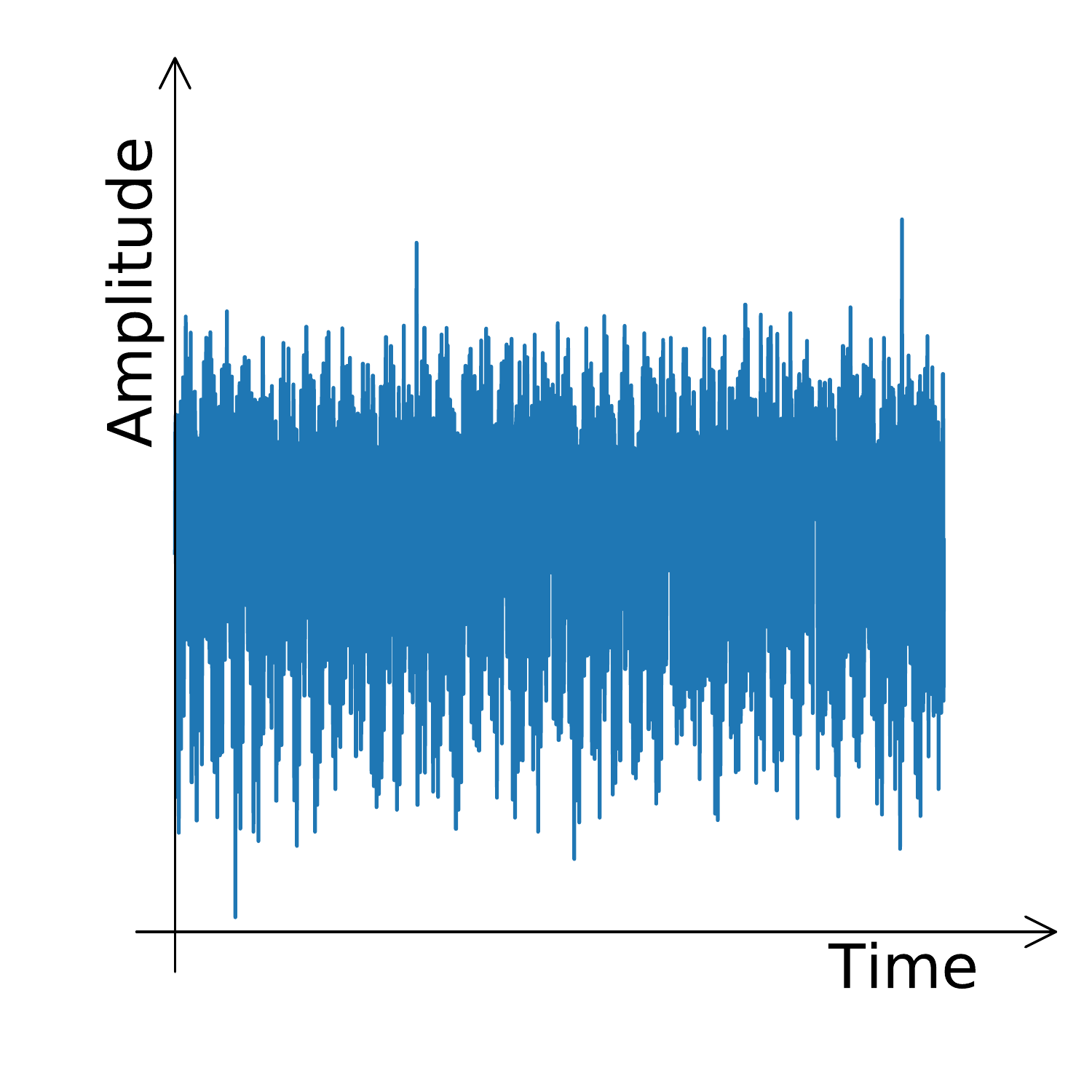}}
    \parbox{.19\textwidth}{\center\includegraphics[width=.18\textwidth]{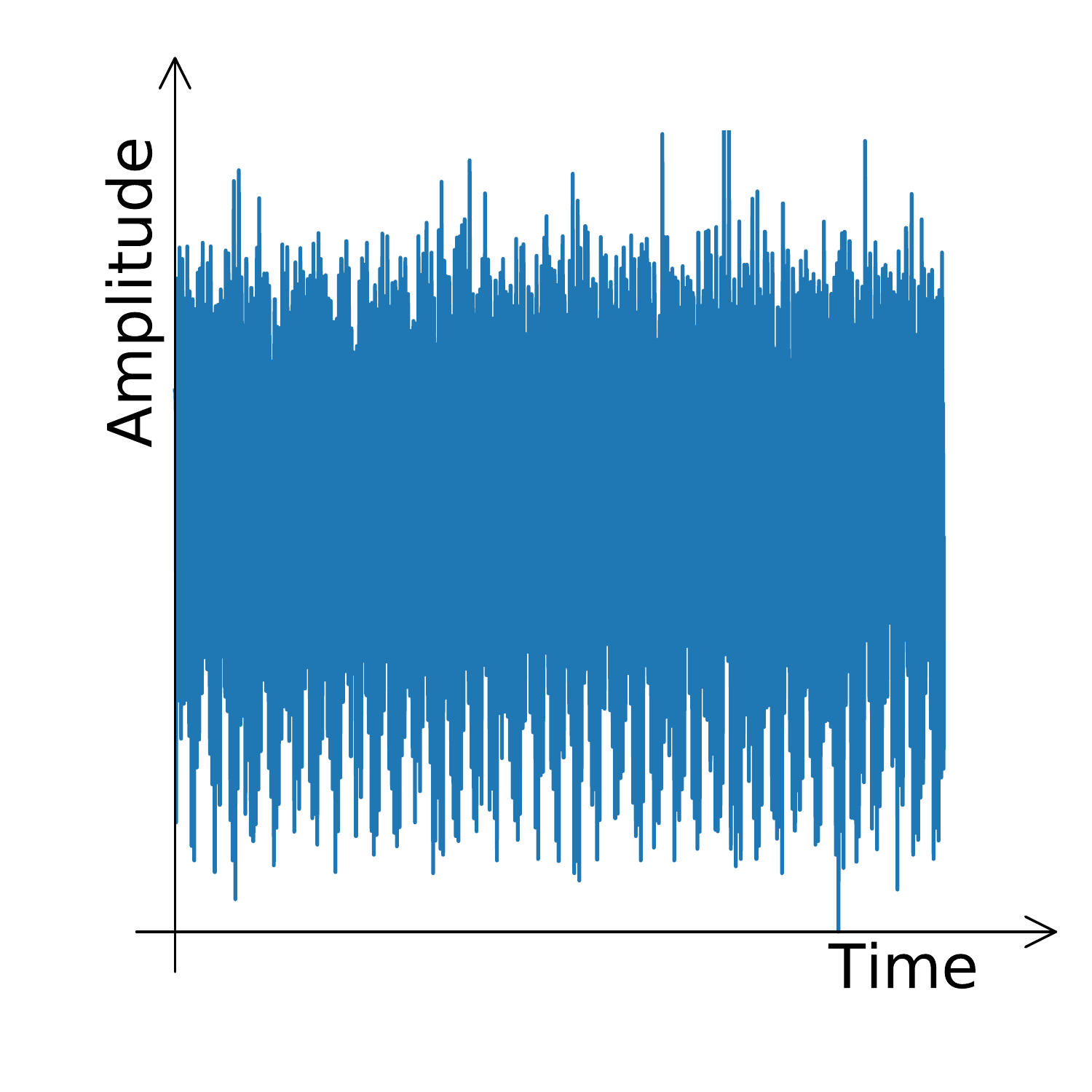}}
    \parbox{.19\textwidth}{\center\includegraphics[width=.18\textwidth]{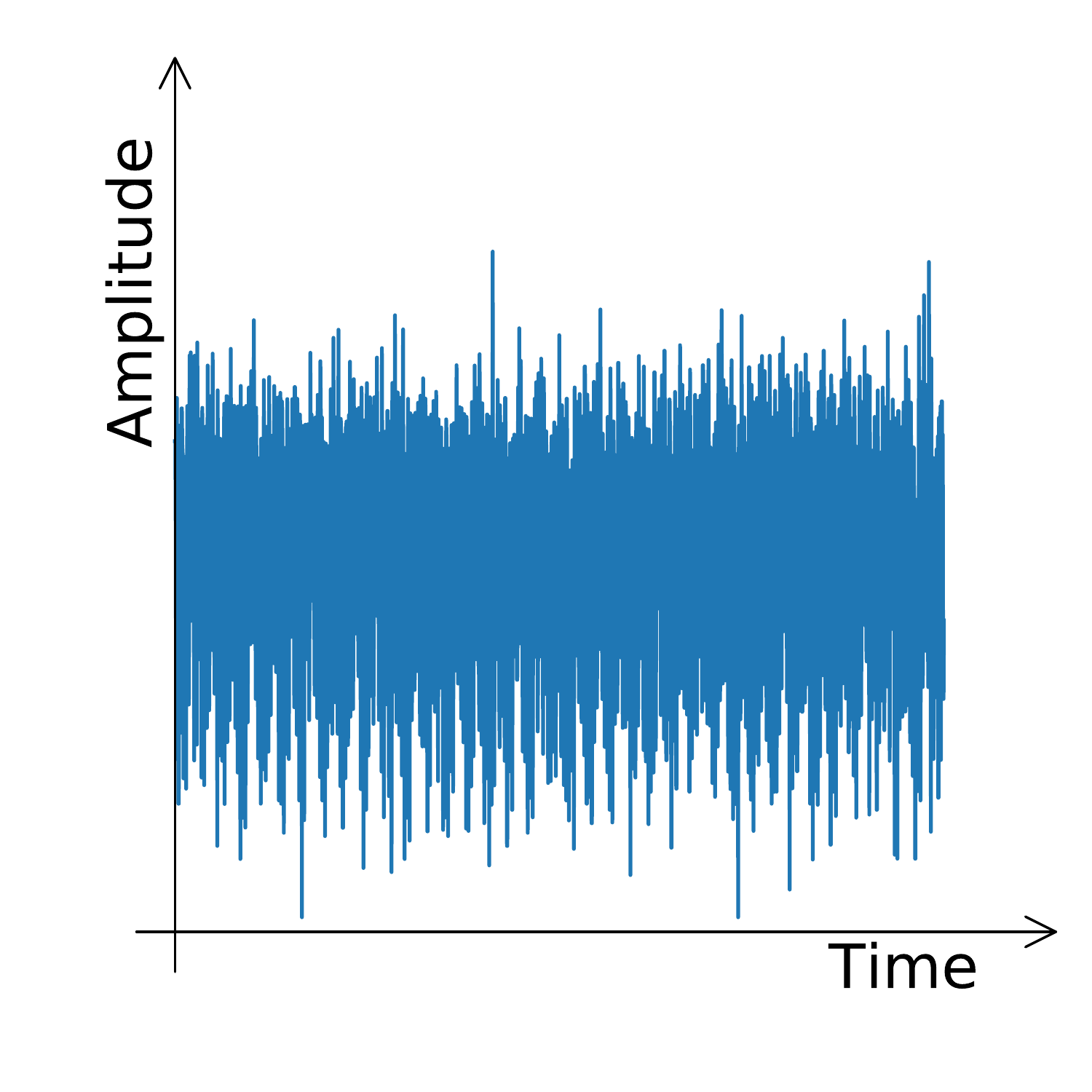}}
    \parbox{.19\textwidth}{\center\includegraphics[width=.18\textwidth]{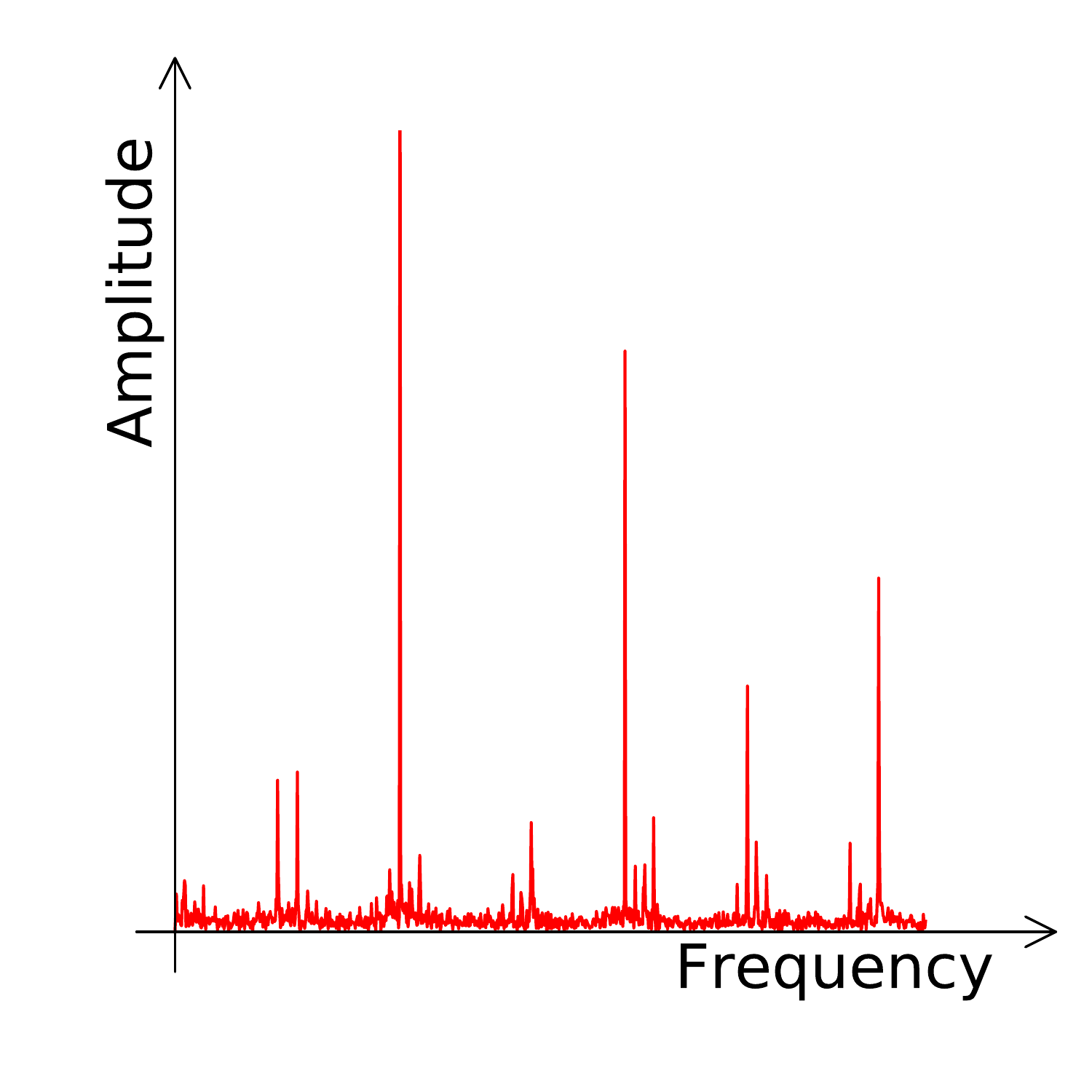}}
    \parbox{.19\textwidth}{\center\includegraphics[width=.18\textwidth]{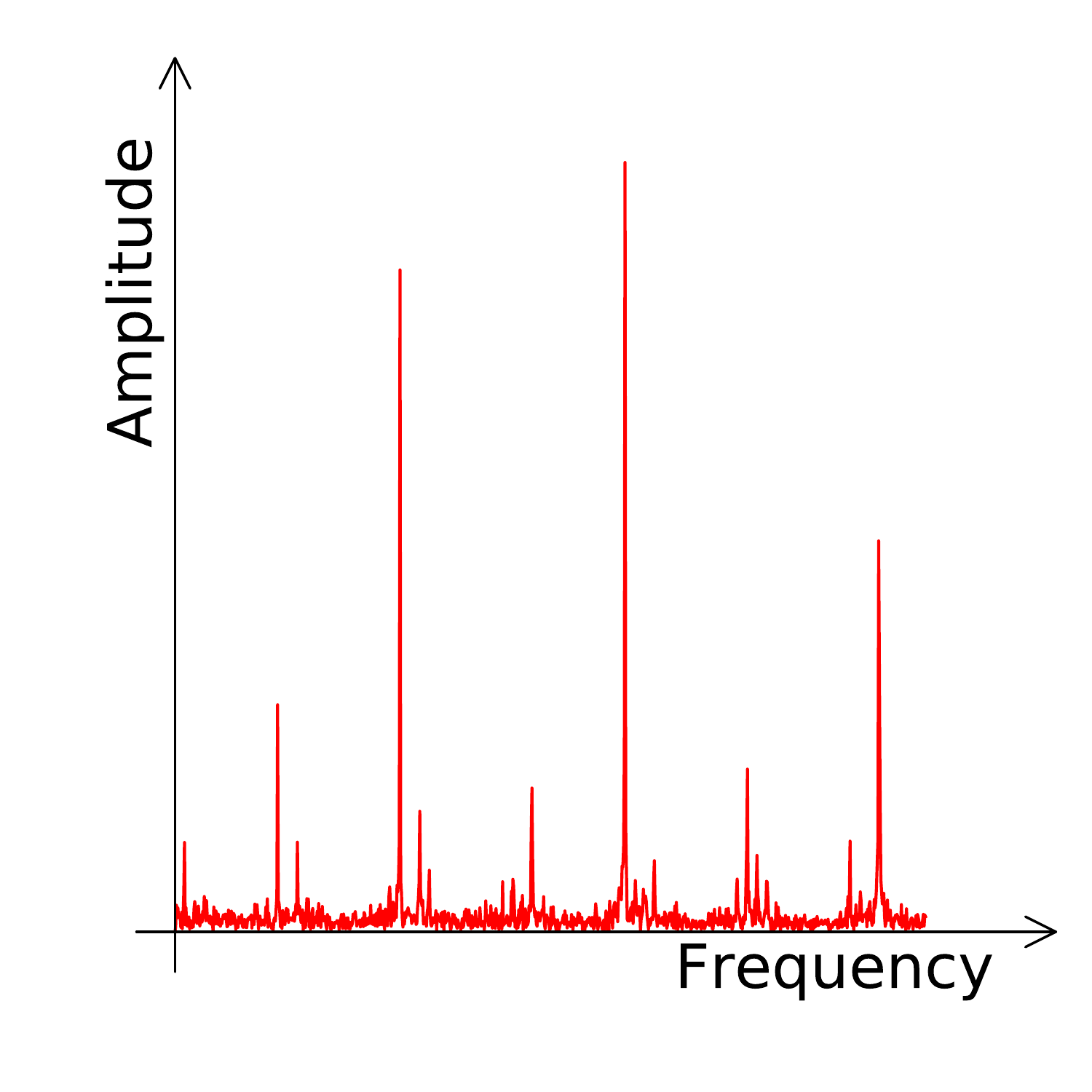}}
    \parbox{.19\textwidth}{\center\includegraphics[width=.18\textwidth]{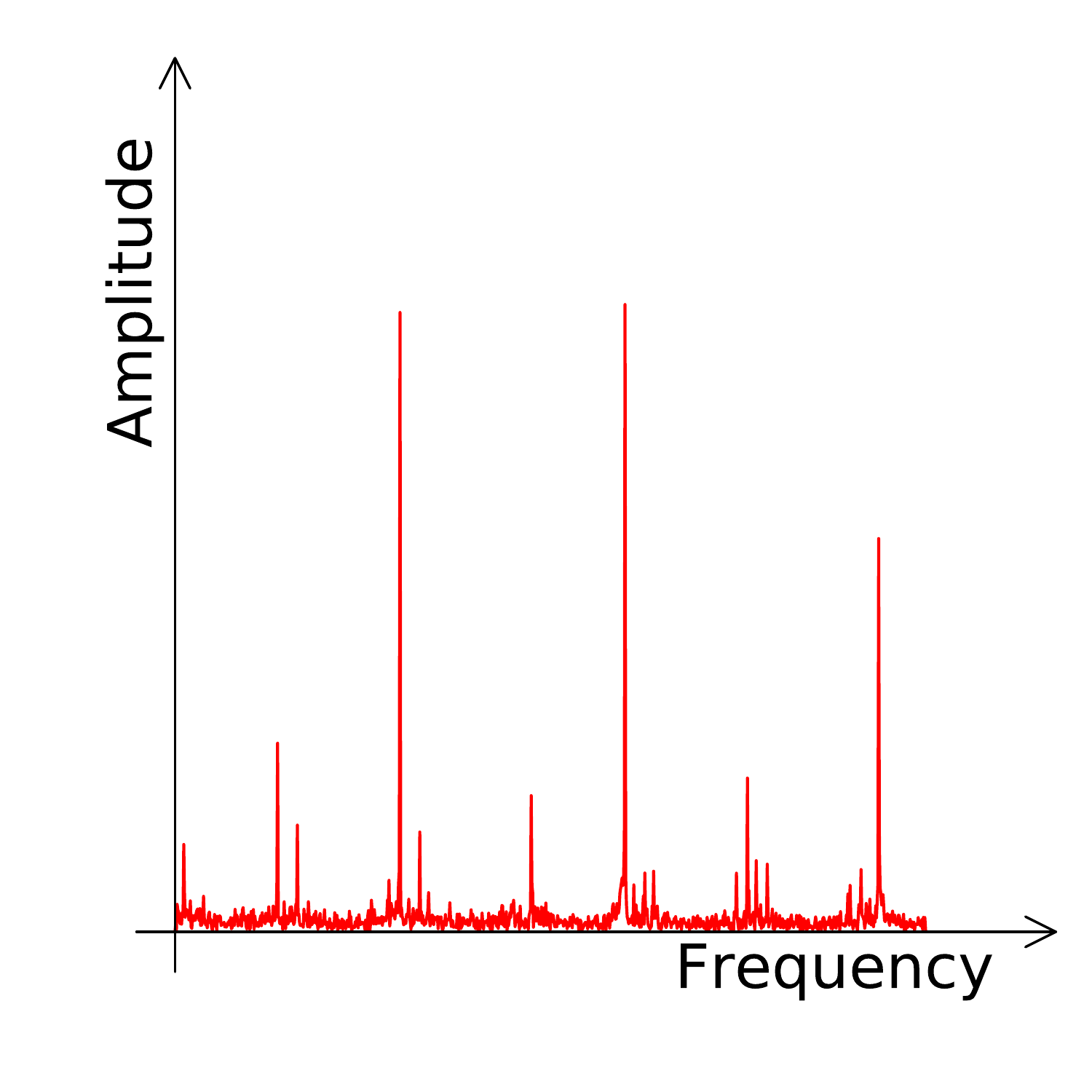}}
    \parbox{.19\textwidth}{\center\includegraphics[width=.18\textwidth]{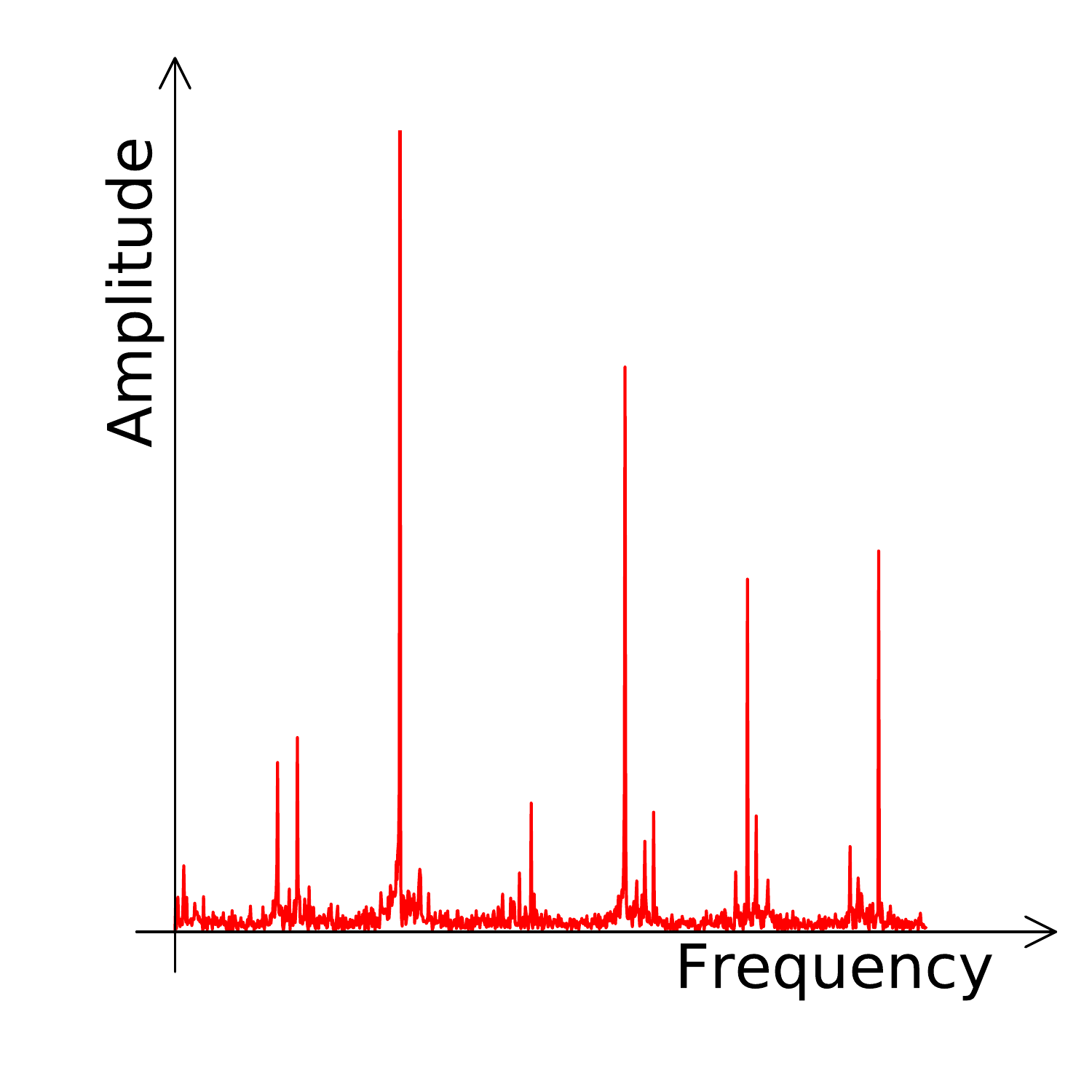}}
    \parbox{.19\textwidth}{\center\includegraphics[width=.18\textwidth]{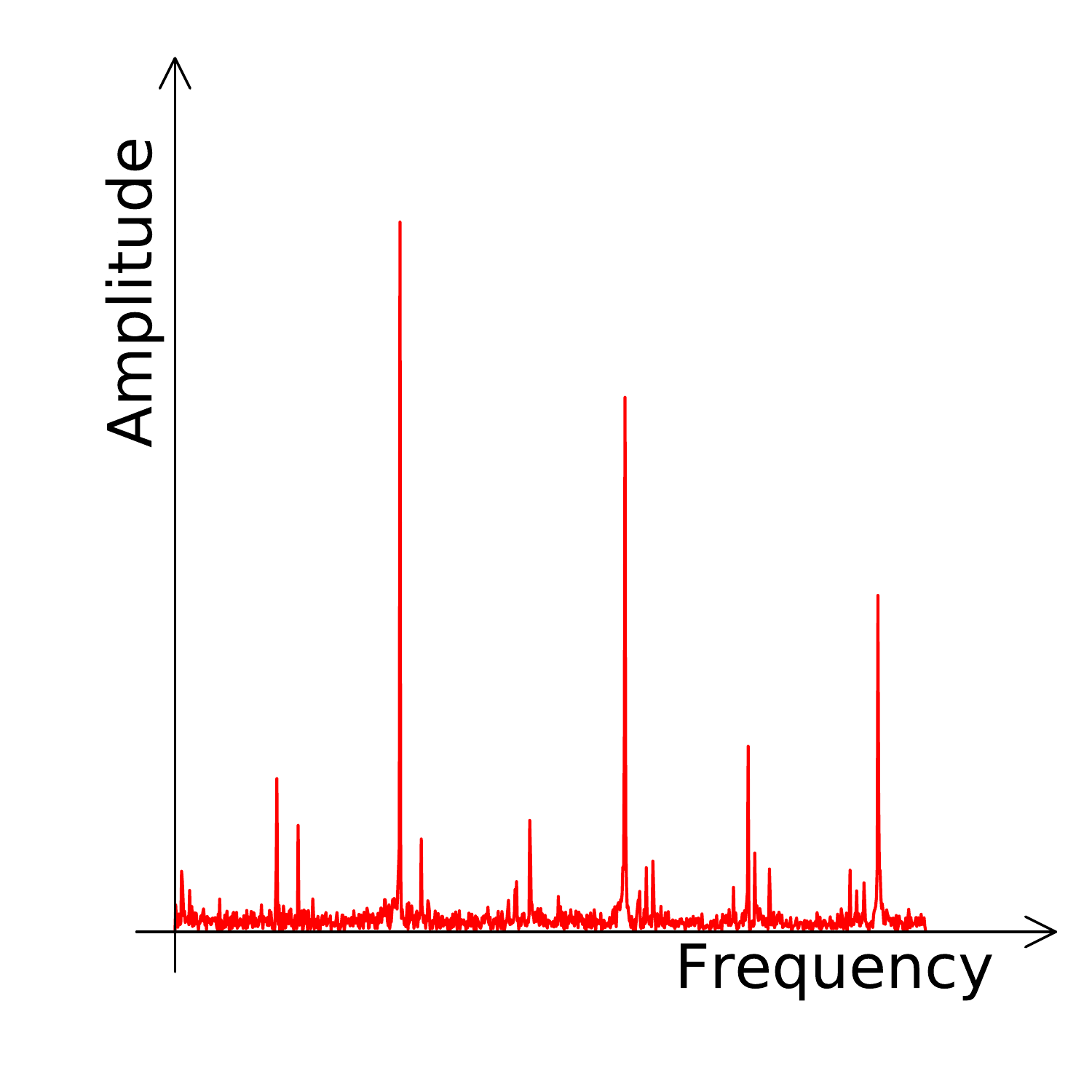}}
    \parbox{.19\columnwidth}{\center\scriptsize(b1)}
    \parbox{.19\columnwidth}{\center\scriptsize(b2)}
    \parbox{.19\columnwidth}{\center\scriptsize(b3)}
    \parbox{.19\columnwidth}{\center\scriptsize(b4)}
    \parbox{.19\columnwidth}{\center\scriptsize(b5)}
  \caption{Vibration signals collected from different users in different gestures. The signals collected from user 1 are illustrated in the upper two rows (a1)$\sim$(a5), while the ones of user 2 are demonstrated in the lower two rows (b1)$\sim$(b5). The users are in different gestures. From left to right: standing, sitting upright, sitting-and-leaning-forward, sitting-and-leaning-backward, and walking.}
  \label{fig:example}
  \end{center}
  \end{figure*}

  A human head has many unique physiological features, e.g., head size, bone shape, facial structure, ratio of fat, etc., and these features can exert impact on the propagation of vibration signals which can be exploited to distinguish users. Furthermore, behavioral biometrics of wearing smart headset could also influence the propagation of the vibration signals; therefore, an authentication mechanism for smart headset is supposed to take into account the behavioral biometrics for realizing behavior-irrelevant. An example to intuitively motivate the design of VibHead is given in Fig.~\ref{fig:example}. We mount a linear motor (which is usually used in off-the-self mobile devices, e.g., iPhone) as well as an IMU sensor (ADXL375) on the surface of a HoloLens headset~\footnote{As will be shown later, we build a prototype of VibHead through add-on motors and IMU sensors instead of build-in ones, since the current smart headset products available in market have no build-in motors and the interface to IMU sensors is usually closed. Fortunately, we believe that the next generation of smart headsets, e.g., PlayStation VR2~\cite{SONY2022}, will have build-in motor with vibrations and brings users for better real-world experience.}. The motor generates vibration at a frequency of $300$Hz, and the sampling frequency of the IMU sensor is $2$kHz. We consider five typical poses in which people are authenticated, i.e., standing, sitting upright, sitting-and-leaning-forward, sitting-and-leaning-backward, and walking. Fig.~\ref{fig:example} shows the vibration signals collected from two different users wearing our device in five different gestures. By comparing the vibration signals in the time and frequency domain, it is easy to find that the signals collected from different users are quite distinct, even when the users are in the same gestures. Hence, it is possible to identify the users by exploiting the vibration response. Nevertheless, it is also illustrated that, for any particular user, its behaviors (or gestures) indeed influence the propagation of vibration; hence, we have to elaborately design recognition method for identifying different users.

\section{System Overview}  \label{sec:overview}
  In this section, we first give an overview on the system architecture of our VibHead system in Sec.~\ref{ssec:sysarch}, and then briefly introduce our threat model in Sec.~\ref{ssec:threat}.
  \subsection{System Architecture} \label{ssec:sysarch}
\begin{figure}[htb!]
      \centering
      \includegraphics[width=0.8\columnwidth]{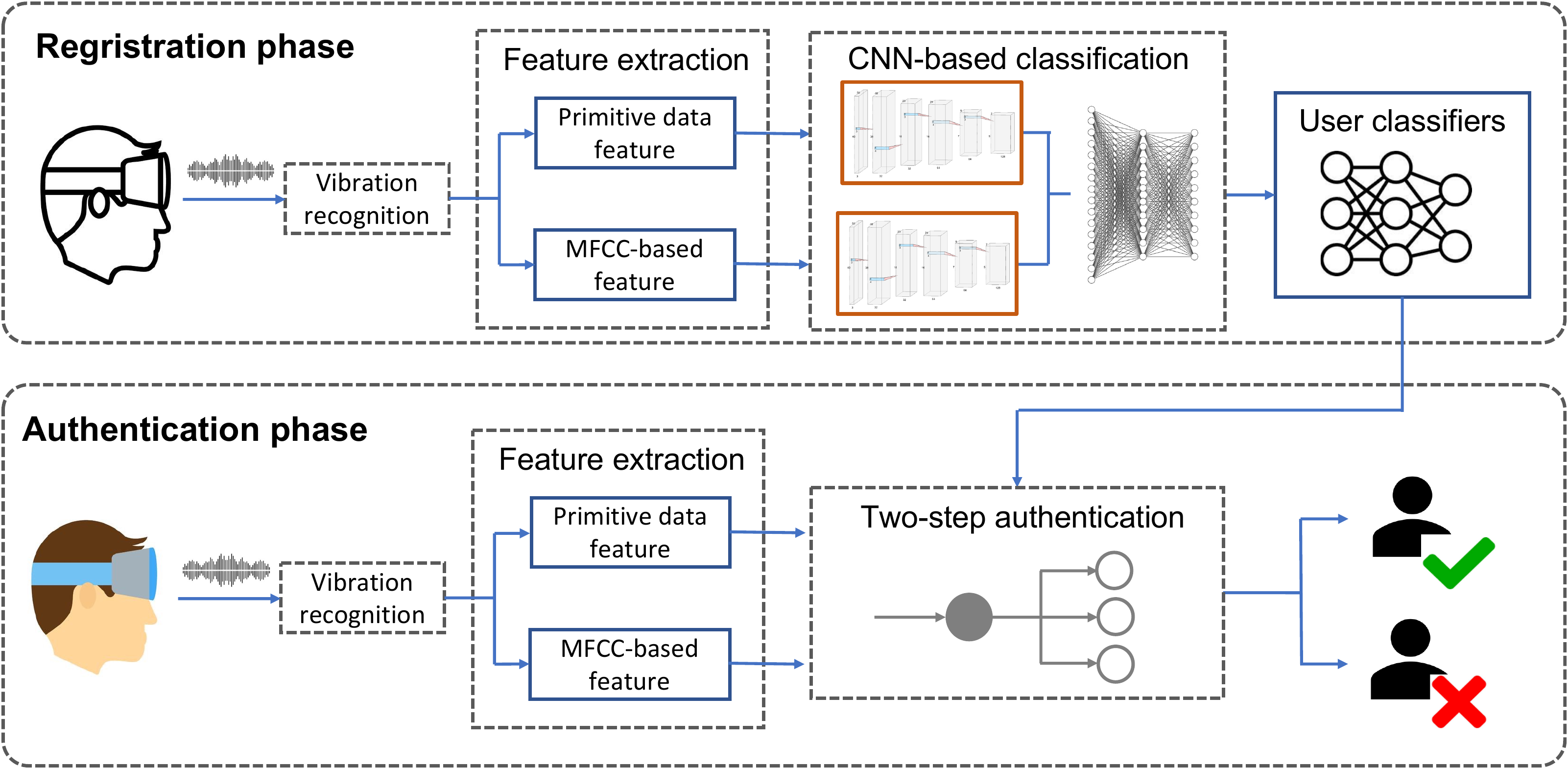}
      \caption{System architecture of VibHead.}
    \label{fig:sysarch}
    \end{figure}

    We design an user authentication system for smart headsets, \textit{VibHead}, aiming at identifying different individuals through learning from vibration signals. The pipeline of our VibHead system is illustrate in Fig.~\ref{fig:sysarch}. Specifically, our VibHead system consists of two phase, i.e., registration phase and authentication phase.

    In the registration phase, a set of classifiers is trained based on collected vibration signals from a group of default legitimate users. In particular, once an user wearing the smart headset requests a registration service through VibHead, we first let the vibration motor mounted on the headset generate a short-piece of vibration signals. Then, the vibration signals propagate via the head, and can be sampled and recognized by IMU sensors. We consider five typical poses the users usually take during authentication, including sitting upright, sitting-and-leaning-forward, sitting-and-leaning-backward, standing and walking \footnote{Although people may take a large variety of gestures in the daily use of smart headsets, the poses which they usually take when being authenticated are relatively limited.}. Hence, our VibHead system needs to guide the users to take the above different gestures in the data collection process. The collected primitive IMU data samples are then imported into a specific-designed CNN-based classification module to train a set of user classifiers. To improve the classification accuracy and accelerate the training process, we additionally extract the MFCC-based features from the primitive data samples. We will show the efficacy of combining the two types of features later in Sec.~\ref{sec:classification}.

    In the authentication phase, when the authentication service is triggered by an unknown login attempt, the vibration signals are generated. The signals propagates through the login user's head and are sampled by the IMU sensors. Then, the features are extracted in the same way as in the registration phase. Based on the user classifiers trained in the registration phase, we design a two-step authentication scheme, through which we make a decision on whether the login user is legitimate or not.

  \subsection{Threat Model} \label{ssec:threat}
    We suppose that the goal of attackers (or illegitimate users) is to bypass the user authentication system of a target smart headset. If the attackers successfully login the smart headset, they can steal private information. We assume that the attackers know how the smart headset authenticates its login users but cannot modify or manipulate the software or data of the headset. We also suppose that, the attackers cannot attack during the registration phase, which is usually performed when the legitimate users first initialize the smart headset.

    We consider impersonation attack in this paper. The attackers attempt to bypass the authentication by pretending to be legitimate users. Specifically, the attackers have observed the registration process of the legitimate users, including how the legitimate users wear the target smart headset and take the corresponding gestures for authentication (including standing, sitting upright, sitting-and-leaning-forward, sitting-and-leaning-backward, and walking). The attackers then wear the target smart headset and start the authentication. They may mimic the legitimate users' different gestures to deceive the authentication system to grant them authority.

\section{Vibration Collection and Feature Extraction} \label{sec:vibfeature}

  \subsection{Vibration Generation and Collection}
\begin{figure}[htp]
      \centering
      \includegraphics[width=0.6\columnwidth]{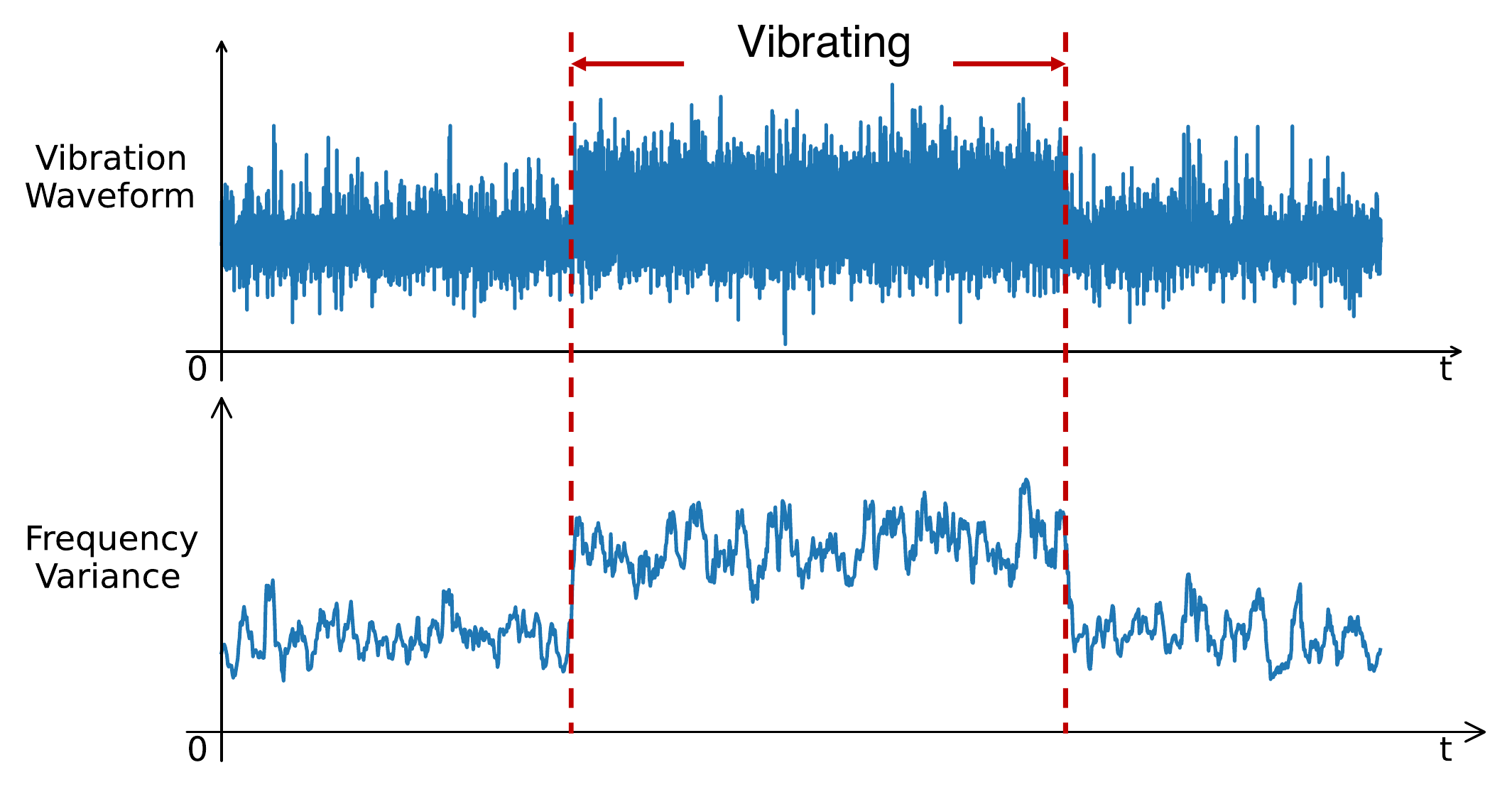}
      \caption{Signal Synchronization. The vibration is sampled by an IMU sensor at a frequency of $2$~kHz (as shown in the upper row), and the frequency variance is calculated by a $50$~ms window with a step size of $5$~ms (as illustrated in the lower row).}
    \label{fig:waveform}
    \end{figure}

    In our VibHead system, an user is asked to wear the smart headset and we use a linear motor to generate vibrations at a frequency of $300$~Hz. The wave travels through the headset and the user's head, and then can be detected by an IMU sensor (specifically the build-in three-axes accelerometer). Considering the propagation latency of the vibration signals, we need to synchronize the motor and the IMU sensor such that the vibration signals can be properly captured. In our VibHead system, we adopt a sliding window to serve the above goal. For example, as shown in Fig.~\ref{fig:waveform}, for a piece of signals collected by the accelerometer (at a frequency of $2$~kHz) where the middle part is the impulse of vibration, we use a $50$~ms sliding window shifting $5$~ms each time. It is apparent that we can recognize the impulse of vibration according to the jump and drop of the frequency variance. It is worthy to note that, the time length of the vibration could influence the classification/authentication performance of VibHead. As will be shown later in Sec.~\ref{sec:classification} and Sec.~\ref{sec:authen}, we usually set the time length of the vibration, i.e. $T$,  to be $400 \sim 1000$ ms. 

  \subsection{Feature Extraction}
    As mentioned in Sec.~\ref{ssec:sysarch}, both primitive data features and MFCC-based features are utilized for classifier training. On one hand, we direct utilize the raw data sample collected by the accelerometer of the IMU sensors to train the classifiers. Each raw data sample (or primitive data feature) can be represented by a $3$-tuple which consists of the readings on the three axes of the accelerometer.  On the other hand, to further improve the efficacy of our user classification, MFCCs, which has been widely used to represent the short-term power spectrum of acoustic signals~\cite{MurtyY-SPL06} effectively, are also exploited for the classifier training. Its mel-scale filter banks not only are optimized for human speech and perception frequency, but also can be utilized to characterize the vibration propagation through the medium of a human head, which is influenced by the human's behavioral and physiological characteristics.

\begin{figure*}[htb!]
    \begin{center}
      \parbox{.32\textwidth}{\center\includegraphics[width=.32\textwidth]{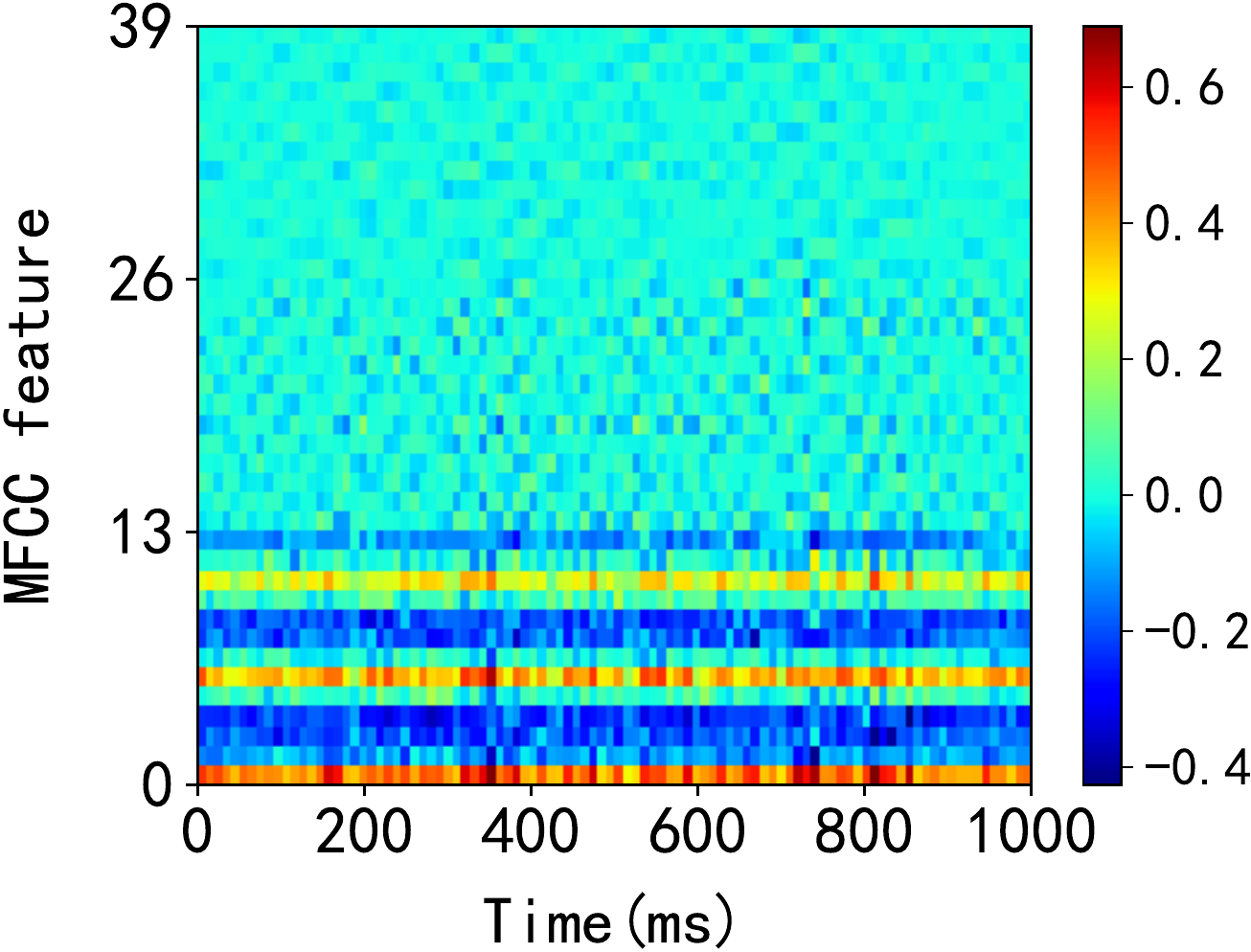}}
      \parbox{.32\textwidth}{\center\includegraphics[width=.32\textwidth]{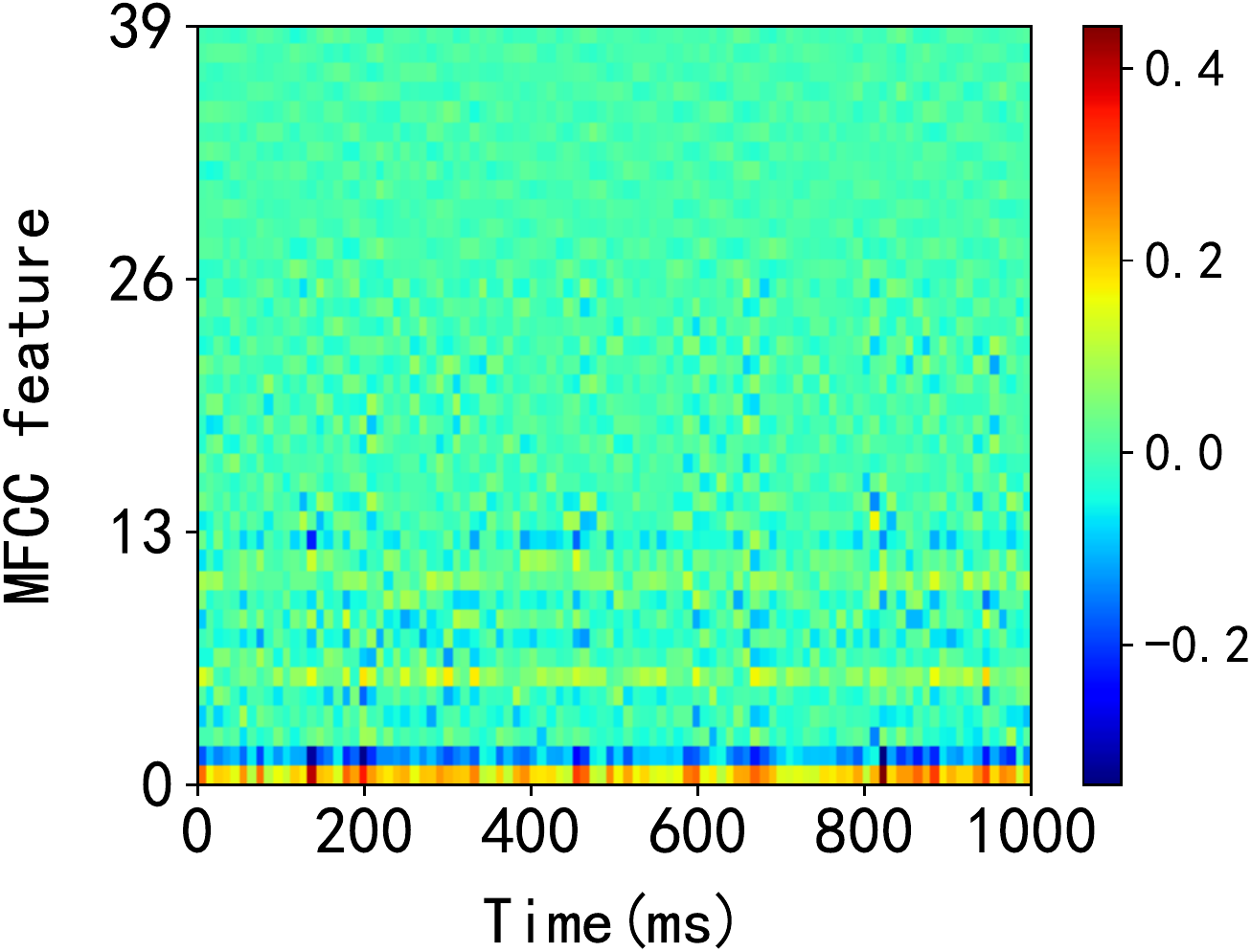}}
      \parbox{.32\textwidth}{\center\includegraphics[width=.32\textwidth]{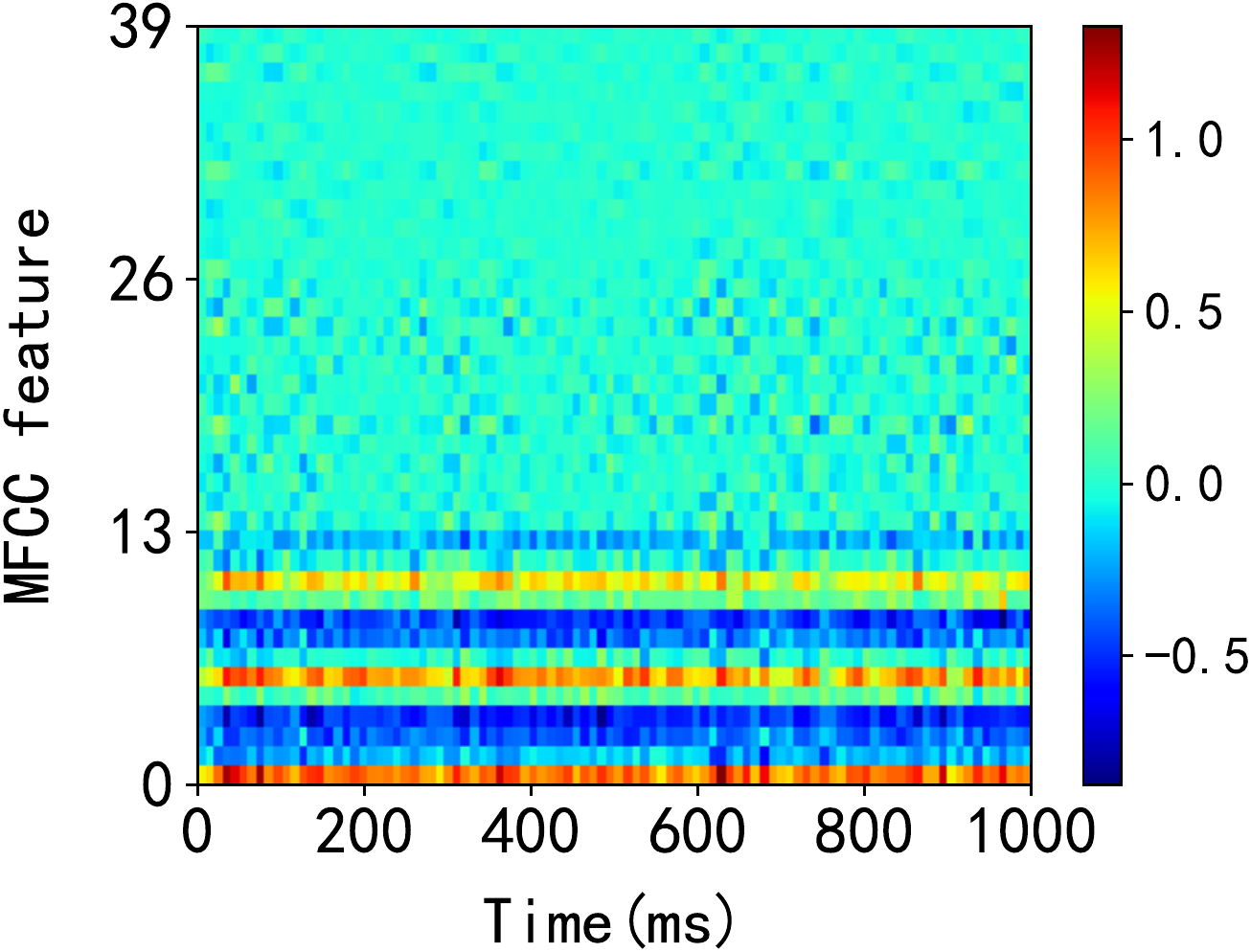}}
      \parbox{.32\columnwidth}{\center\scriptsize(a1) User 1 (x-axis)}
      \parbox{.32\columnwidth}{\center\scriptsize(a2) User 1 (y-axis)}
      \parbox{.32\columnwidth}{\center\scriptsize(a3) User 1 (z-axis)}
      \parbox{.32\textwidth}{\center\includegraphics[width=.32\textwidth]{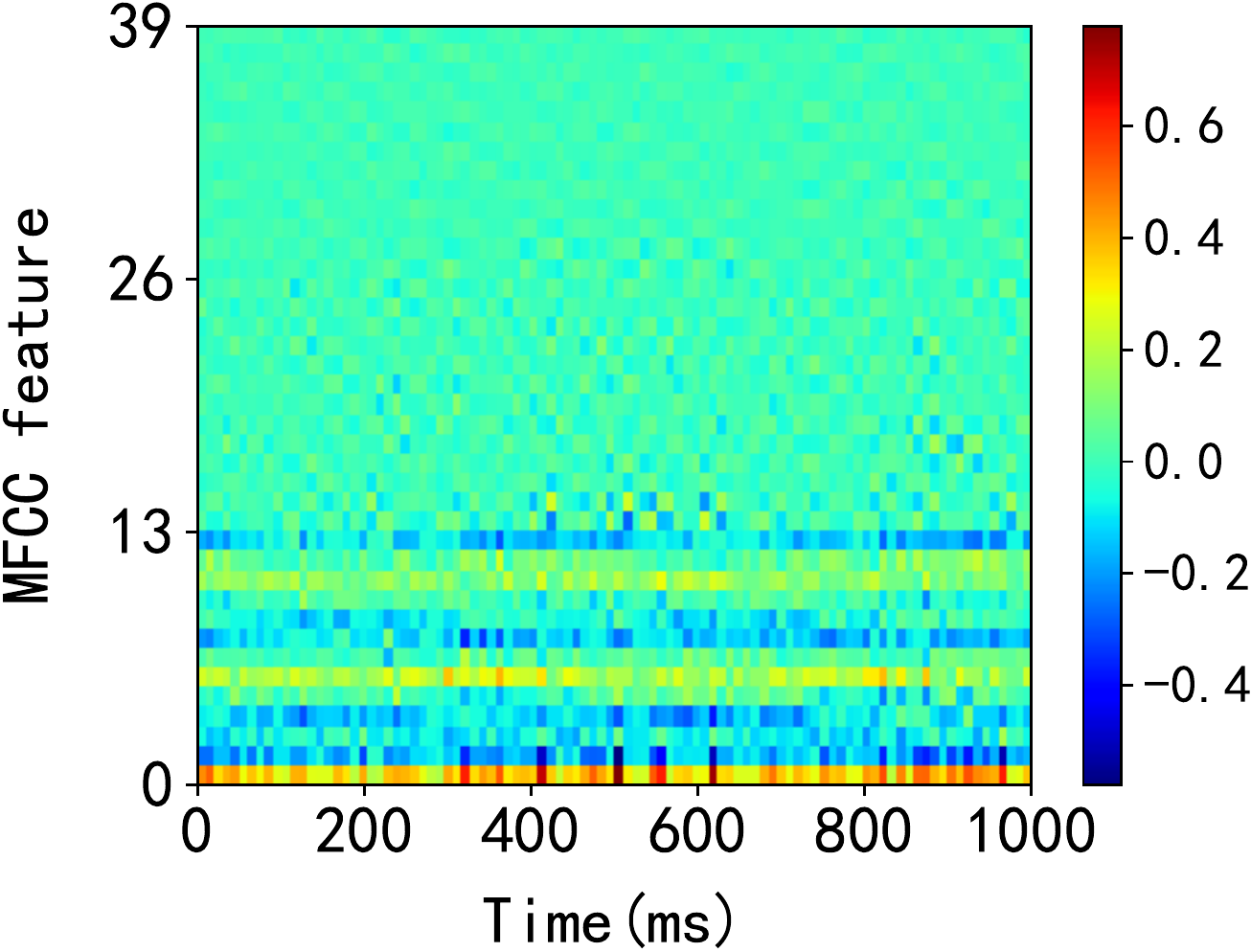}}
      \parbox{.32\textwidth}{\center\includegraphics[width=.32\textwidth]{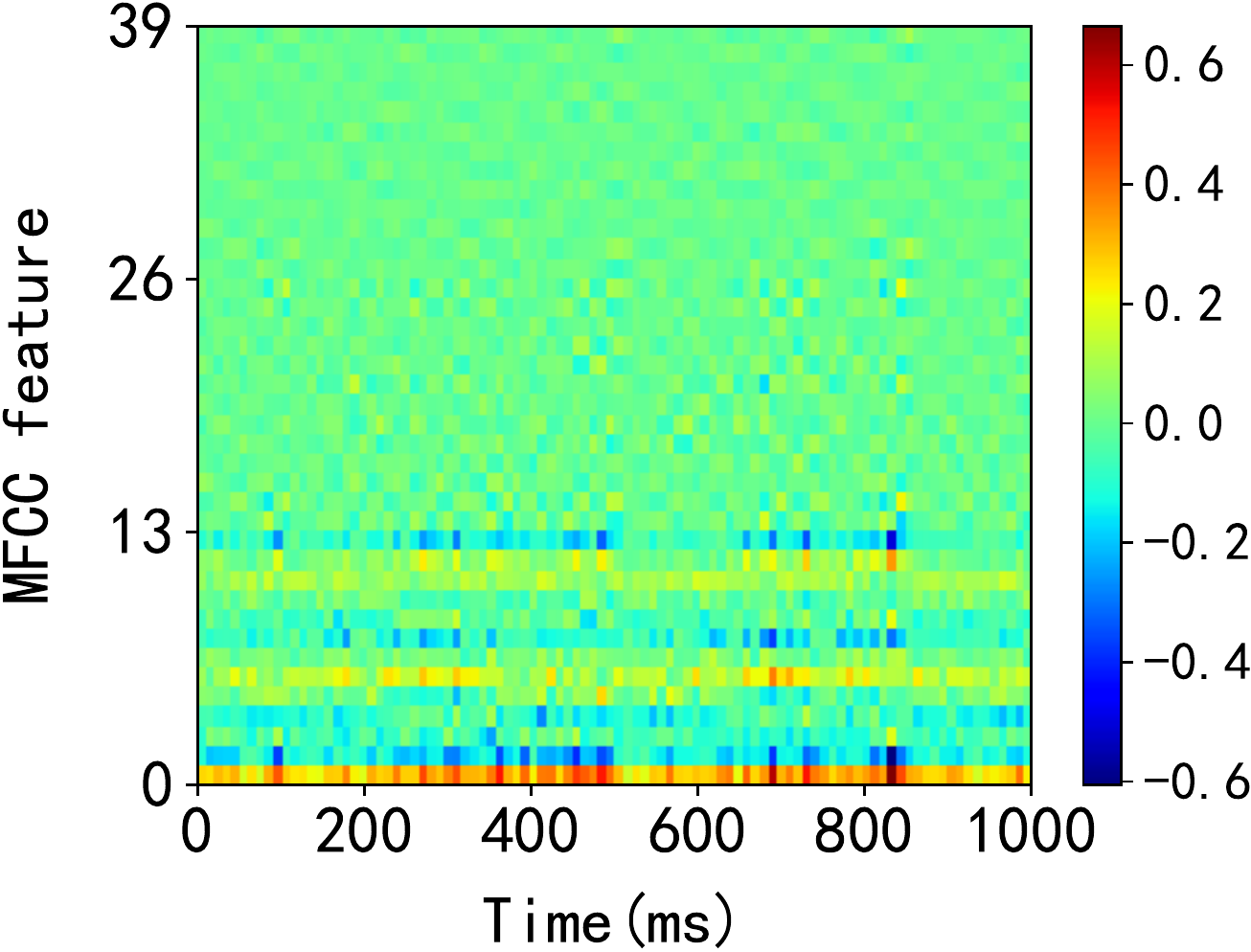}}
      \parbox{.32\textwidth}{\center\includegraphics[width=.32\textwidth]{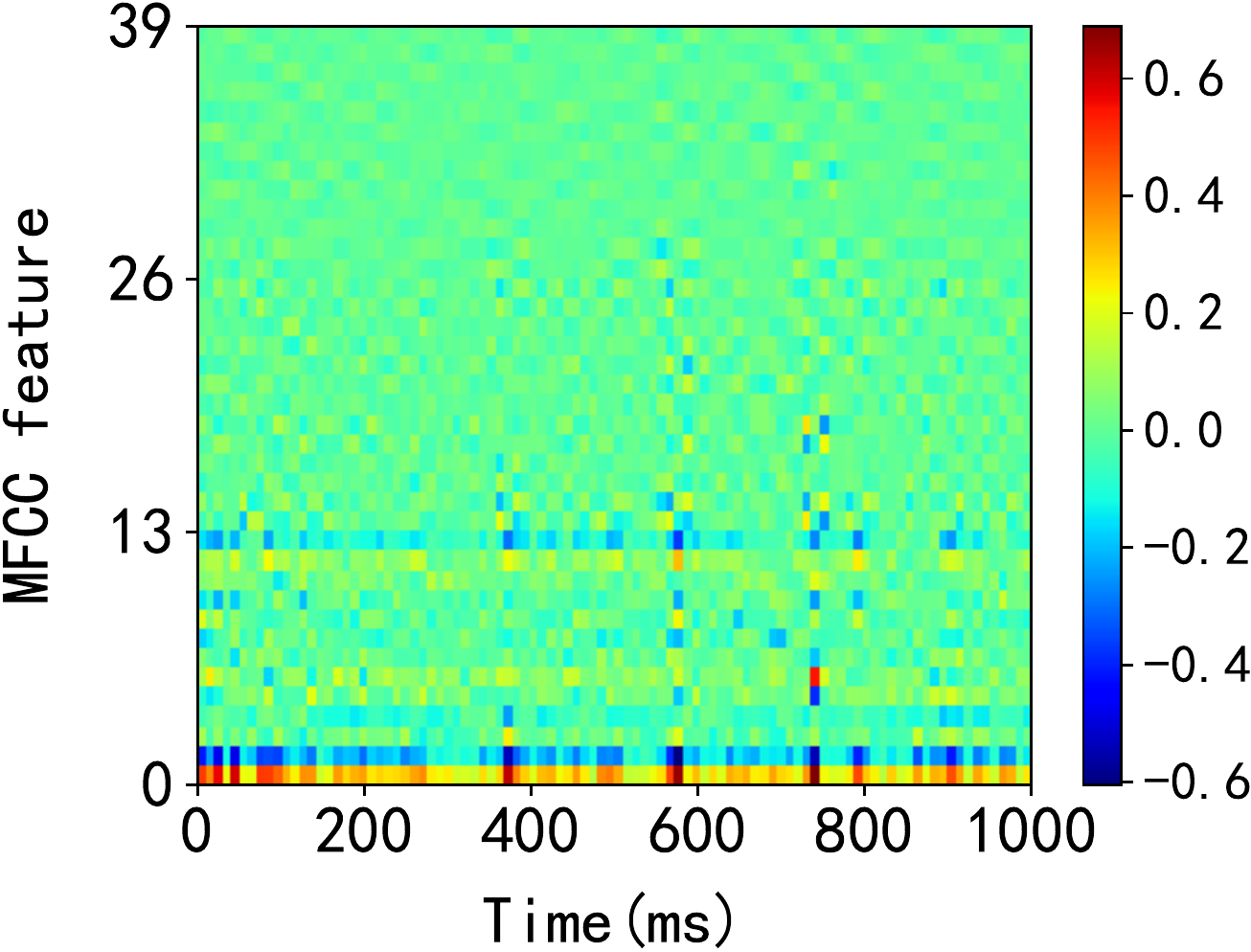}}
      \parbox{.32\columnwidth}{\center\scriptsize(b1) User 1 (x-axis)}
      \parbox{.32\columnwidth}{\center\scriptsize(b2) User 2 (y-axis)}
      \parbox{.32\columnwidth}{\center\scriptsize(b3) User 3 (z-axis)}
      \parbox{.32\textwidth}{\center\includegraphics[width=.32\textwidth]{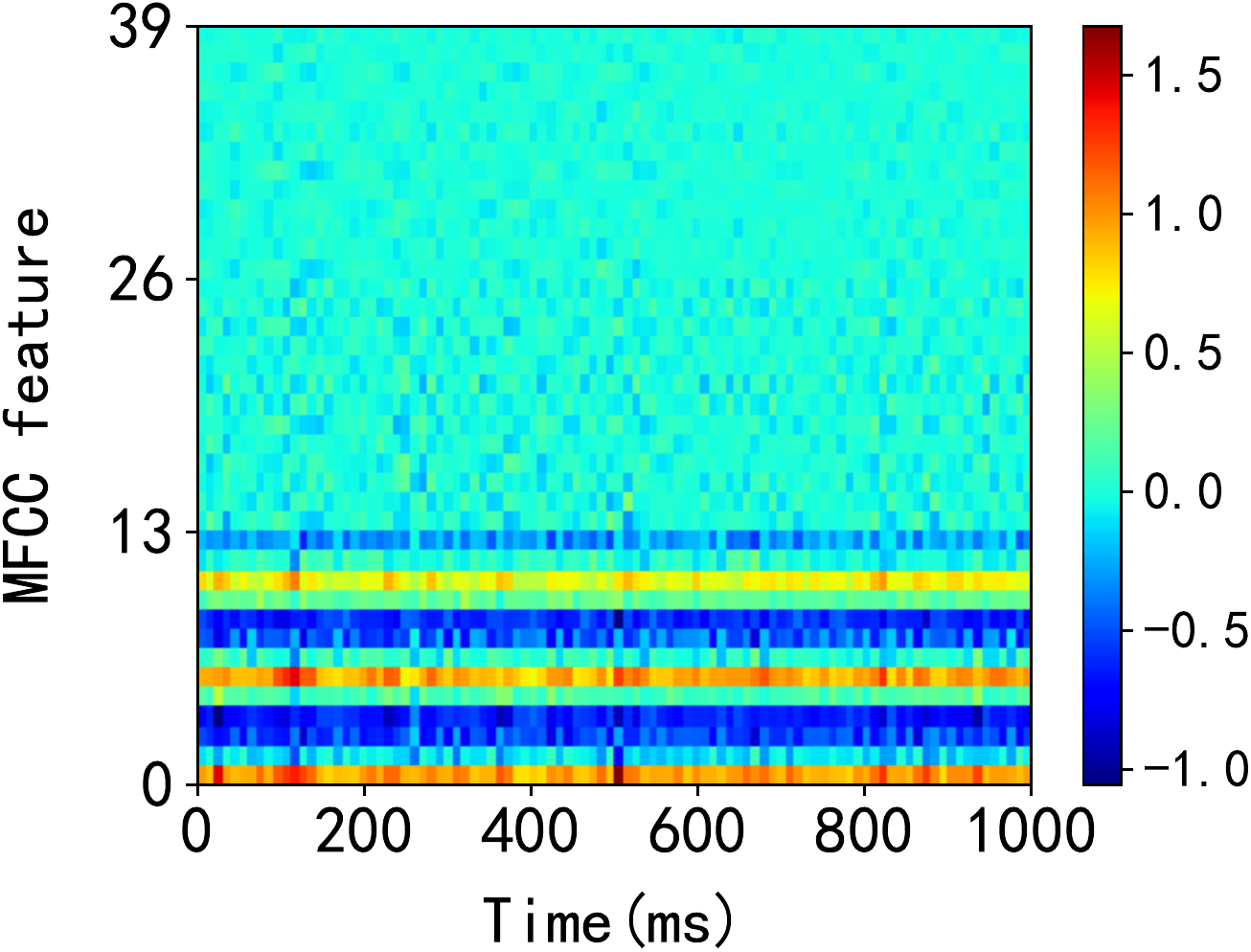}}
      \parbox{.32\textwidth}{\center\includegraphics[width=.32\textwidth]{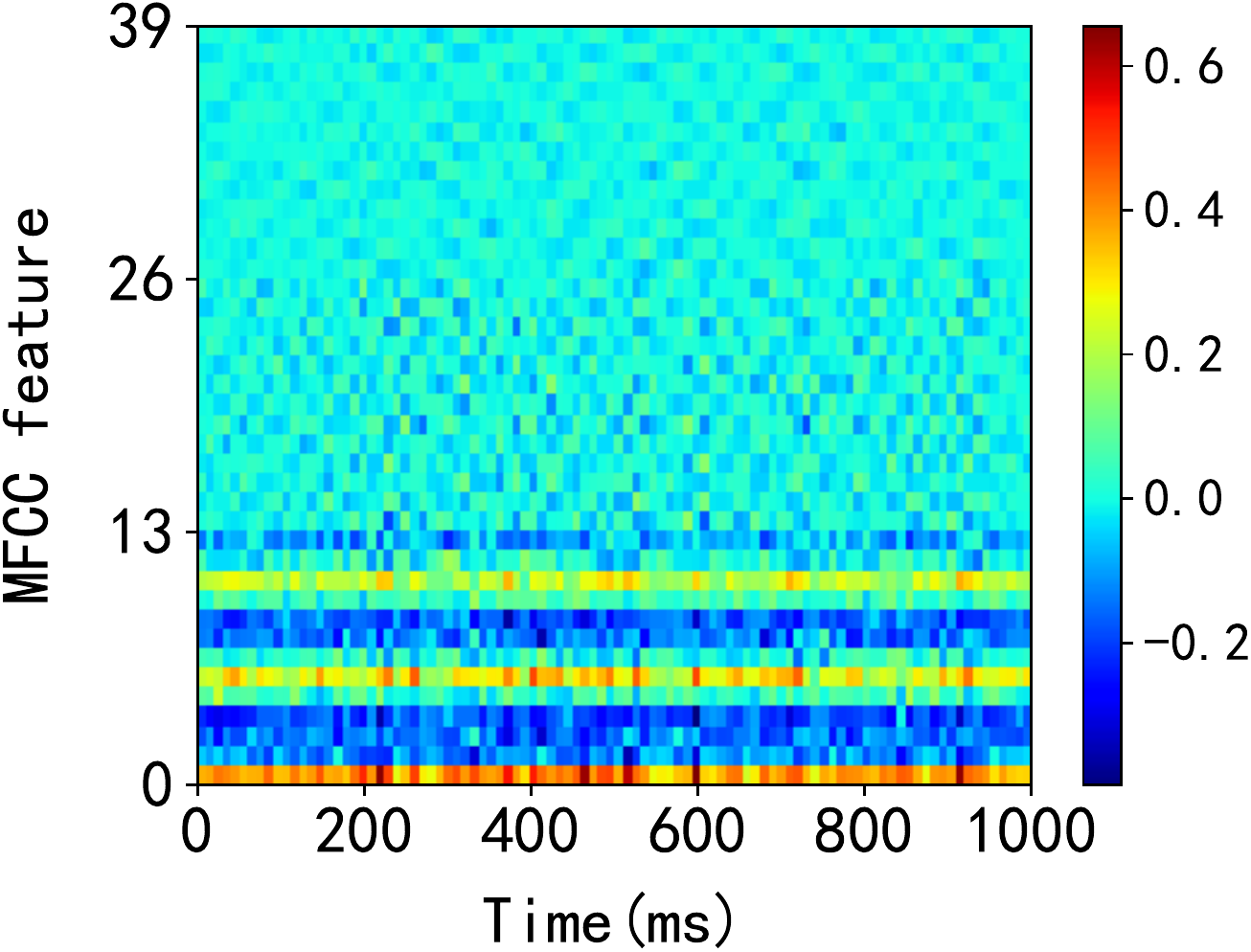}}
      \parbox{.32\textwidth}{\center\includegraphics[width=.32\textwidth]{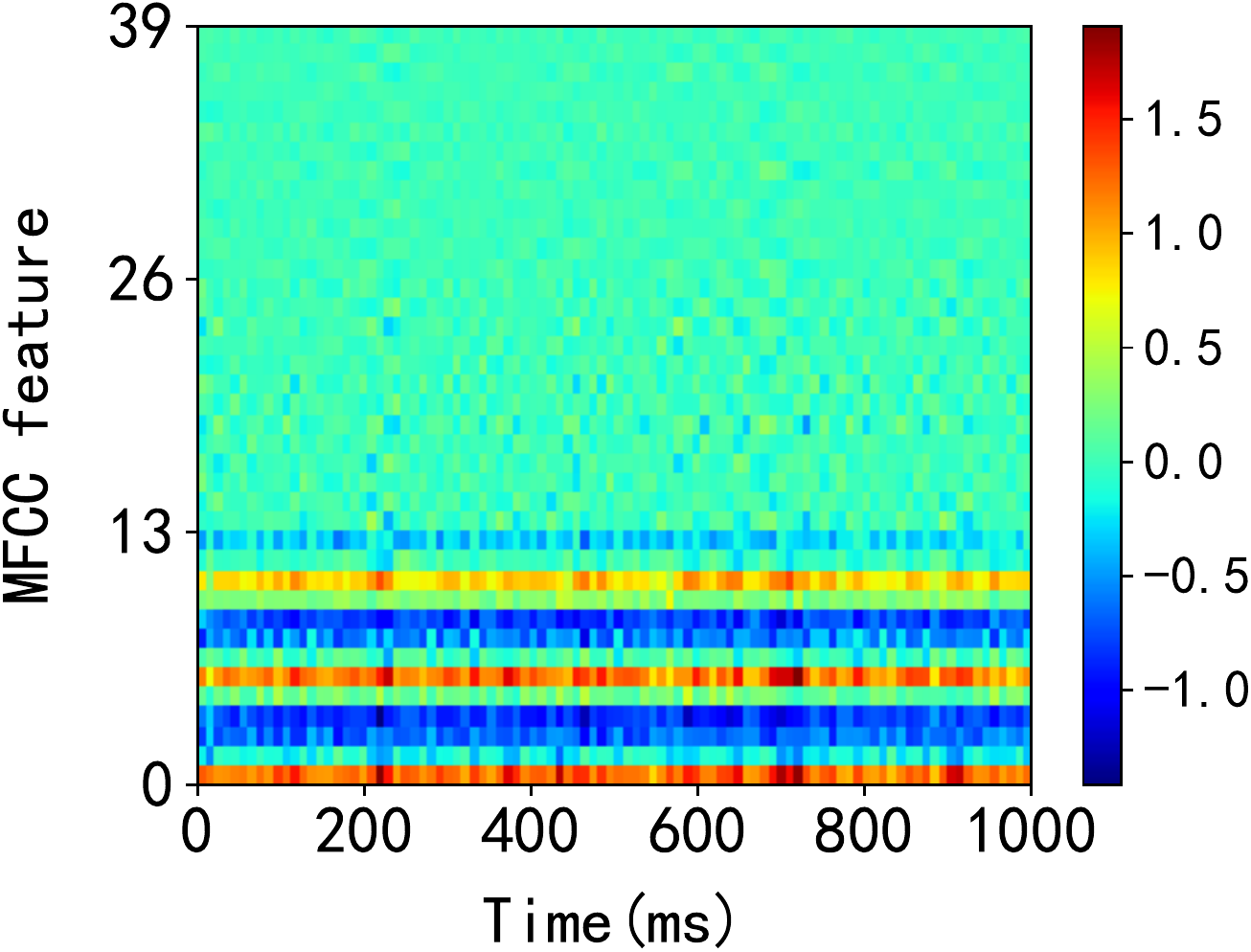}}
      \parbox{.32\columnwidth}{\center\scriptsize(c1) User 3 (x-axis)}
      \parbox{.32\columnwidth}{\center\scriptsize(c2) User 3 (y-axis)}
      \parbox{.32\columnwidth}{\center\scriptsize(c3) User 3 (z-axis)}
    \caption{MFCC-based features of three different users in the gesture of standing. The motor generates vibration at a frequency of $300$Hz, while the IMU sensor samples the vibration signal at a frequency of $2$kHz. The time length of the vibration signals is $1$s.}
    \label{fig:mfcc}
    \end{center}
    \end{figure*}

    Specifically, we extract the MFCC-based features from the (raw) vibration signals collected by the accelerometer. We adopt a $25$~ms Hamming window shifting $10$~ms each time for signal framing. The number of filterbank channels is set to $40$. For each window, a $39$-dimensional MFCC-based feature (consisting of the first 12 cepstral coefficients, the frame energy coefficient, and the corresponding deltas and delta-deltas) is calculated. In Fig.~\ref{fig:mfcc}, we illustrate the MFCC-based features extracted from the vibration signals collected from two different users. Since the vibration signals are collected by a three-axes accelerometer, we extract the MFCC-based features for the three axes, respectively. It is found that, even for different users in the same gesture, their MFCC-based features are distinguishable.

\section{User Classification}  \label{sec:classification}
\begin{figure}[htb!]
  \centering
    \includegraphics[width=0.8\columnwidth]{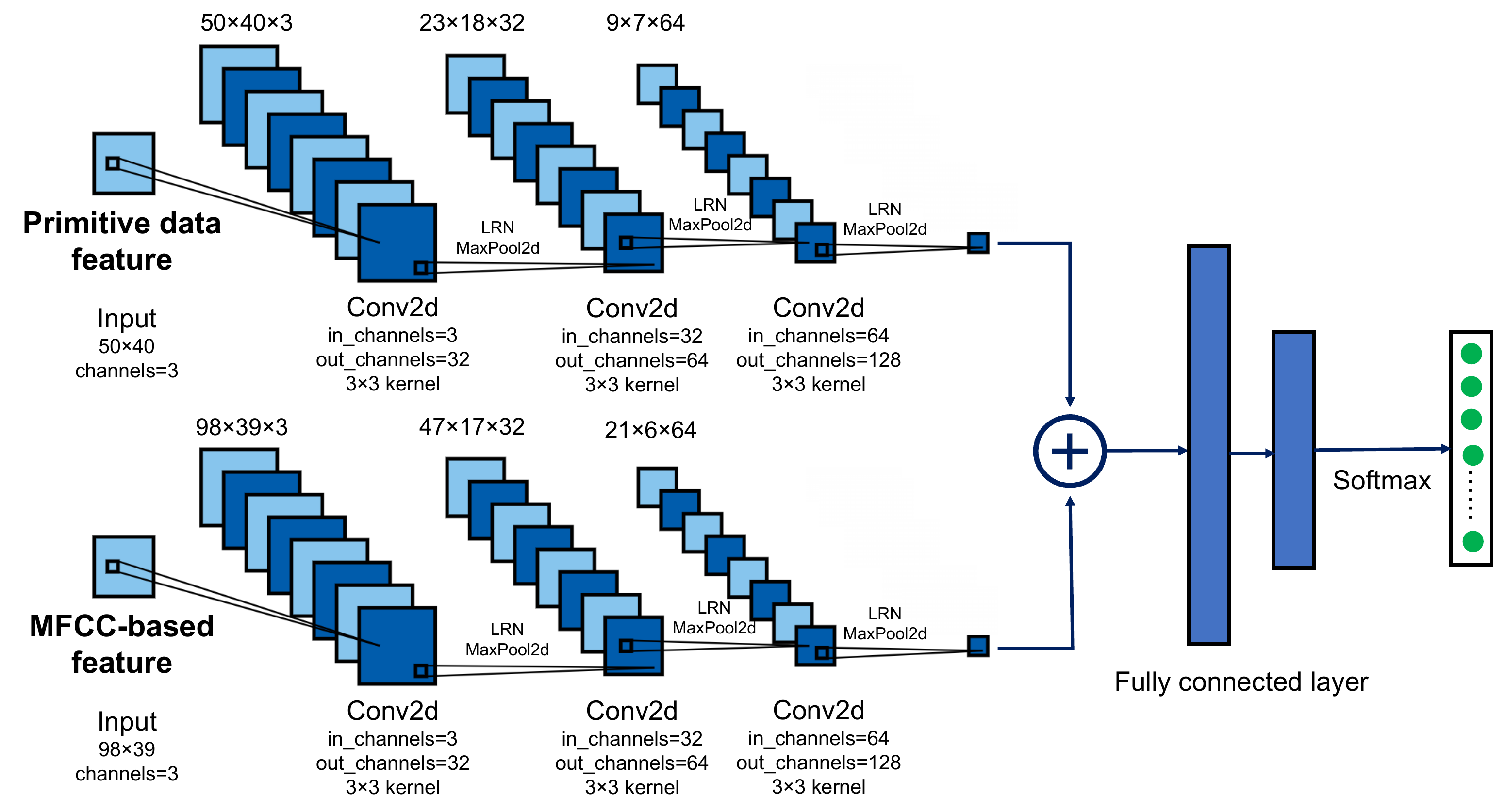}
  \caption{CNN-based user classification in VibHead. We assume that the time length of the input data samples is $1$s and the size of the primitive data feature and the one of the MFCC-based feature are $50 \times 40$ and $98 \times 39$, respectively.}
  \label{fig:cnn}
  \end{figure}

  Our VibHead system applies a CNN-based model for user classification (see Fig.~\ref{fig:cnn}. The primitive data feature and the MFCC-based feature extracted from the vibration signals are first imported to two separated encoders. The encoders share the same structure but have different coefficients. Specifically, each of the encoders has three convolutional layers with Relu being the activation function. The kernel size is $3 \times 3$ and strides is $1$ for each of the convolutional layers. The numbers of the out channels of the three convolutional layers are $32$, $64$ and $128$, respectively. Each convolutional layer is followed by a batch normalization operation to speed up the training convergence as well as a MaxPooling layer to deal with the over-fitting problem and the network complexity.

  The output of the two encoders are then concatenated. For example, when $T=1s$, the sizes of the outputs of the encoders are $32 \times 128 \times 2 \times 1$ and $32 \times 128 \times 8 \times 1$, respectively. We obtain a $32 \times 128 \times 10 \times 1$ matrix by concatenating in the third dimension. The resulting  concatenated matrix is finally fed to a fully connected layer with softmax as the classification function. Let $F(x; \theta)$ denote the CNN-based user classifier, where $x$ represents data sample and $\theta$ denotes the set of unknown parameters of the model. If the aim of the classifier $F$ is to distinguish $N$ users, the output of the final softmax layer, $\mathbf{v} = F(x; \theta)$, is a $K$-dimensional probability vector $\mathbf{v} = (v_1, v_2, \cdots, v_K)$, where $v_i$ denotes the probability at which we predict the data sample $x$ is ``labeled'' as user $i$. We adopt a cross-entropy function to define the loss between our predictions and the labels of training data.

\begin{figure*}[htb!]
  \begin{center}
    \parbox{.45\textwidth}{\center\includegraphics[width=.4\textwidth]{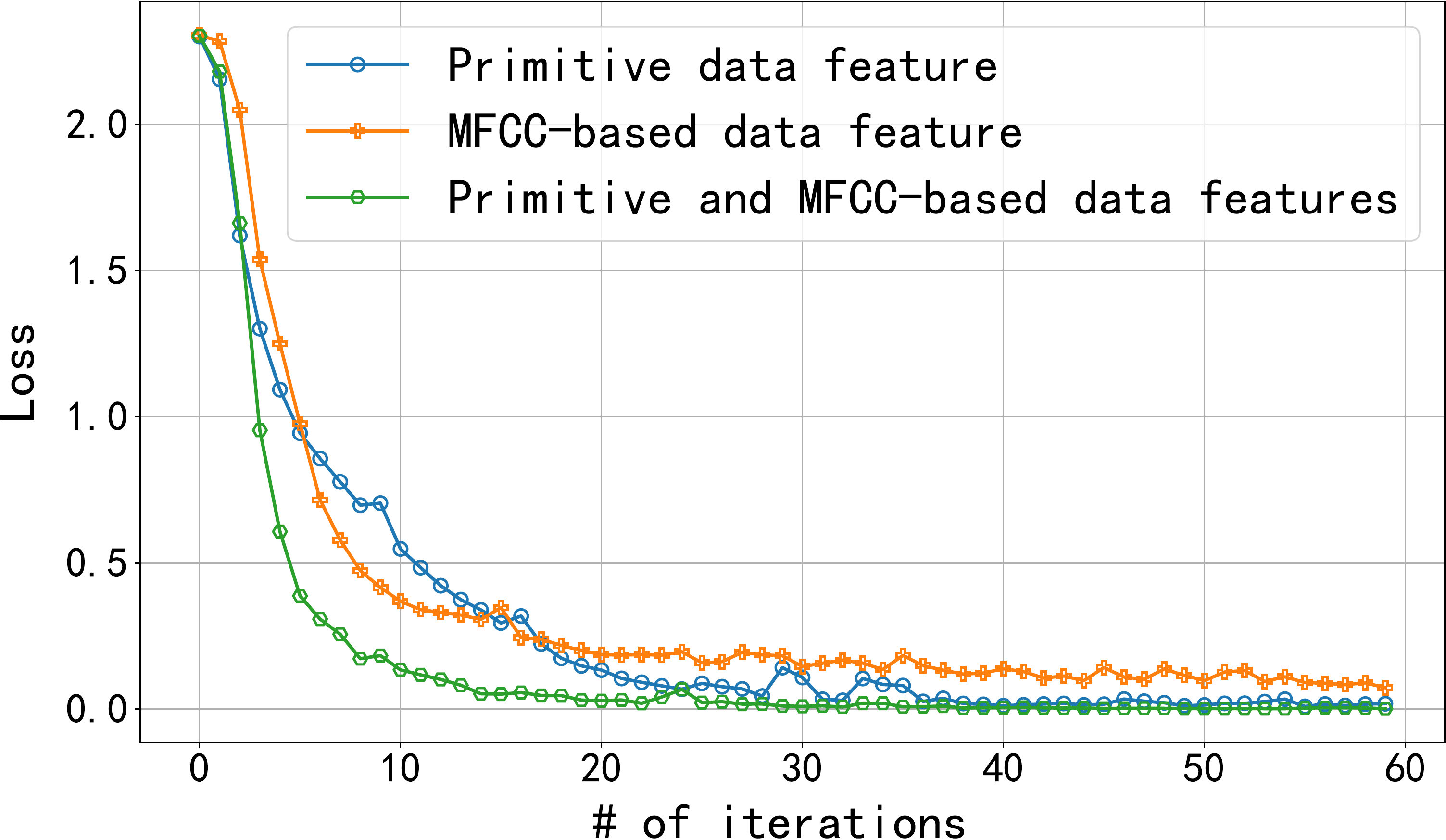}}
    \parbox{.45\textwidth}{\center\includegraphics[width=.4\textwidth]{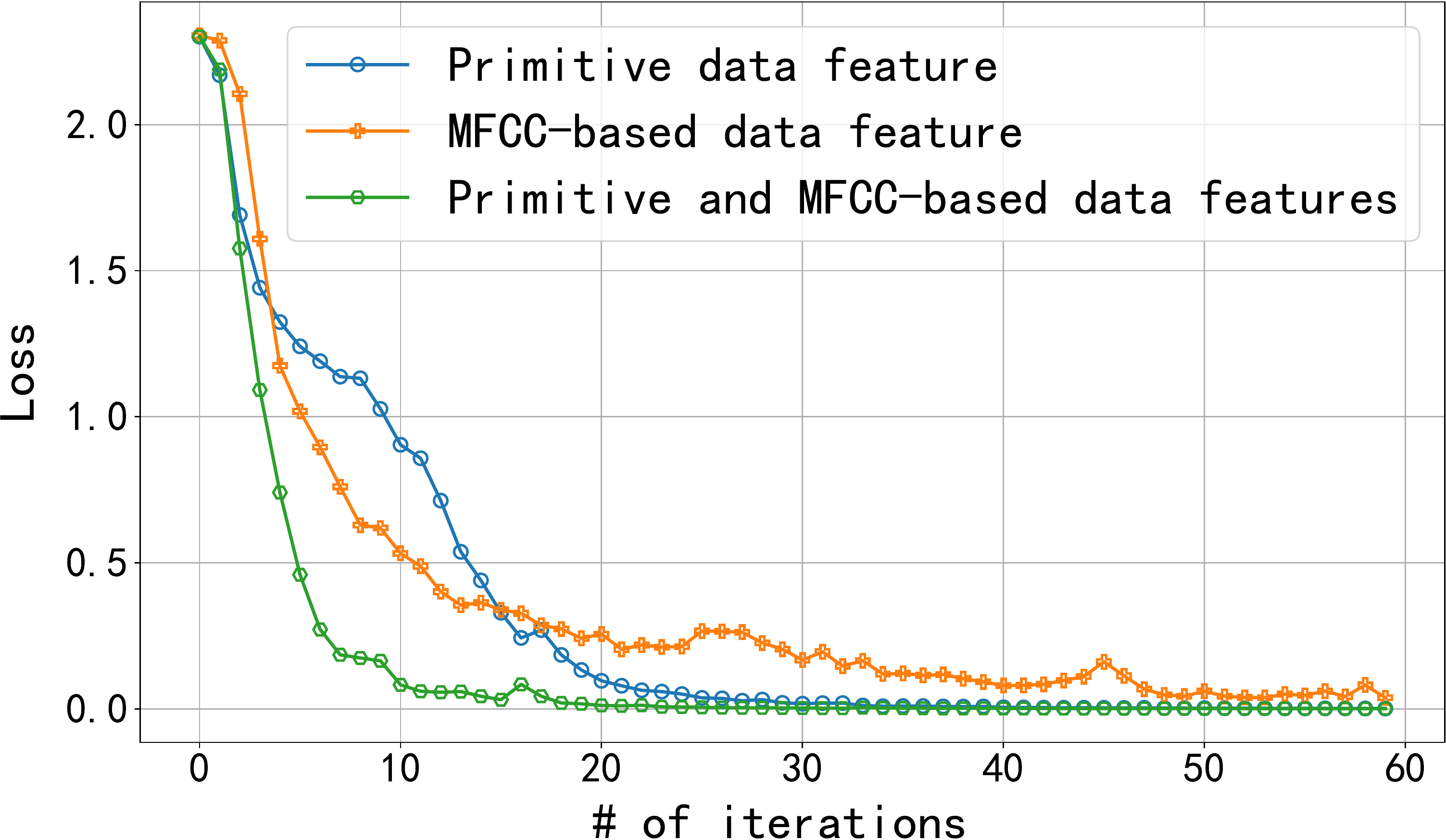}}
    \parbox{.45\columnwidth}{\center\scriptsize(a) $T=400$ms}
    \parbox{.45\columnwidth}{\center\scriptsize(b) $T=600$ms}\vspace{2ex}
    \parbox{.45\textwidth}{\center\includegraphics[width=.4\textwidth]{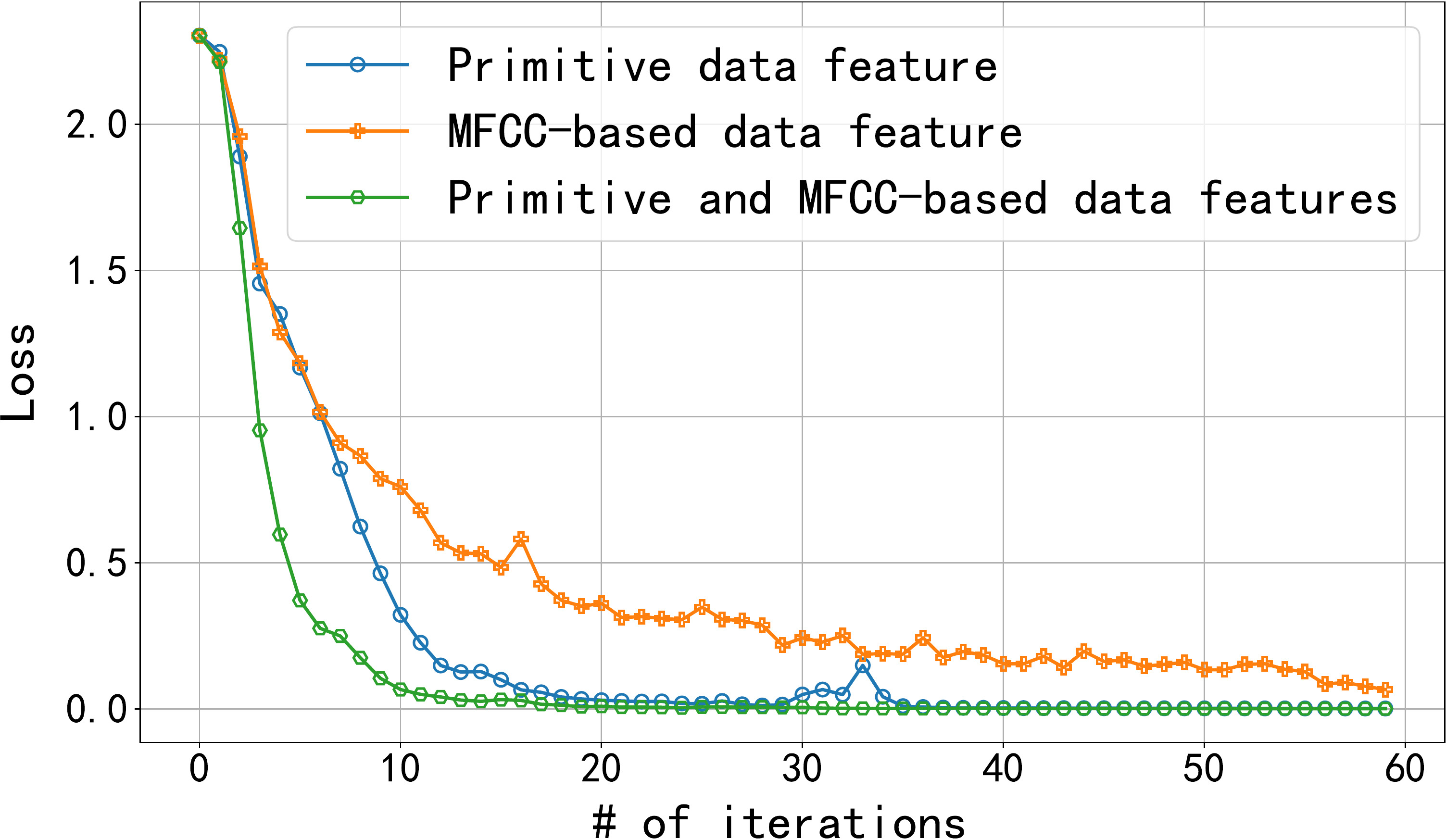}}
    \parbox{.45\textwidth}{\center\includegraphics[width=.4\textwidth]{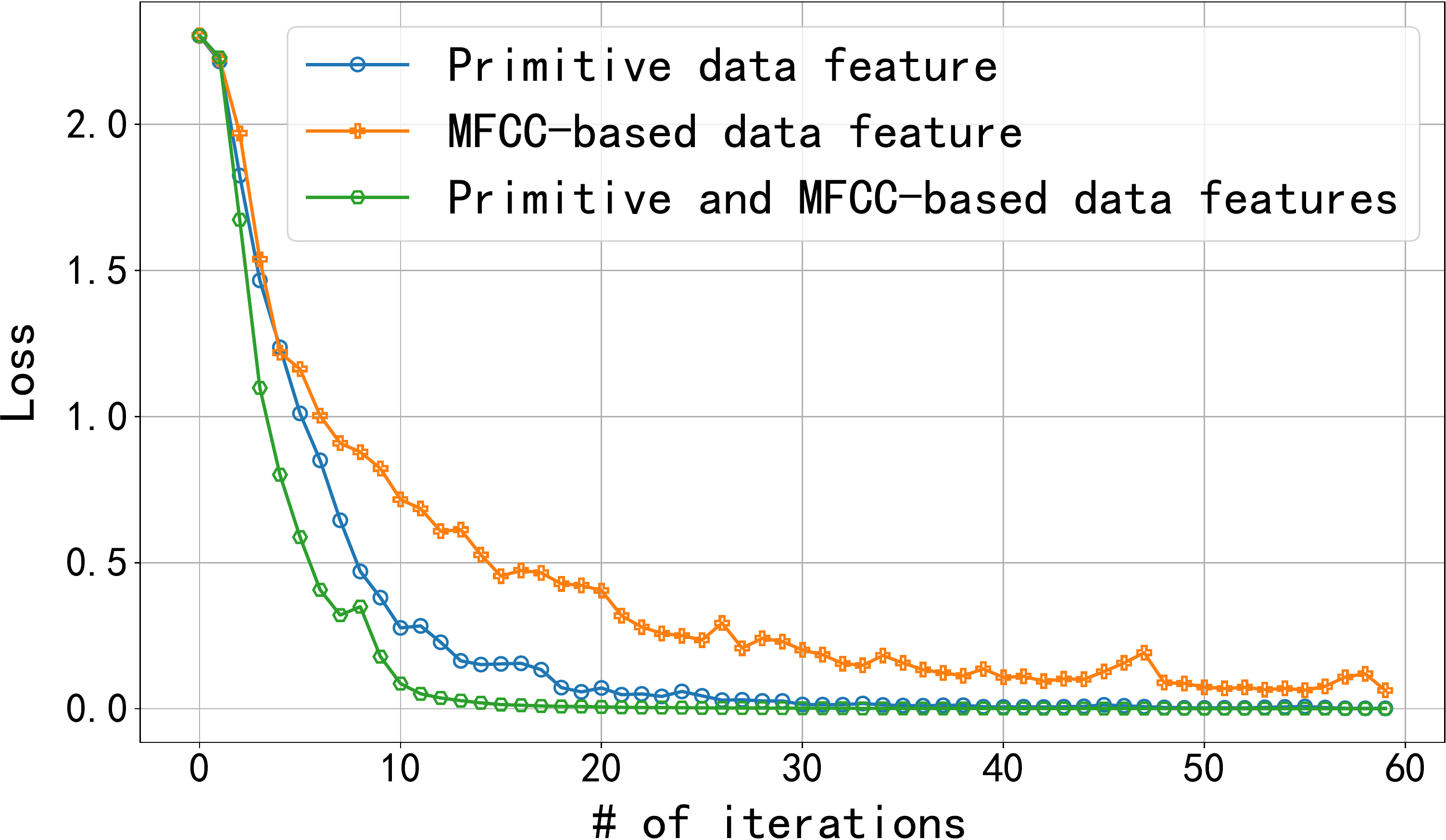}}
    \parbox{.45\columnwidth}{\center\scriptsize(c) $T=800$ms}
    \parbox{.45\columnwidth}{\center\scriptsize(d) $T=1000$ms}
  \caption{Convergence of loss function with different time lengths of data samples.}
  \label{fig:loss}
  \end{center}
  \end{figure*}

  In Fig.~\ref{fig:loss}, we reveal how the during of the vibration (or the number of data points in each training sample), i.e., $T$, influences the convergence of the loss function, and we also illustrate how the two types of features work together to train the CNN-based classification model. Specifically, we collect data from $10$ volunteers in the five different gestures (i.e., standing, sitting upright, sitting-and- leaning-forward, sitting-and-leaning-backward, and walking). For each worker, we collect $60$ data samples, i.e., $12$ samples for each gesture. The time during of vibration, i.e., $T$, is varied from $400$ms to $1$s. As shown in Fig.~\ref{fig:loss}, when taking into account both types of features (i.e., the primitive data feature and the MFCC-based feature), we can obtain a higher convergence rate. For example, it is demonstrated in Fig.~\ref{fig:loss} (b), that the loss function converges in about $20$ iterations if both the two types of features are involved; whereas we need around $30$ iterations (or even more) to ensure the convergence if only one type of features is considered. It is also observed in Fig.~\ref{fig:loss}, that if a longer time duration is adopted, we achieve convergence in fewer iterations (and thus a higher convergence rate). For example, when $T = 400$ms, the loss function achieves convergence in about $20$ iterations, while the convergence can be ensured in $10$ iterations when we increase $T$ to $1$s.

\section{User Authentication}  \label{sec:authen}
  In our VibHead system, we design a two-step authentication mechanism. Assuming that there are $N$ legitimate users $\mathcal N=\{1,2,\cdots,N\}$, we use the above method to build $N+1$ classifiers $\mathcal F = \{F_0, F_1, F_2, \cdots, F_N\}$, where $F_0$ is the classifier trained by the data of all the $N$ legitimate users, while $F_i$ represents the one trained by the data collected from the legitimate users except user $i$ (in other words, $F_i$ can be used to identify users $\mathcal{N}_i = \{1, 2, \cdots, i-1, i, \cdots, N\}$). We also suppose $\mathcal F_i = \{F_j \mid j \in \mathcal N_i\} \subset \mathcal F$, i.e., $\mathcal{F}_i$ denotes a subset of classifiers that can be used to identify the user $i$. When an authentication service is invoked by a login user wearing a smart headset equipped with VibHead, we let the motor mounted on the headset generate vibration signals, and the vibration signals propagate through the user's head and can be sampled by an IMU sensor. We denote the data samples collected by the accelerometer as $x$. Then, the goal of our authentication scheme is to identify if $x$ is collected from a legitimate user.
  
\begin{figure}[htb!]
    \centering
    \includegraphics[width=.5\textwidth]{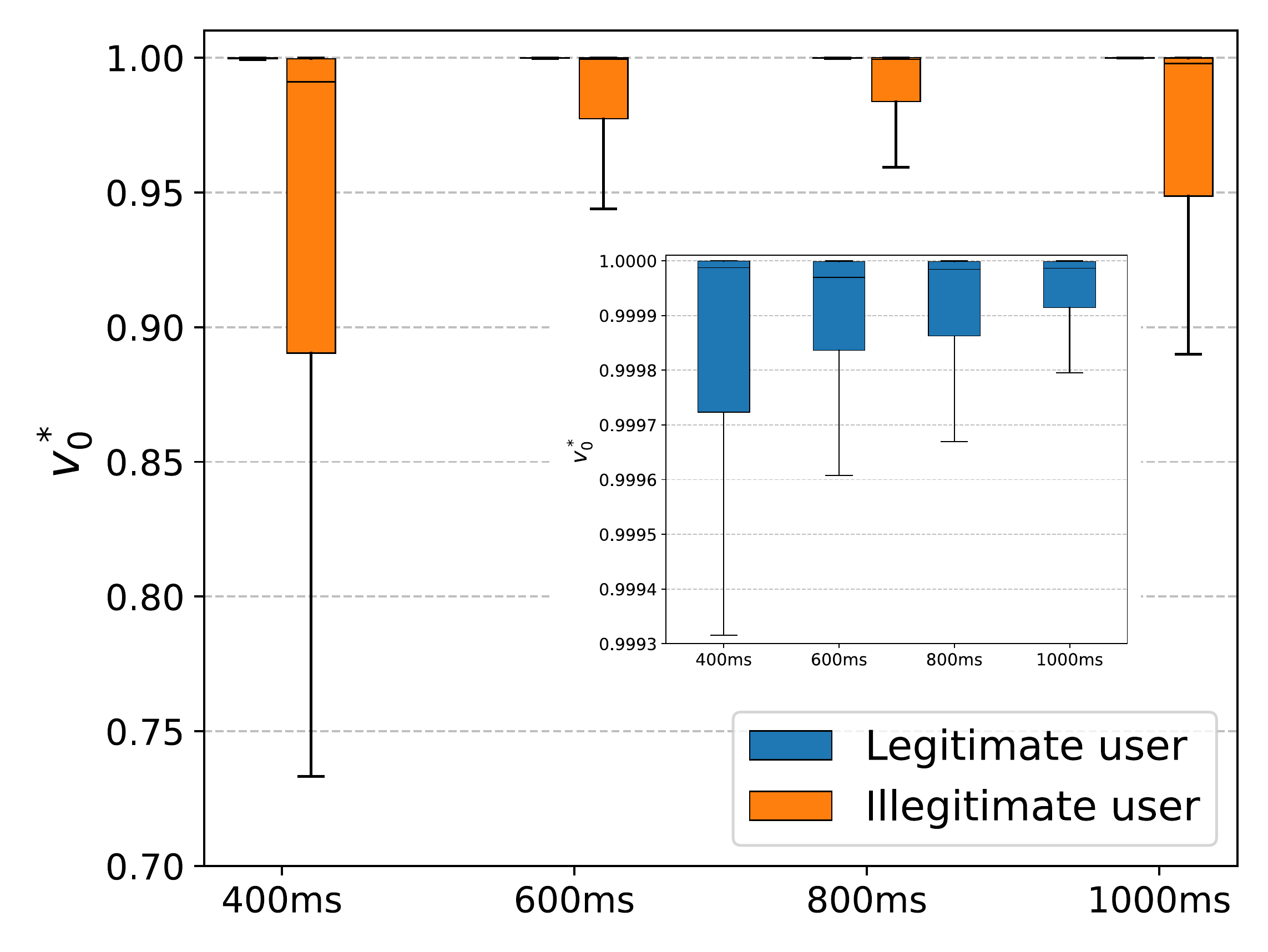}
    \caption{Statistical quantities of $v^*_0$ for legitimate and illegitimate users. We zoom in on the statistical quantities of the legitimate users since their quartile range is too narrow to be clearly observed.} 
  \label{fig:alpha}
  \end{figure}

  In the first step of our authentication scheme, we use the classifier $F_0$ to identify the user of login. We let $\mathbf{v}_0 = F_0(x)$ denote the label of the login user output by $F_0$. As shown in Sec.~\ref{sec:classification}, $\mathbf{v}_0$ is actually an $N$-dimensional probability vector, and the $i$-th element of $\mathbf{v}_0$, i.e., $v_{0,i}$, represents the probability at which the login user is identified as legitimate user $i$. Let $i^*_0 = \arg\max_{i \in \mathcal N} \{v_{0,i}\}$ and $v^*_0 = v_{i^*_0} = \max_{i \in \mathcal N} \{v_{0,i}\}$. In Fig.~\ref{fig:alpha}, we reveal the statistical quantities of $v^*_0$ for $10$ legitimate users and $10$ illegitimate users. For each of the users, we have $40$ test data samples across the different five gestures. We also vary the time during of the data samples from $400$ms to $1$s. It is found that the legitimate users have a very narrow quartile range, whereas the illegitimate user has values of $v^*_0$ much more scattered than the legitimate user. Furthermore, the mean value of $v^*_0$ for the illegitimate user is noticeably smaller than the one for the legitimate user. Therefore, we set a threshold $\alpha$ such that a login user is identified as a legitimate user (particularly user $i^*_0$) if $v^*_0 \geq \alpha$. 
  %

\begin{figure}[htb!]
    \centering
    \includegraphics[width=0.5\columnwidth]{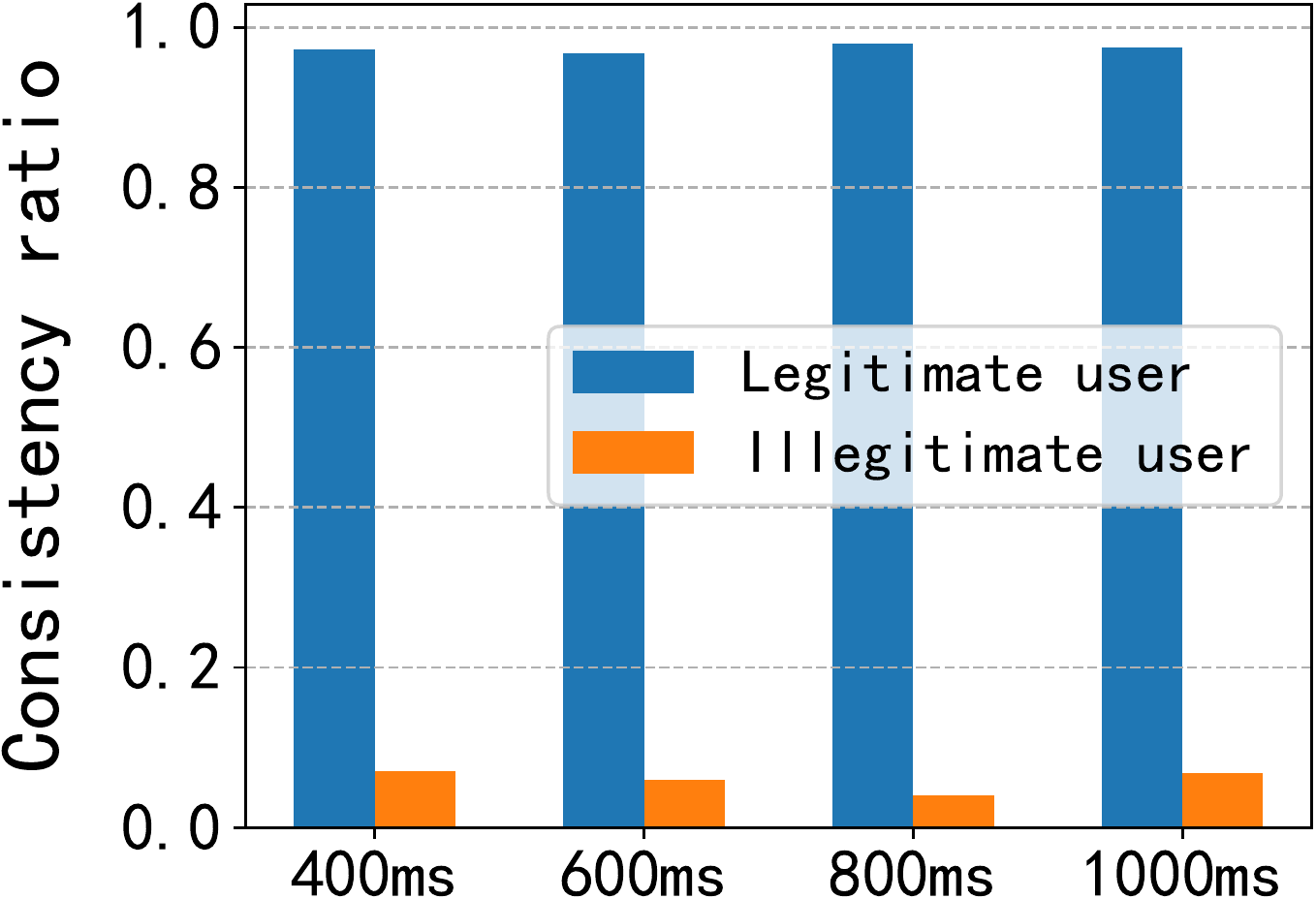}
    \caption{The consistency ratios of login users. We take a legitimate user and an illegitimate user as example, and randomly choose $40$ test data samples for each of the users.} 
  \label{fig:second1}
  \end{figure}

\begin{figure}[htb!]
    \centering
    \includegraphics[width=0.5\columnwidth]{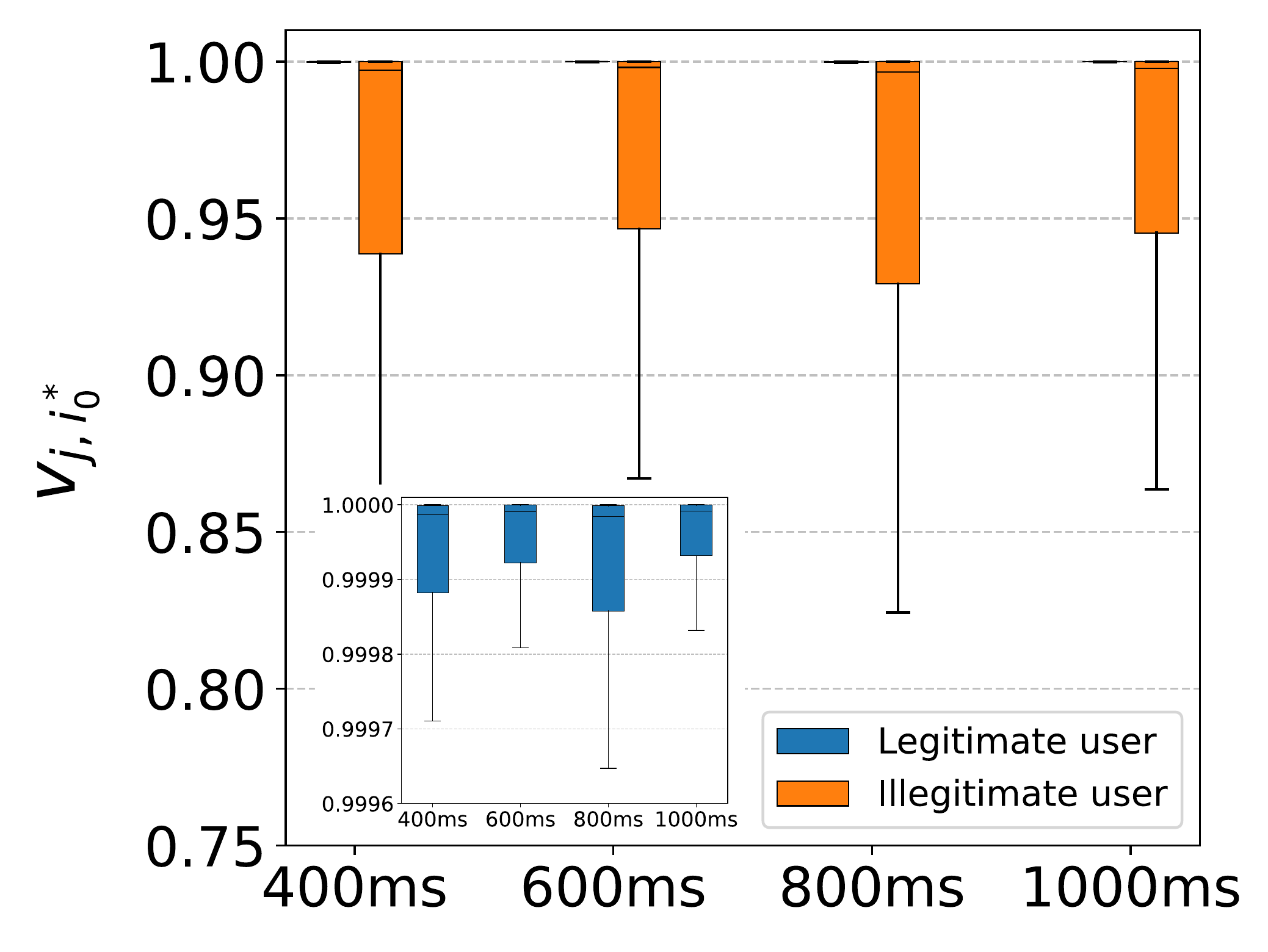}
    \caption{The statistical quantities of $v_{j, i^*_0}$ for $\forall F_j \in \mathcal{F}_{i^*_0}$. We zoom in on the quantities of the legitimate users, as their interquartile ranges are too narrow to be clearly observed.} 
  \label{fig:second2}
  \end{figure}

  The preliminary authentication result of the above step can be further refined in the second step as follows. Recall that $i^*_0 \in \mathcal N$ is the identification result of the first step. For any classifier $F_j \in \mathcal F_{i^*_0}$, we calculate an an vector of $(N-1)$-dimensions $\mathbf{v}_j = F_j(x)$ and thus denote by $i^*_j$ the prediction result obtained through $F_j$. Let $v^*_j = v_{i^*_j}$ denote the maximum element of $\mathbf{v}_j$. In other words, the login user is predicted by classifier $F_j$ as user $i^*_j$ with probability $v^*_j$. In the second step of our authentication scheme, the login user is identified as a legitimate user (particularly user $i^*_0$), if $i^*_j = i^*_0$ holds with $v^*_j \geq \beta$ for any $F_j \in \mathcal F_{i^*_0}$, where $\beta$ is a predefined threshold. The efficacy of the second step of our authentication scheme is revealed in Fig.~\ref{fig:second1} and Fig.~\ref{fig:second2}. In Fig.~\ref{fig:second1}, we show the so-called ``consistency ratio'', i.e., proportion of the test data samples that satisfy the condition $i^*_j = i^*_0$ for $\forall F_j \in \mathcal F_{i^*_0}$. It is illustrated that legitimate users usually always have identical prediction results through $F_0$ and $\mathcal{F}_{i^*_0}$, while it is much more likely that illegitimate users have different prediction results. Furthermore, we show the statistical quantities of $v_{j, i^*_0}$ (i.e., the probability at which the prediction result of $F_j \in \mathcal{F}_{i^*_0}$ is $i^*_0$) in Fig.~\ref{fig:second2}. It is obvious that the values of $v_{j,i^*_0}$ for the legitimate users are very close to $1$, and the quartile range is very narrow. In contrast, the illegitimate users have their values of $v_{j,i^*_0}$ considerably spread in a large range. In a nutshell, compared to illegitimate users, each classifier $F_j \in \mathcal{F}_{i^*_0}$ has much higher ``confidence'' to recognize login legitimate users as $i^*_0$ (i.e., the prediction results of $F_0$). Hence, by elaborately selecting a value for $\beta$, we can differentiate the legitimate users from the illegitimate ones.

  %

\section{Evaluation}  \label{sec:evaluation}
  In this section, we evaluate our VibHead system through extensive experiments. We first describe the setup and the methodology of our experiments in Sec.~\ref{ssec:setup}. We then focus on analyzing the performance of the user classification module of our VibHead system in Sec.~\ref{ssec:clasacc}. We finally report the authentication performance of our VibHead system in Sec.~\ref{ssec:authperf}.

  \subsection{Experiment Setup and Methodology} \label{ssec:setup}
    The prototype of VibHead is implemented with Microsoft HoloLens 2, which is one of the most popular VR/AR headsets in market. It is worth noting that, most VR/AR headset products available in market have no build-in vibration motor and the interface to the readings of the build-in accelerometer is usually closed. Hence, we mount a linear vibration motor used in iPhone 7 Plus on the strap of the headset and attach an IMU sensor (ADXL375) to the headset for receiving vibration signals.

    To collect dataset for evaluation, we recruited $20$ volunteers (including $8$ females and $12$ males) in our experiments. Among the $20$ volunteers, $10$ of them were registered in our VibHead system as default legitimate users, while the rest $10$ volunteers played the role of illegitimate users and attempted to attack our VibHead system. We consider five different gestures that users usually take when using smart headset (especially when registering for initialization or calling authentication service)., i.e., standing, sitting-upright, sitting-and-leaning-forward, sitting-and-leaning-backward, and walking. Specifically, we collected $100$ data sample for each volunteer and $20$ samples for each gesture. To facilitate our experiments, we set the total duration of each data sample be $T=1$s, and we can clip the samples with different lengths (e.g., $T=400, 600, 800$ms), when evaluating our VibHead with signals of different lengths. Among the data collected from the legitimate users, we used $60\%$ of them as training data and $40\%$ as test data. For each legitimate user, we collected the training data and the test data on different days. When evaluating the authentication performance of our VibHead, the test data of the legitimate users and all of the data collected from the illegitimate users are used. Considering smart headsets are usually used indoors, the data collection is conducted in an indoor environment. We empirically let $\alpha = 0.9$ and $\beta=0.8$.

    The following metrics are adopted to quantitatively analyze the efficacy of our VibHead system:
    \begin{itemize}
      \item \textbf{Accuracy}: The probability that a legitimated user is correctly identified.
      \item \textbf{False Acceptance Rate} (FAR): The probability that an illegitimate user is authenticated as a illegitimate user.
      \item \textbf{False Rejection Rate} (FRR): The probability that a legitimate user is authenticated as an illegitimate user.
    \end{itemize}

    We also implement two traditional machine learning models, i.e., \textit{Support Vector Machine} (SVM) and \textit{Random Forest} (RF),  as benchmarks, through \textit{Scikit-learn}~\cite{Pedregosa-JMLR11}. The two methods can be used for user classification; nevertheless, we may choose different hyper-parameters (i.e., $\alpha$ and $\beta$), when adapting our two-step authentication scheme to the two classification models for comparable performance. In particular, we let $\alpha = 0.65$ and $\beta = 0.6$ for the SVM model, and $\alpha=0.52$ and $\beta=0.48$ for the RF model.
    %

  \subsection{Classification Accuracy} \label{ssec:clasacc}
\begin{figure*}[htb!]
    \begin{center}
      \parbox{.32\textwidth}{\center\includegraphics[width=.3\textwidth]{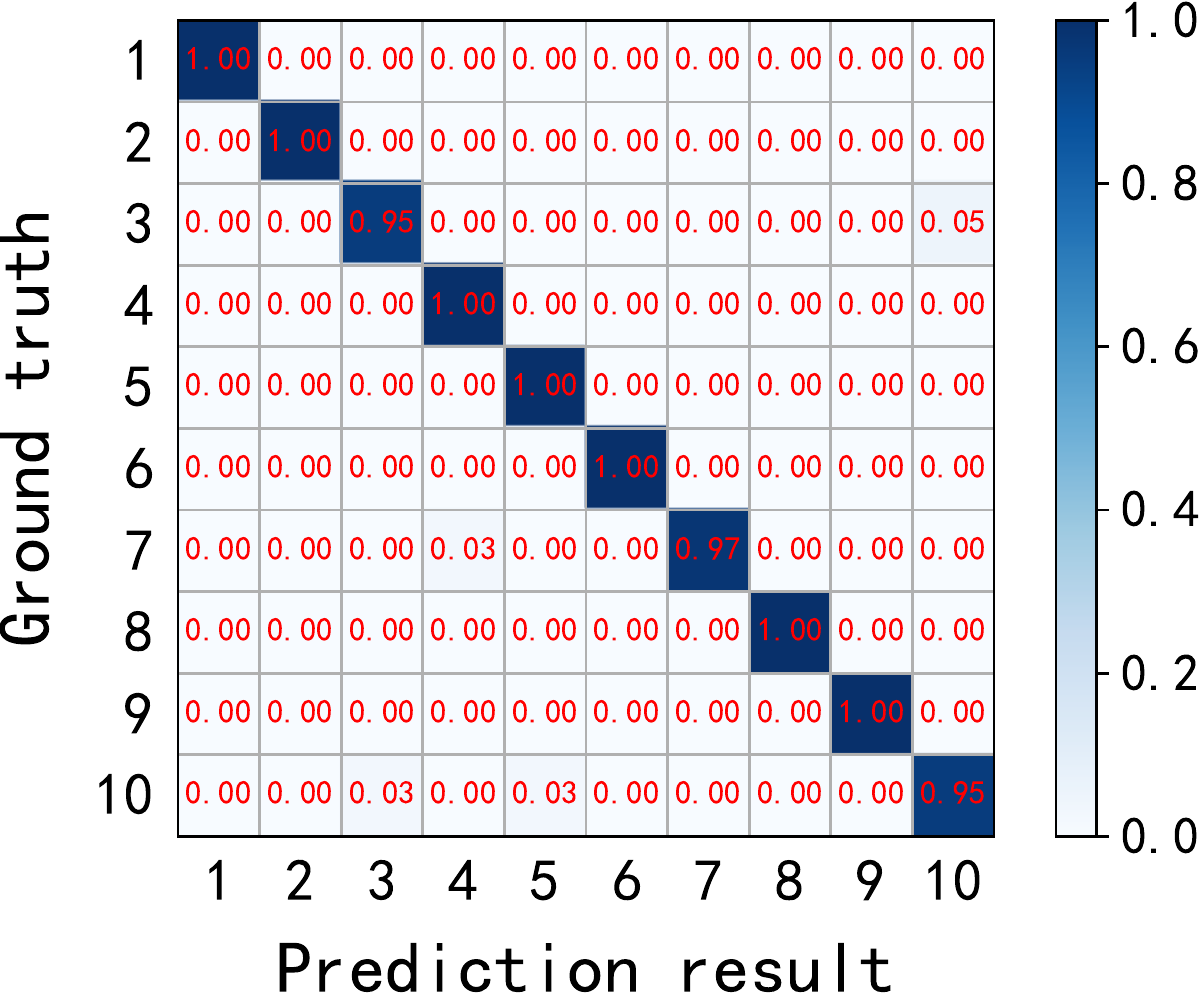}}
      \parbox{.32\textwidth}{\center\includegraphics[width=.3\textwidth]{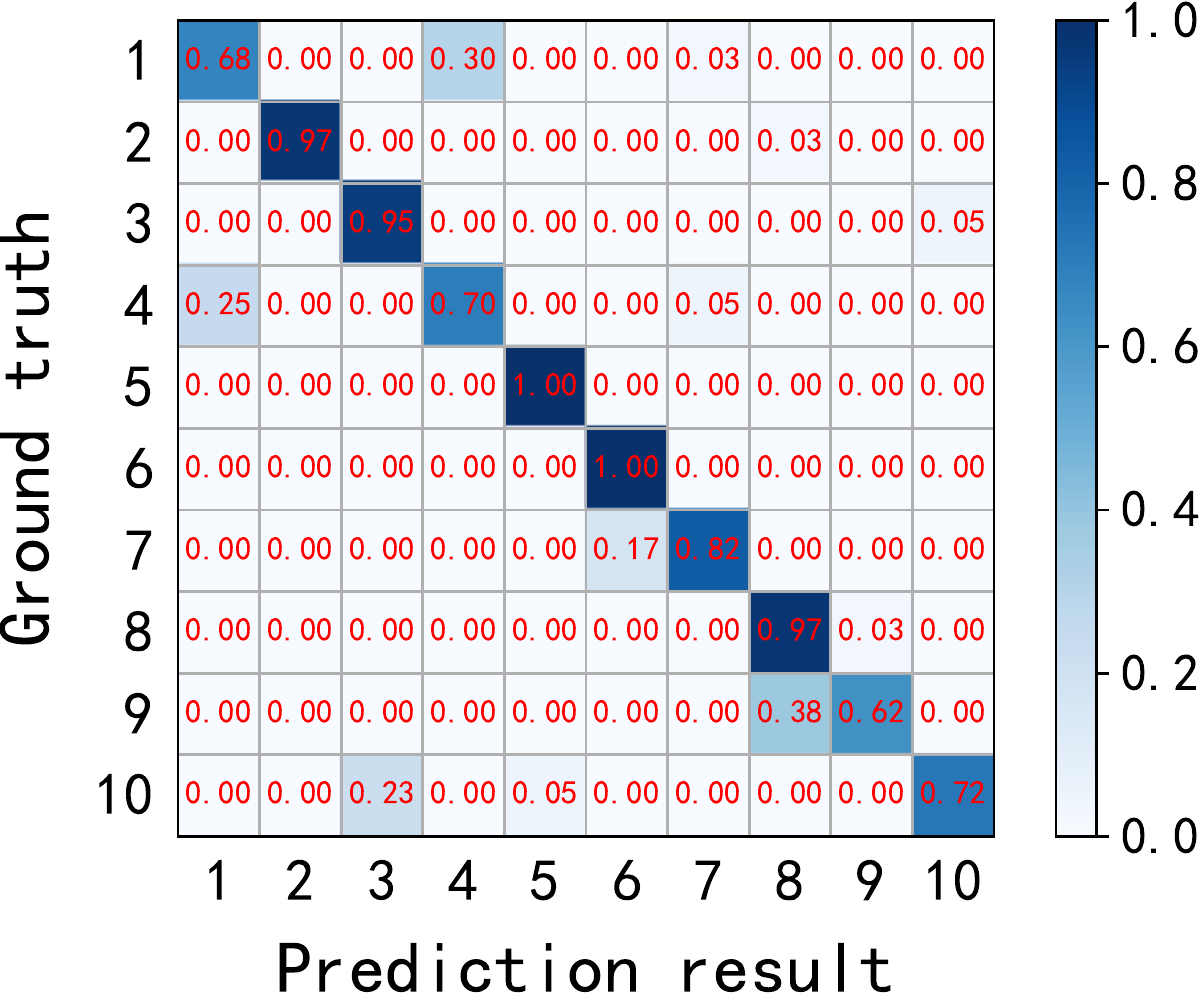}}
      \parbox{.32\textwidth}{\center\includegraphics[width=.3\textwidth]{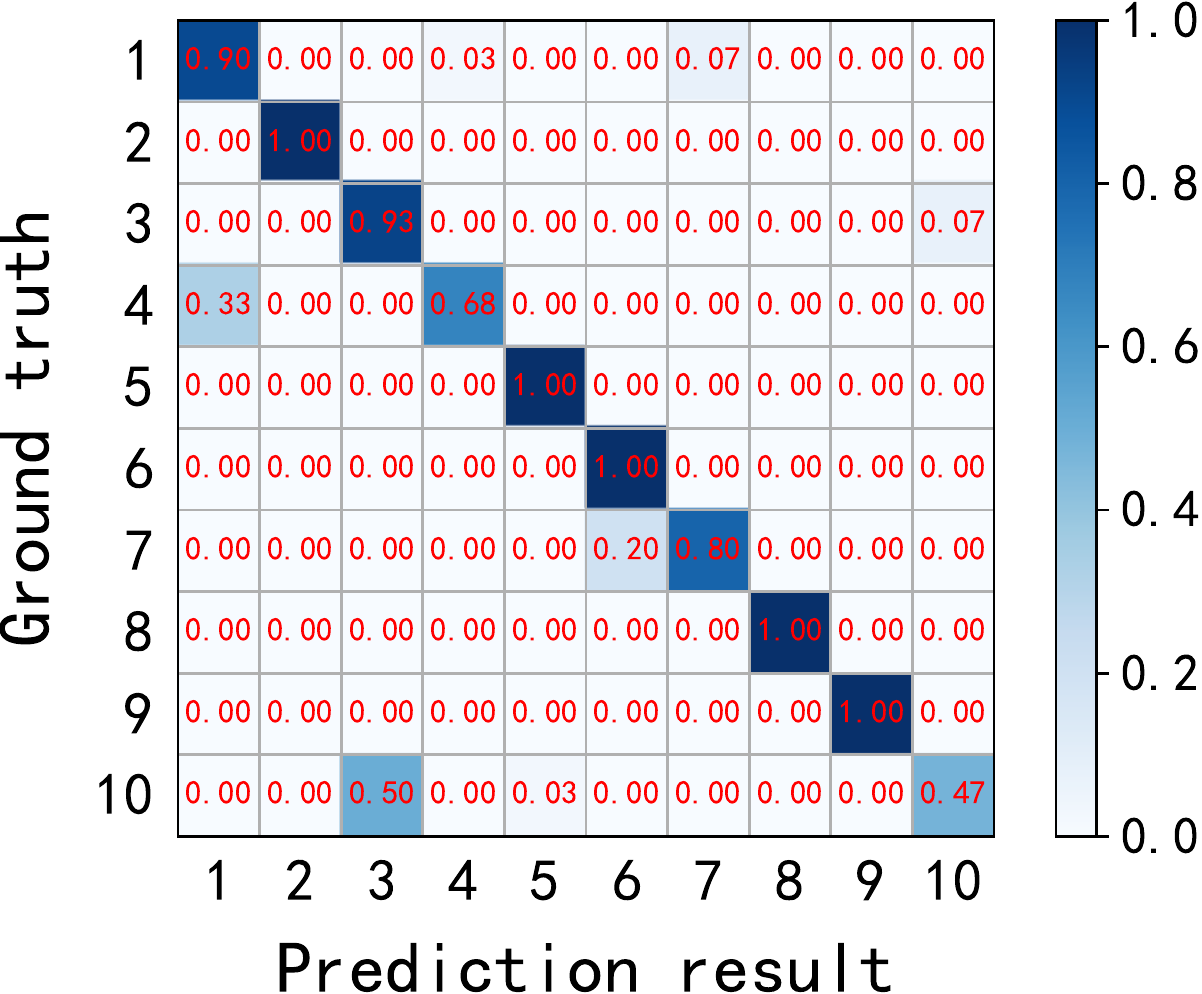}}
      \parbox{.32\columnwidth}{\center\scriptsize(a1) VibHead with $T=400$}
      \parbox{.32\columnwidth}{\center\scriptsize(a2) SVM with $T=400ms$}
      \parbox{.32\columnwidth}{\center\scriptsize(a3) RF with $T=400ms$}
      \parbox{.32\textwidth}{\center\includegraphics[width=.3\textwidth]{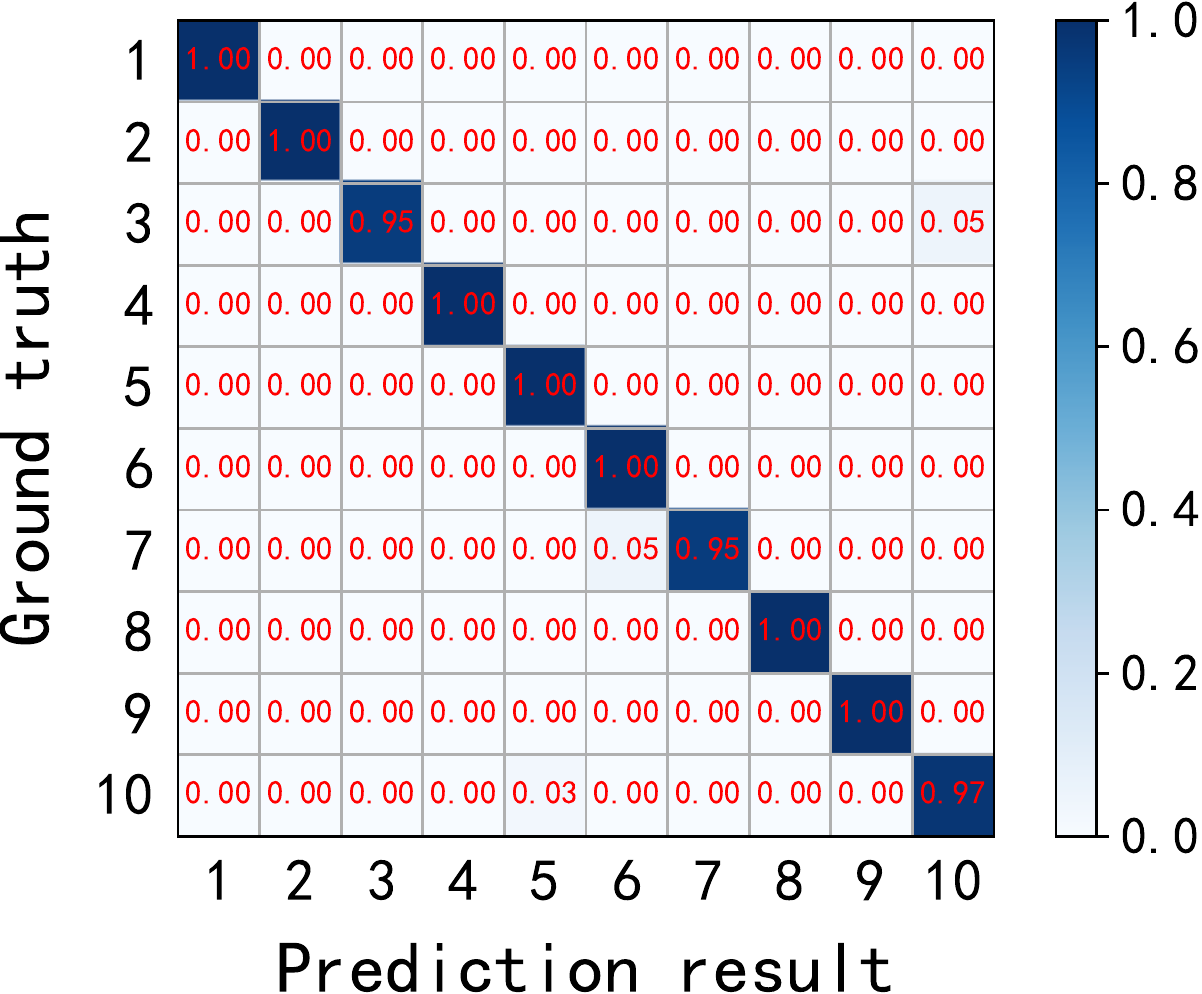}}
      \parbox{.32\textwidth}{\center\includegraphics[width=.3\textwidth]{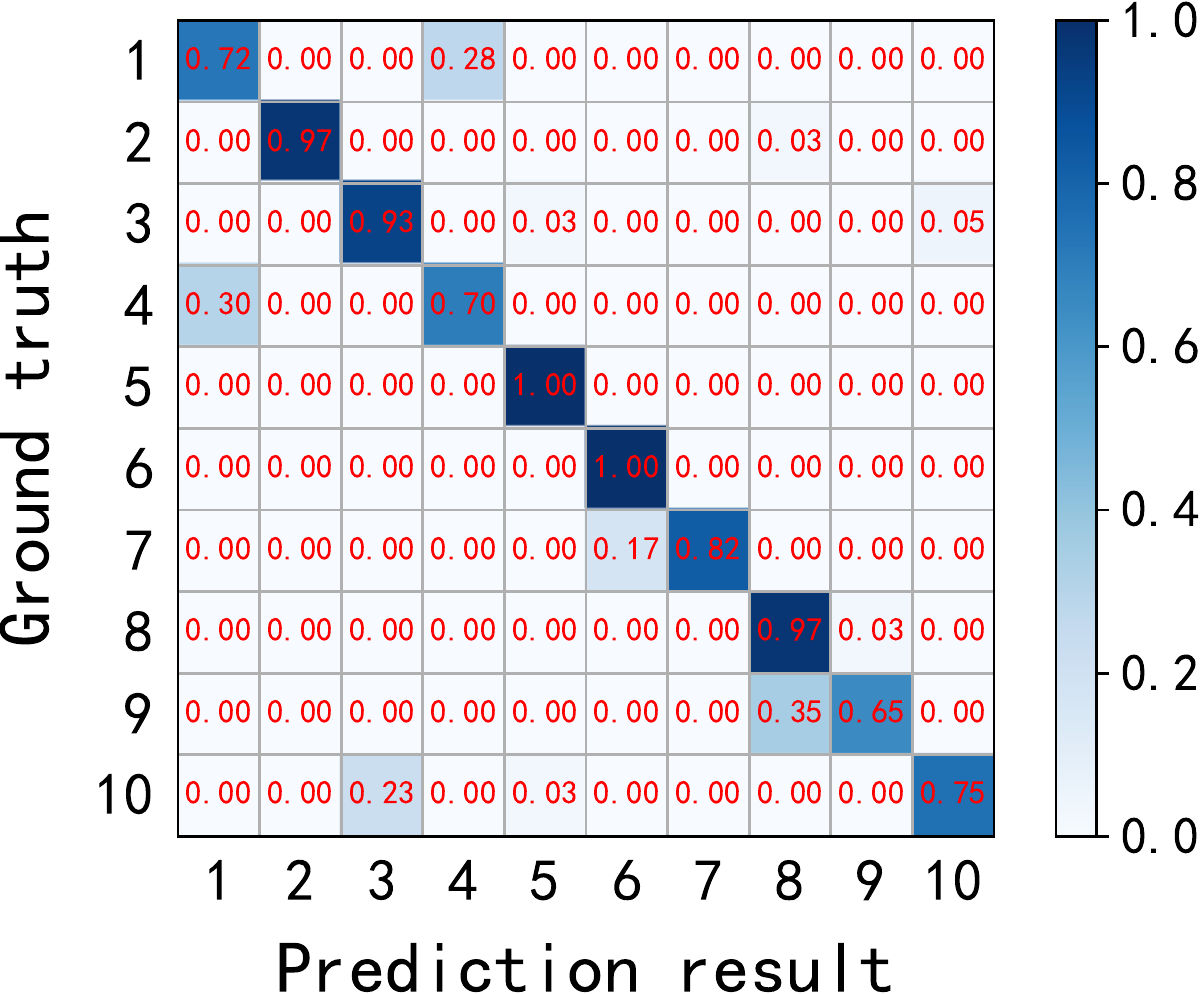}}
      \parbox{.32\textwidth}{\center\includegraphics[width=.3\textwidth]{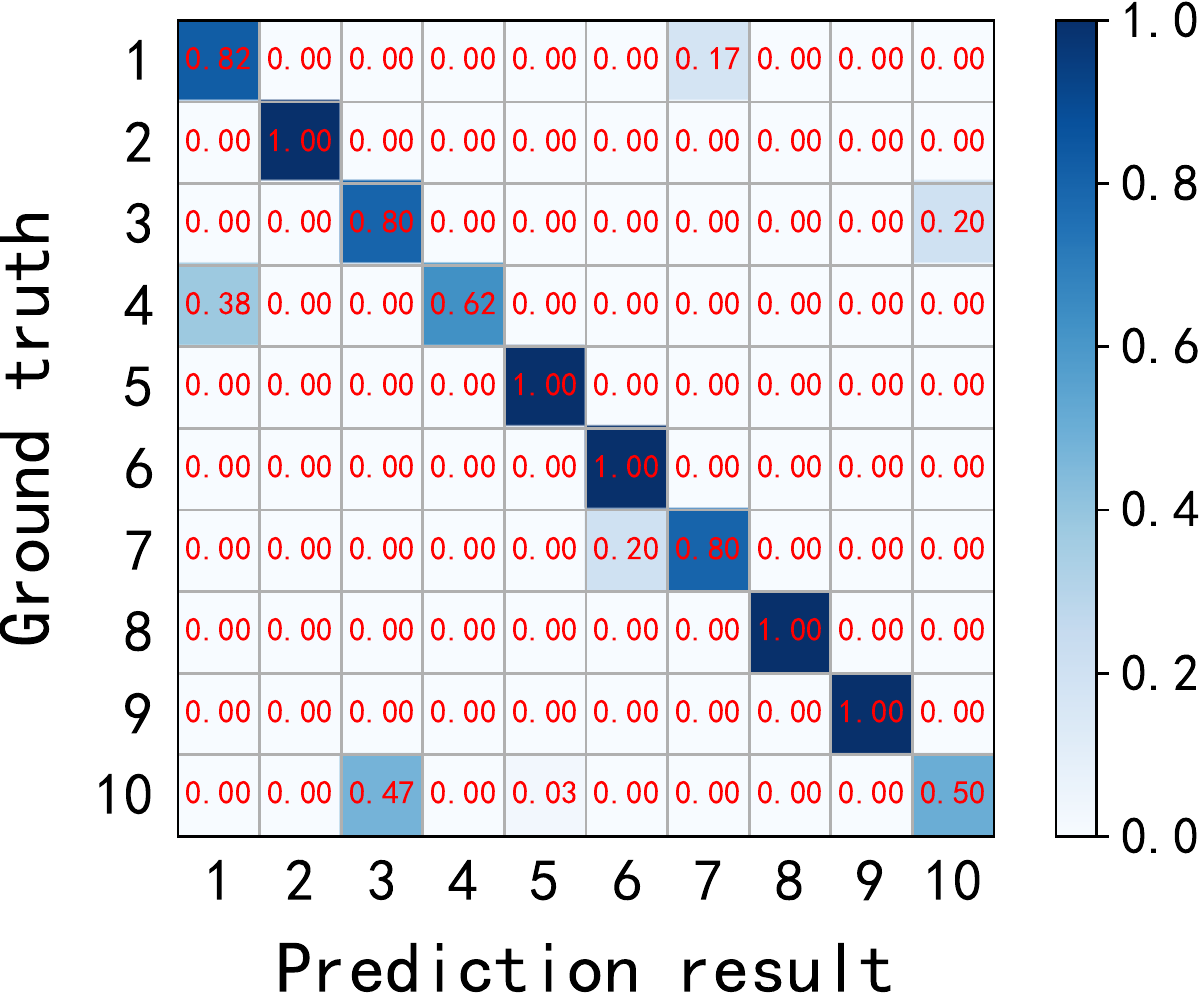}}
      \parbox{.32\columnwidth}{\center\scriptsize(b1) VibHead with $T=600ms$}
      \parbox{.32\columnwidth}{\center\scriptsize(b2) SVM with $T=600ms$}
      \parbox{.32\columnwidth}{\center\scriptsize(b3) RF with $T=600ms$}
      \parbox{.32\textwidth}{\center\includegraphics[width=.3\textwidth]{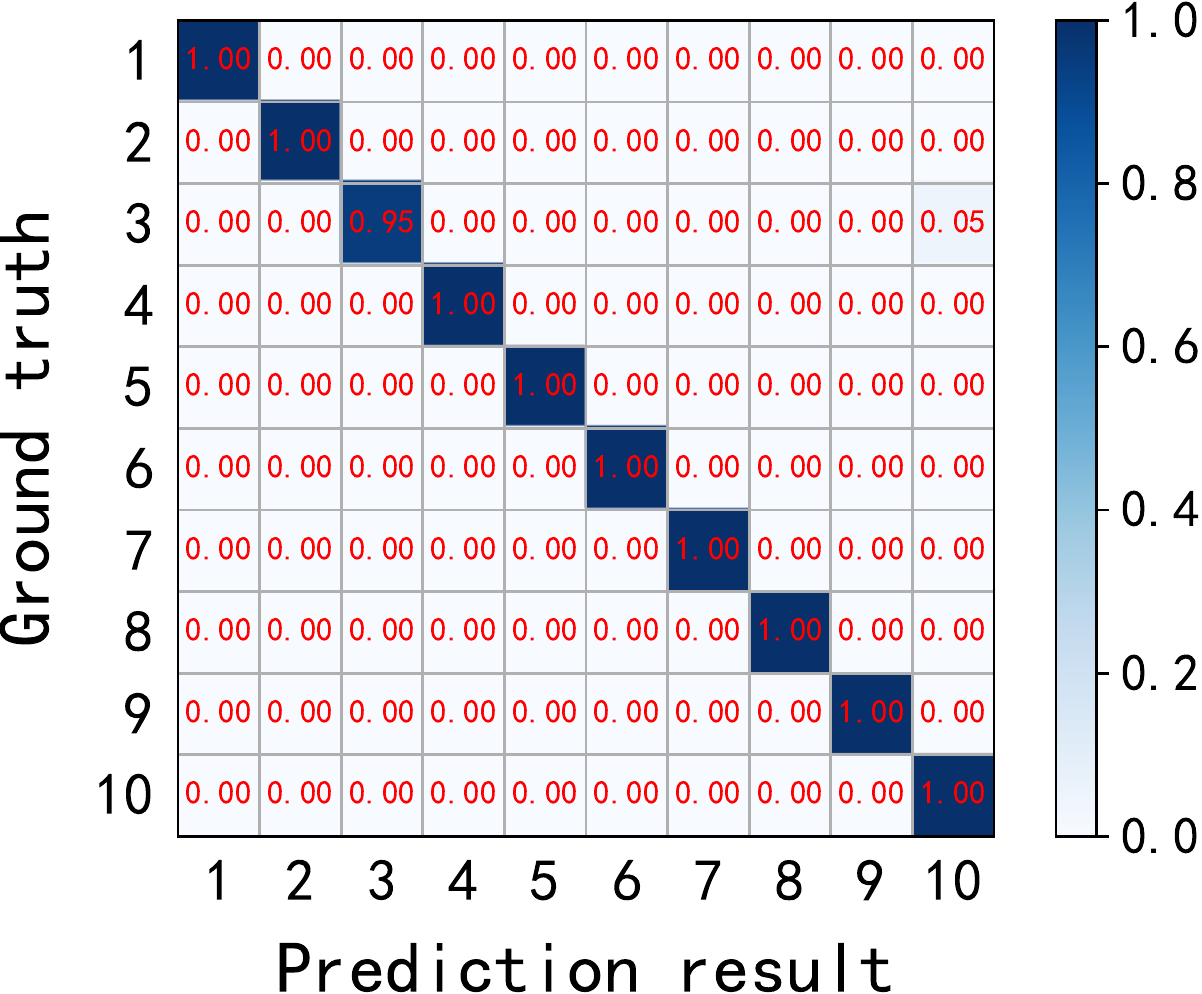}}
      \parbox{.32\textwidth}{\center\includegraphics[width=.3\textwidth]{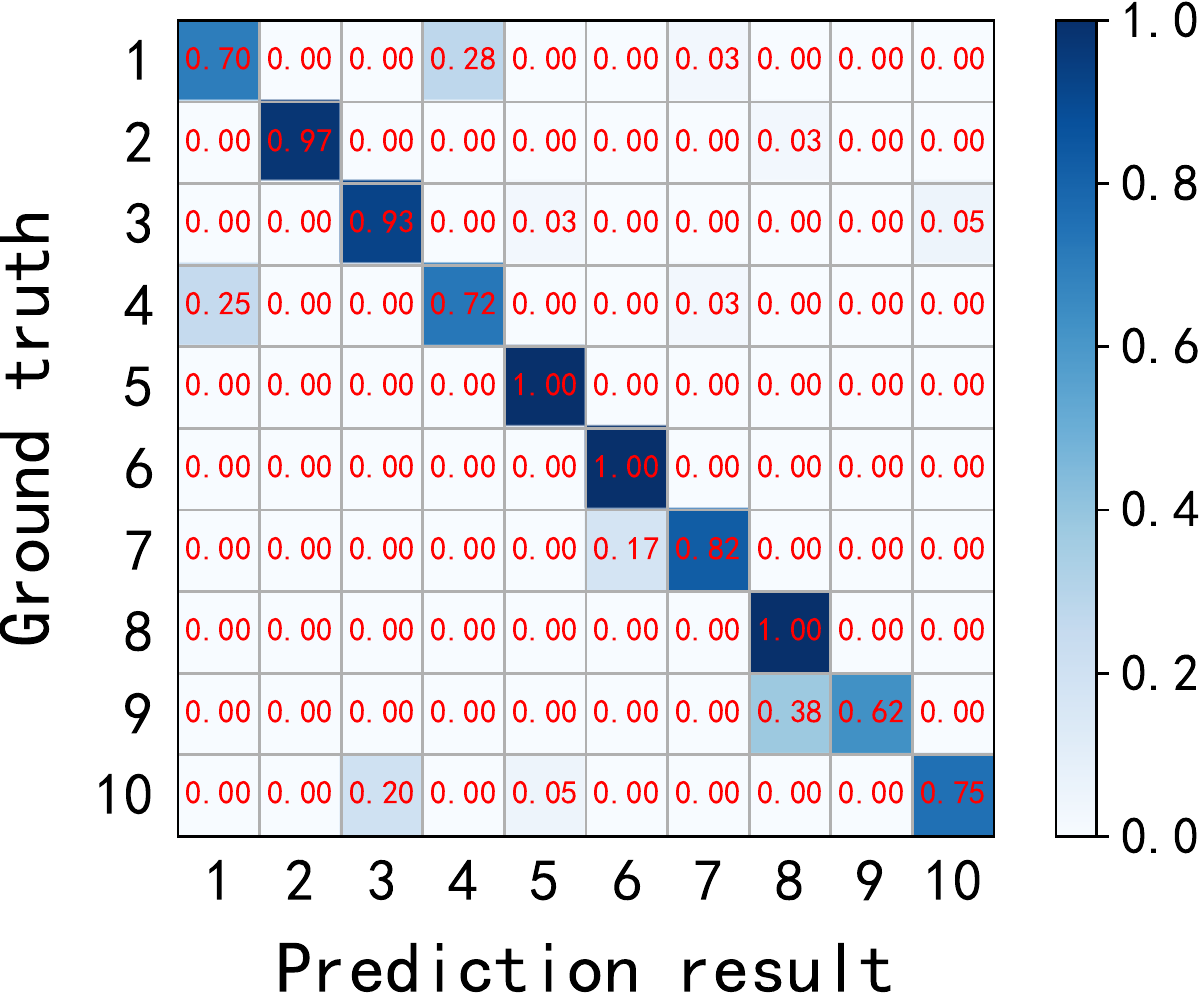}}
      \parbox{.32\textwidth}{\center\includegraphics[width=.3\textwidth]{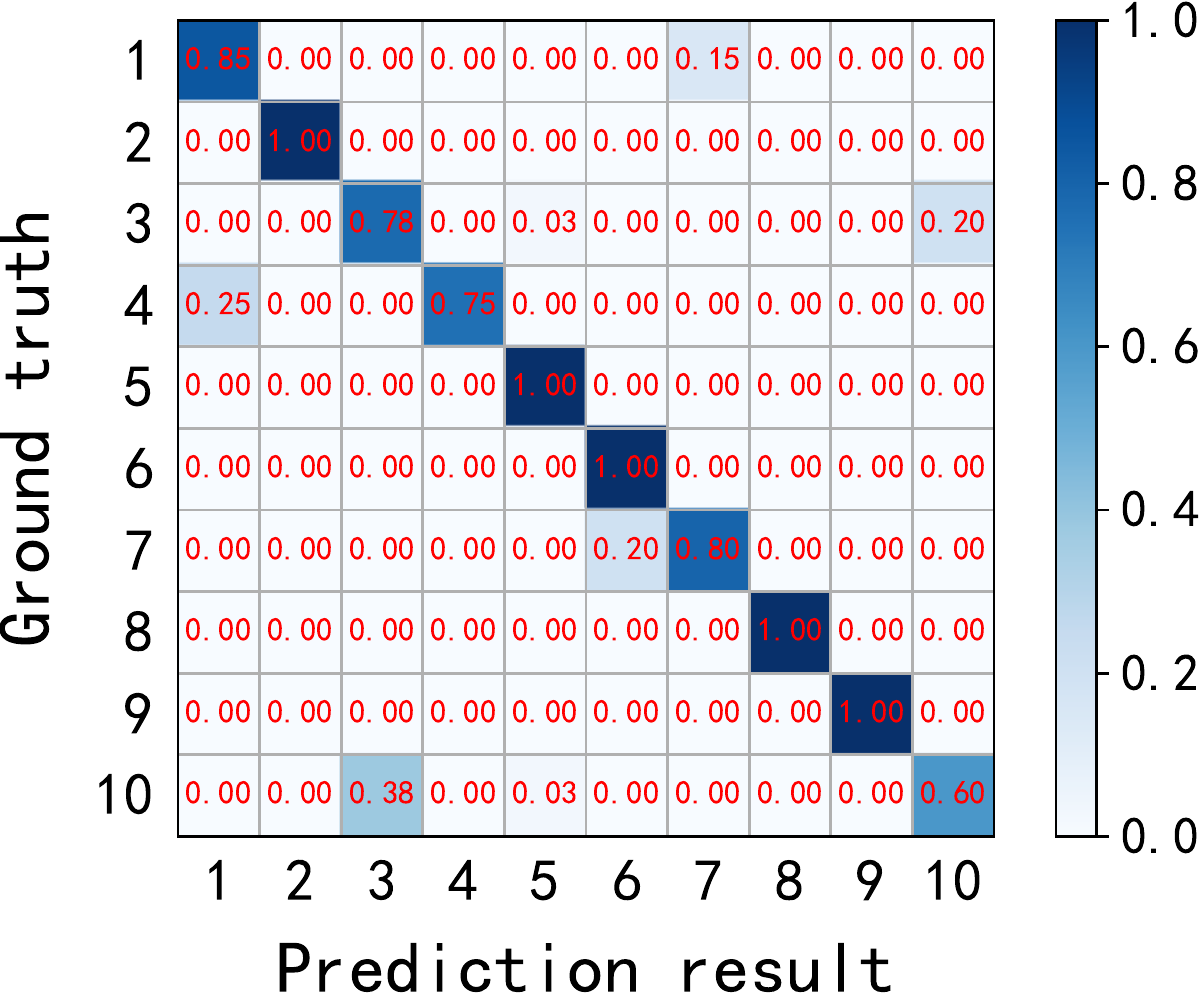}}
      \parbox{.32\columnwidth}{\center\scriptsize(c1) VibHead with $T=800ms$}
      \parbox{.32\columnwidth}{\center\scriptsize(c2) SVM with $T=800ms$}
      \parbox{.32\columnwidth}{\center\scriptsize(c3) RF with $T=800ms$}
      \parbox{.32\textwidth}{\center\includegraphics[width=.3\textwidth]{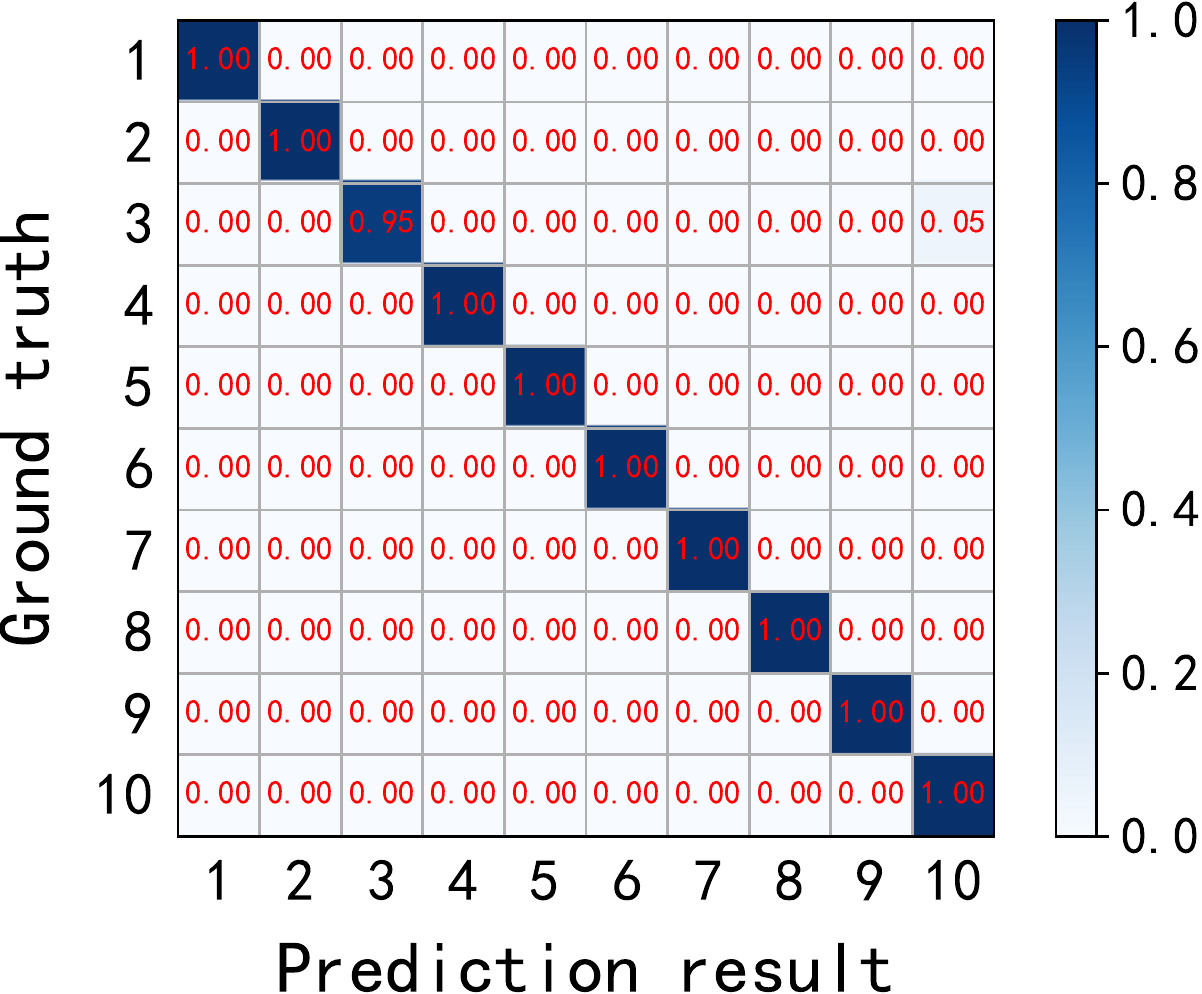}}
      \parbox{.32\textwidth}{\center\includegraphics[width=.3\textwidth]{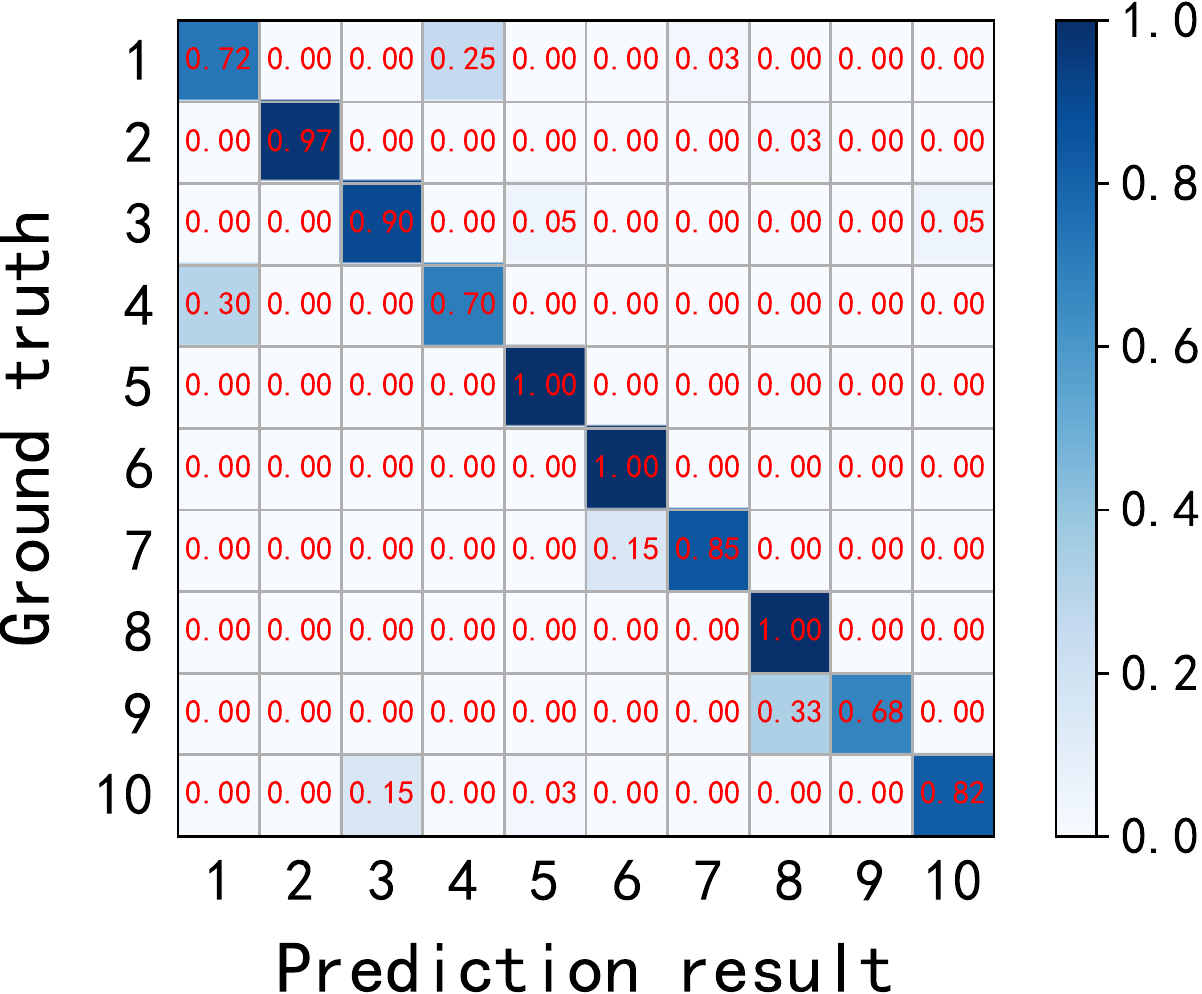}}
      \parbox{.32\textwidth}{\center\includegraphics[width=.3\textwidth]{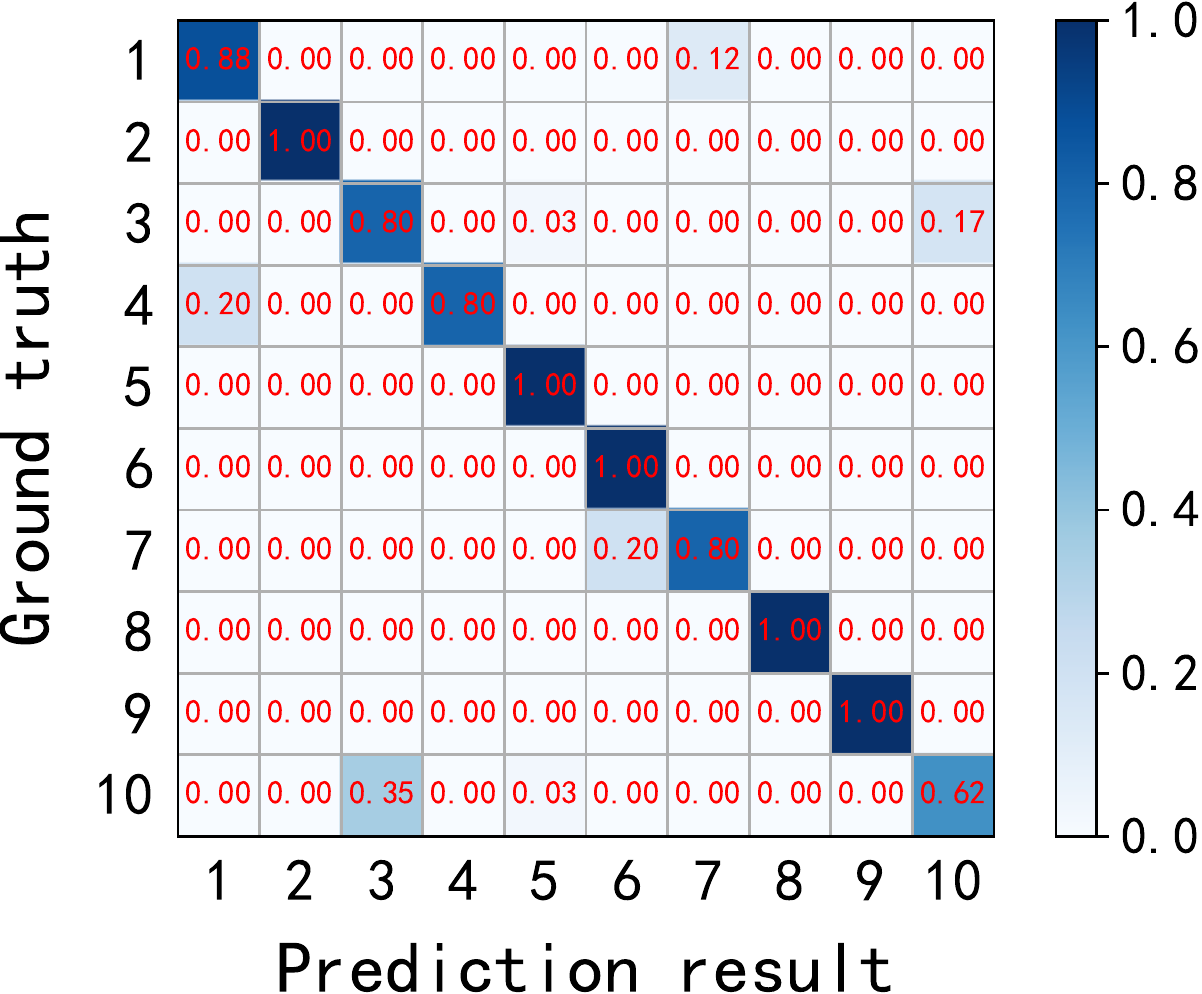}}
      \parbox{.32\columnwidth}{\center\scriptsize(d1) VibHead with $T=1s$}
      \parbox{.32\columnwidth}{\center\scriptsize(d2) SVM with $T=1s$}
      \parbox{.32\columnwidth}{\center\scriptsize(d3) RF with $T=1s$}
    \caption{Confusion matrices of user classification.}
    \label{fig:classconfmat}
    \end{center}
    \end{figure*}

    We first report the classification results of VibHead and the other competing methods, e.g., SVM and RF, respectively. The confusion matrices are illustrated in Fig.~\ref{fig:classconfmat}. It is observed that, our VibHead system almost perfectly identifies the legitimated users. Even with $T=400$ms, the classification accuracy is higher than $0.95$. When $T=1s$, almost all the legitimate users can be correctly identified such that the accuracy further approaches $1$. In contract, the SVM and RF models both have lower classification accuracy than VibHead. For example, when $T=1s$, for the SVM model, user $9$ is correctly identified only with probability $0.68$, while the classification accuracy of the RF model on user $10$ is only $0.62$.

\begin{figure}[htb!]
      \centering
      \includegraphics[width=0.5\columnwidth]{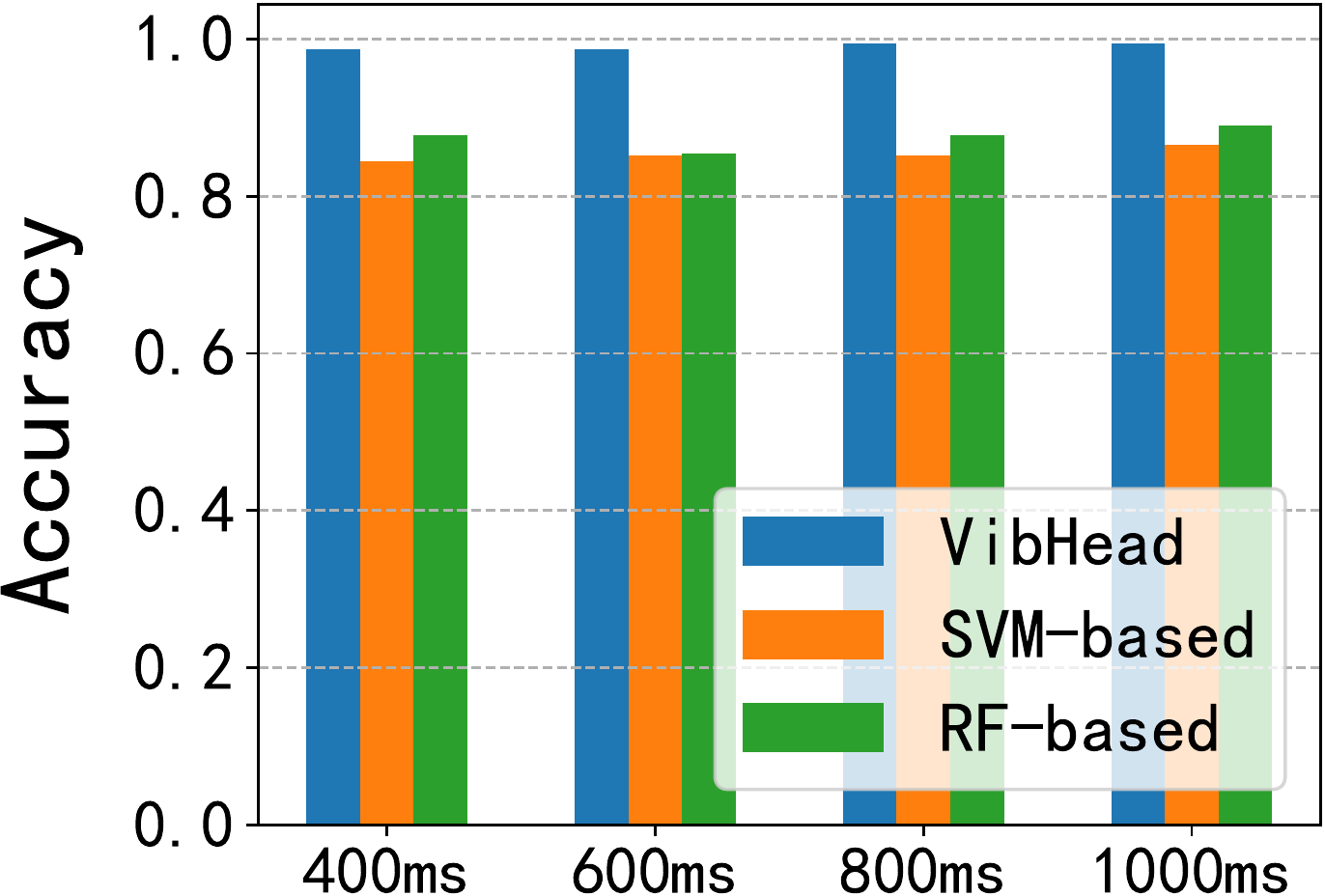}
      \caption{Classification accuracy of different models.} 
    \label{fig:classacc}
    \end{figure}

\begin{figure*}[htb!]
      \begin{center}
        \parbox{.45\textwidth}{\center\includegraphics[width=.4\textwidth]{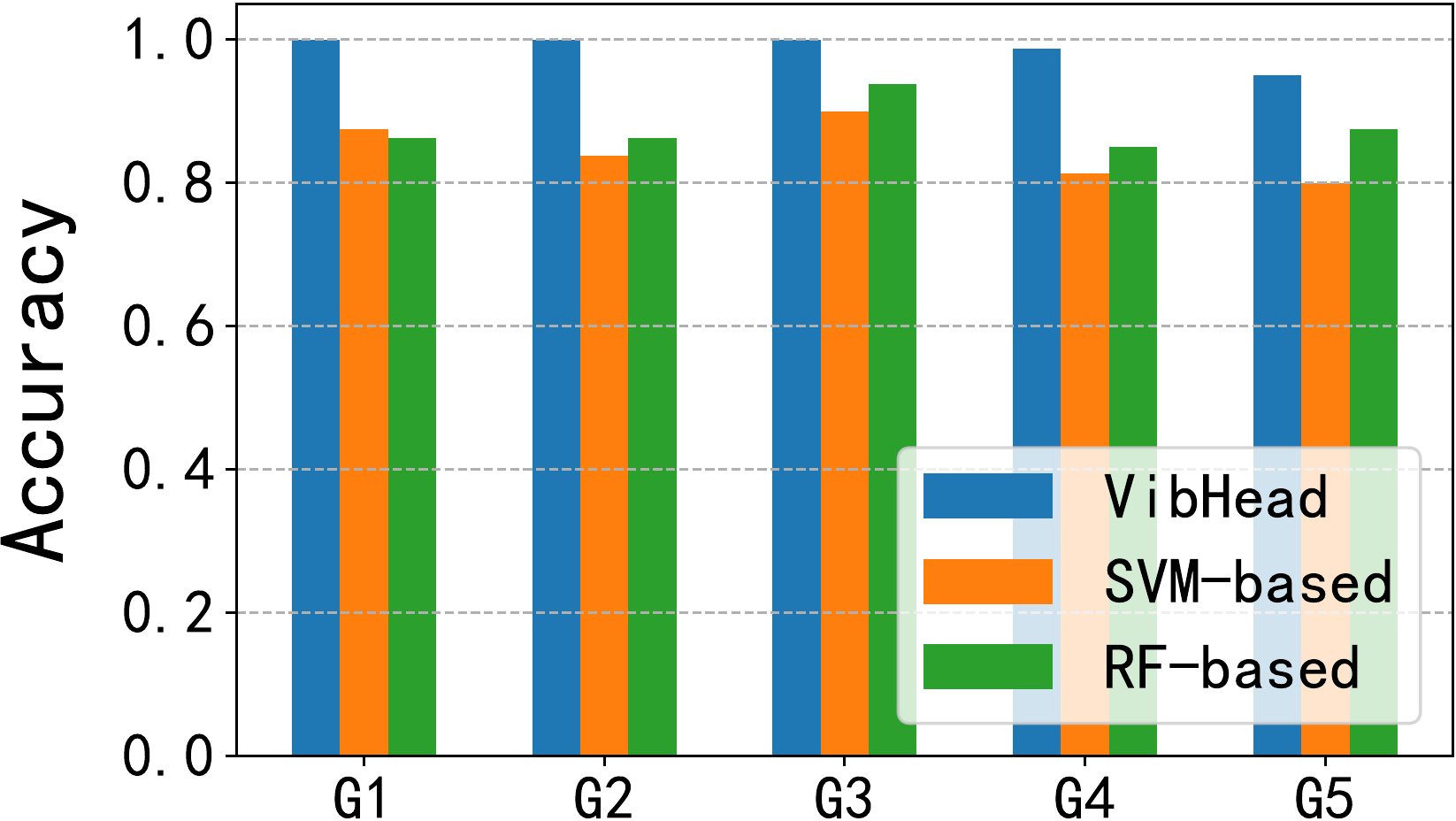}}
        \parbox{.45\textwidth}{\center\includegraphics[width=.4\textwidth]{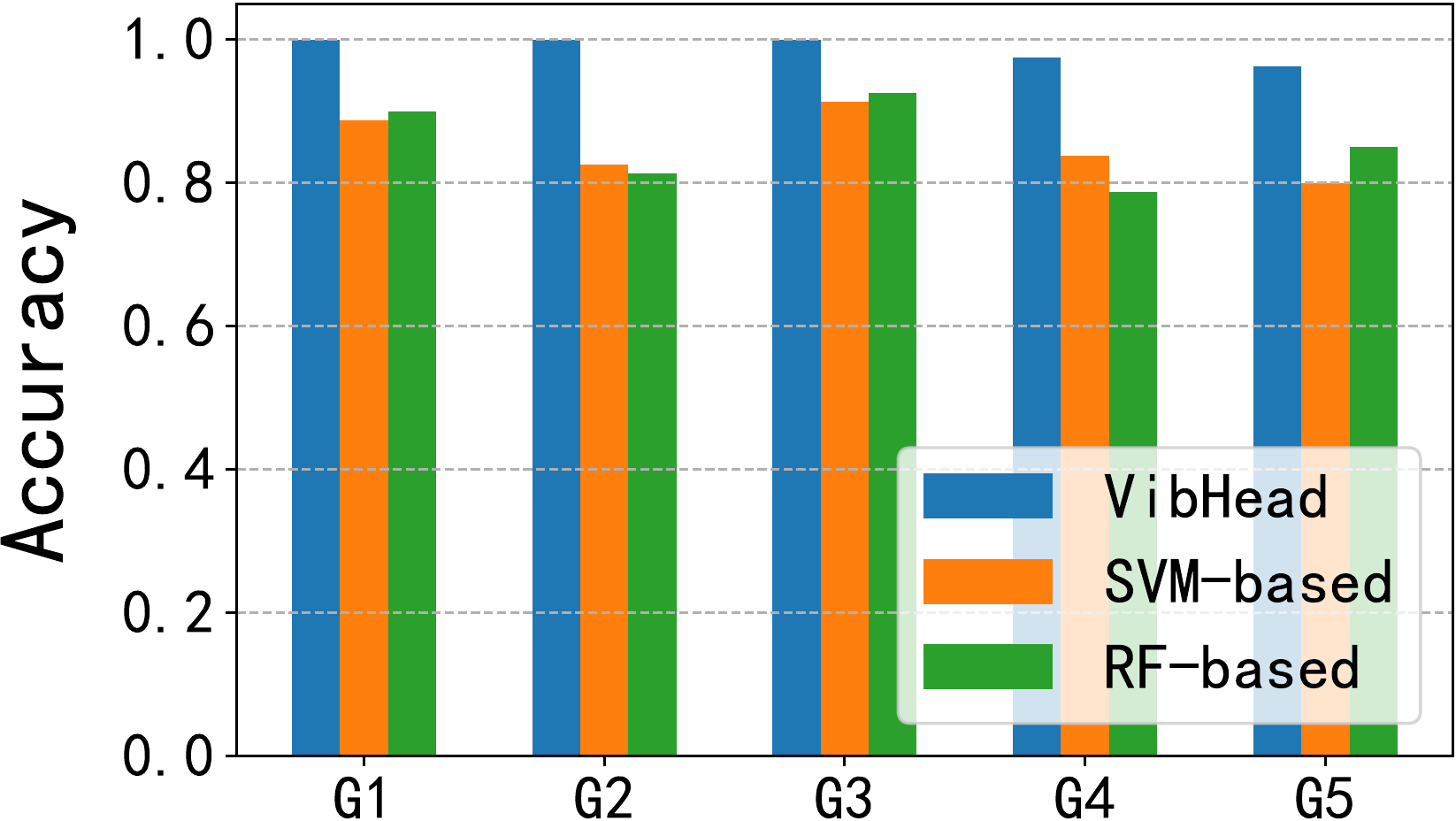}}
        \parbox{.45\columnwidth}{\center\scriptsize(a) $T=400ms$}
        \parbox{.45\columnwidth}{\center\scriptsize(b) $T=600ms$}
        \parbox{.45\textwidth}{\center\includegraphics[width=.4\textwidth]{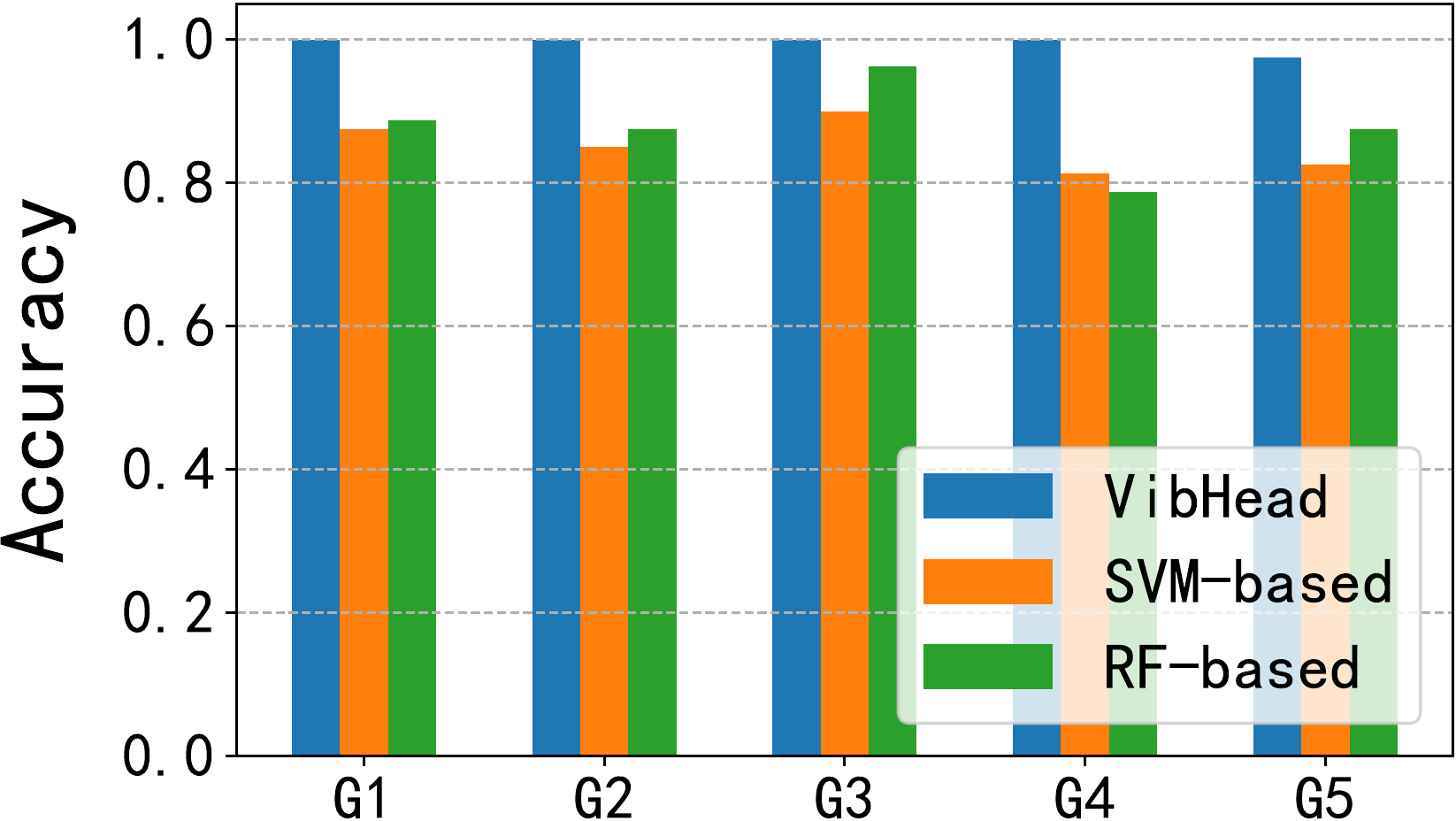}}
        \parbox{.45\textwidth}{\center\includegraphics[width=.4\textwidth]{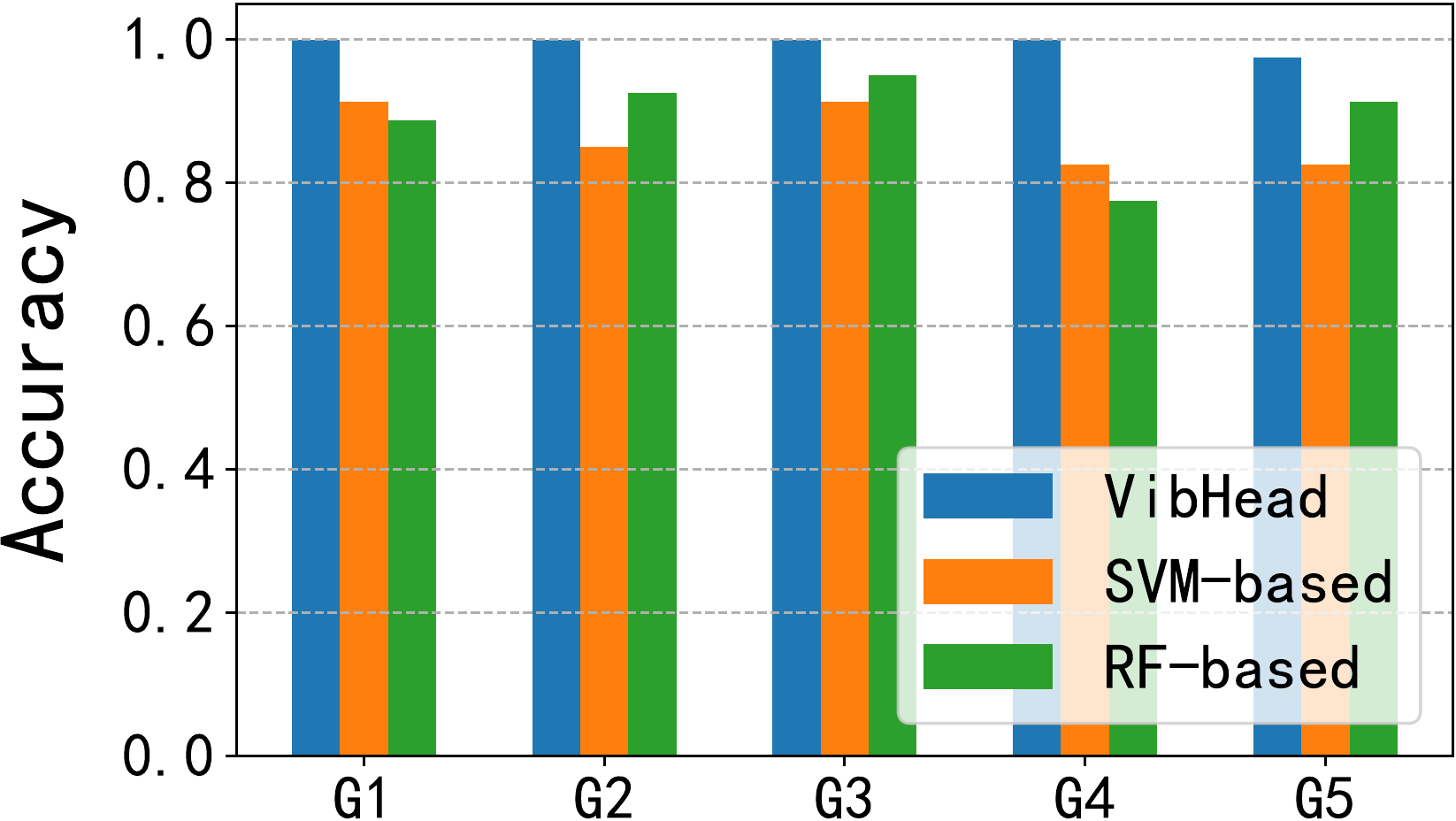}}
        \parbox{.45\columnwidth}{\center\scriptsize(c) $T=800ms$}
        \parbox{.45\columnwidth}{\center\scriptsize(d) $T=1s$}
      \caption{User classification accuracy with different gestures. G1, G2, G3, G4, and G5 stands for the five different gestures, i.e., standing, sitting upright, sitting-and-leaning-forward, sitting-and-leaning-backward, and walking, respectively.}
    \label{fig:classaccges}
    \end{center}
    \end{figure*}

    We also summarize the classification accuracy of the three different models in Fig.~\ref{fig:classacc}. The classification accuracy of VibHead is very close to $1$ even when $T=400$ms, while the one of the RF model is around $0.83$. The accuracy of the SVM model is even worse. It also can be observed that increasing the during of the data samples, i.e., $T$, could further improve the accuracy of our classification model. For example, when $T=1$s, the accuracy is quite close to $1$ such that almost all legitimate users can be correctly recognized.
    The classification accuracy for different gestures is given in Fig.~\ref{fig:classaccges}, where G1, G2, G3, G4, and G5 stands for the five different gestures, i.e., standing, sitting upright, sitting-and-leaning-forward, sitting-and-leaning-backward, and walking, respectively. As shown above, for all gestures, VibHead has higher classification accuracy than the SVM model and the RF model. Specifically, its classification accuracy for the four ``static'' poses (i.e., standing, sitting upright, sitting-and-leaning-forward and sitting-and-leaning-backward) achieves almost $1$. For the walking gesture, the classification accuracy of VibHead decreases; but it is still at least $0.92$. In contrast, the classification accuracy of the SVM model and the RF model for the walking gesture is may less than $0.8$.

\begin{figure}[htb!]
    \begin{center}
      \parbox{.45\textwidth}{\center\includegraphics[width=.4\textwidth]{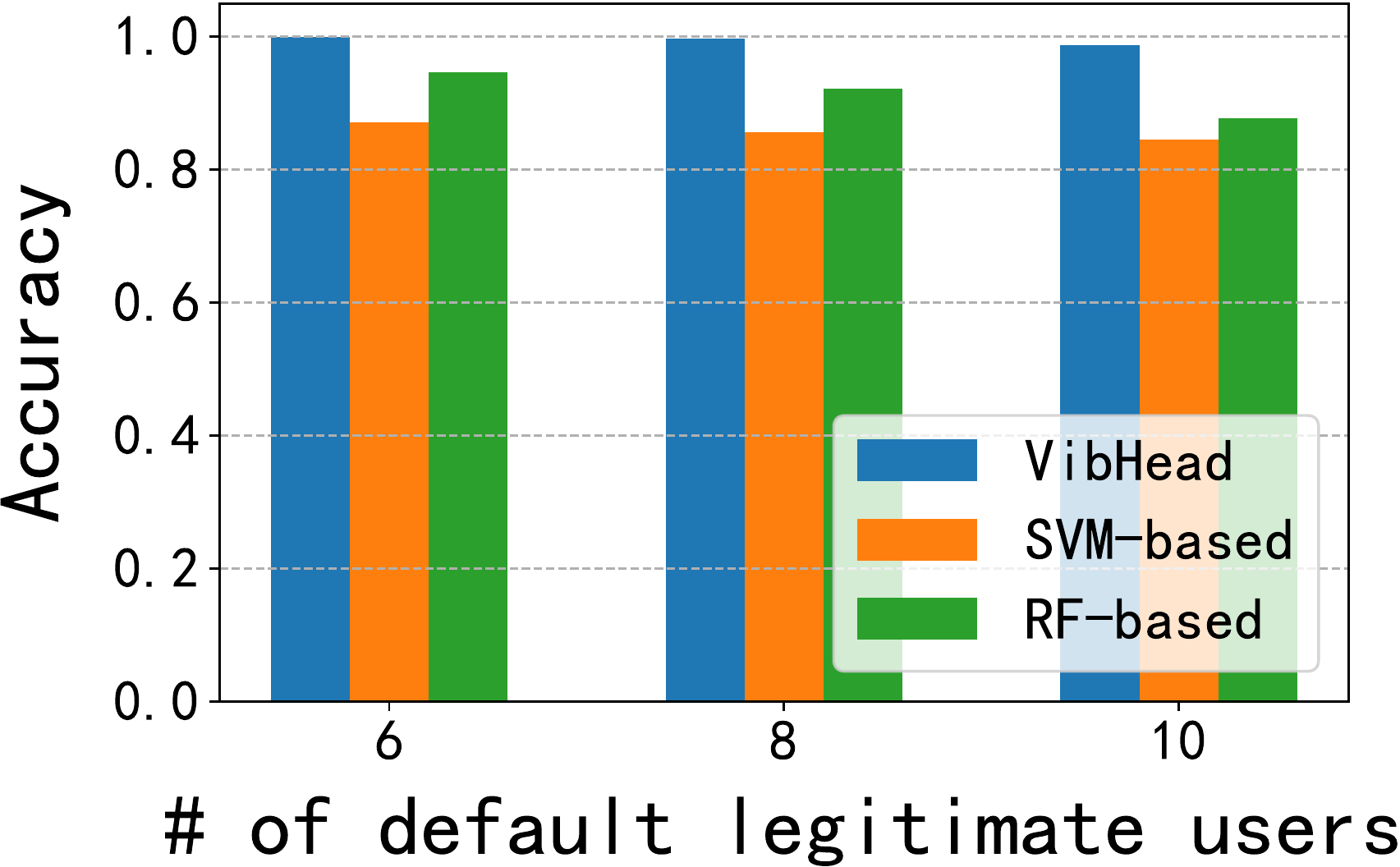}}
      \parbox{.45\textwidth}{\center\includegraphics[width=.4\textwidth]{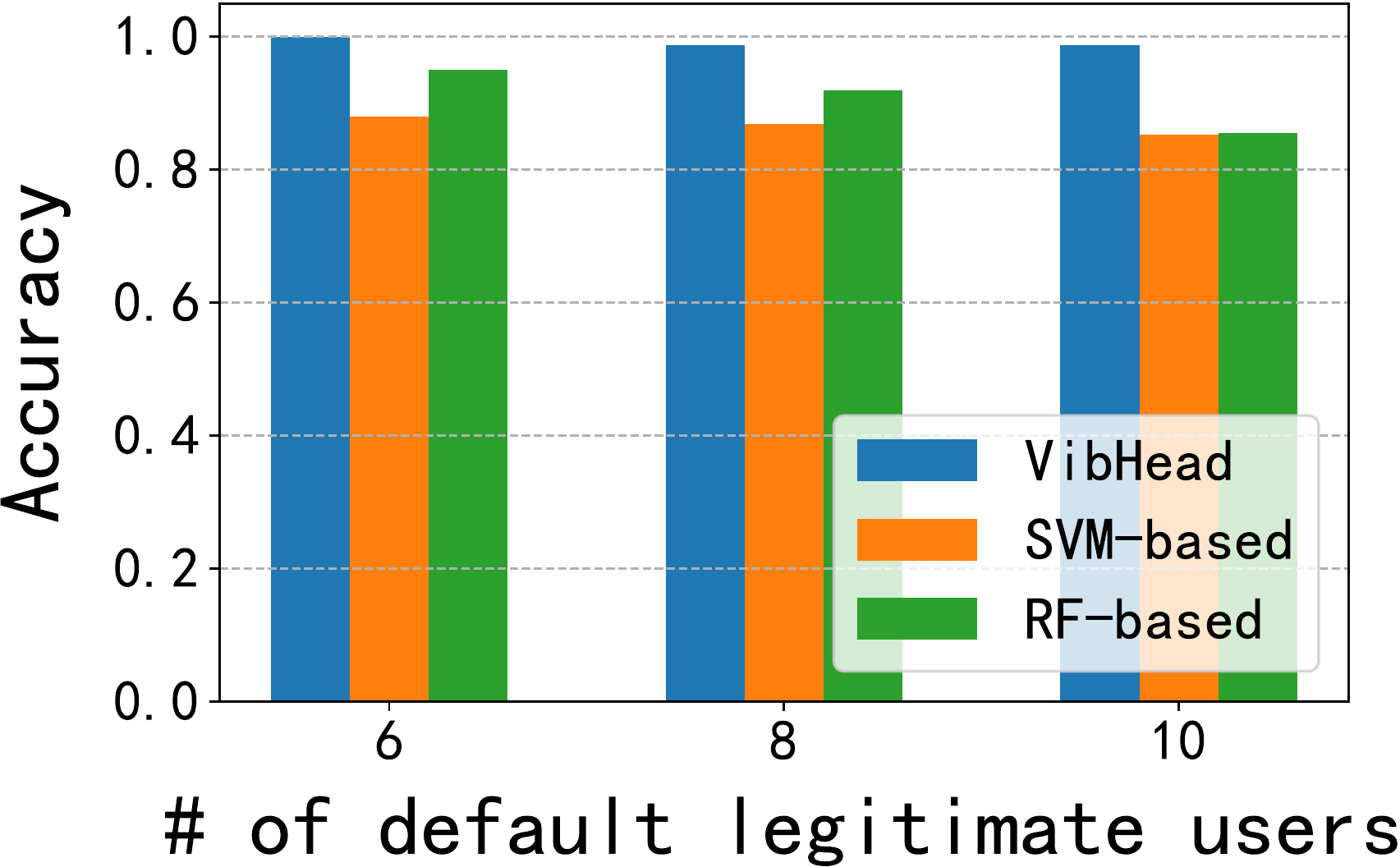}}
      \parbox{.45\columnwidth}{\center\scriptsize(a) $T=400ms$}
      \parbox{.45\columnwidth}{\center\scriptsize(b) $T=600ms$}
      \parbox{.45\textwidth}{\center\includegraphics[width=.4\textwidth]{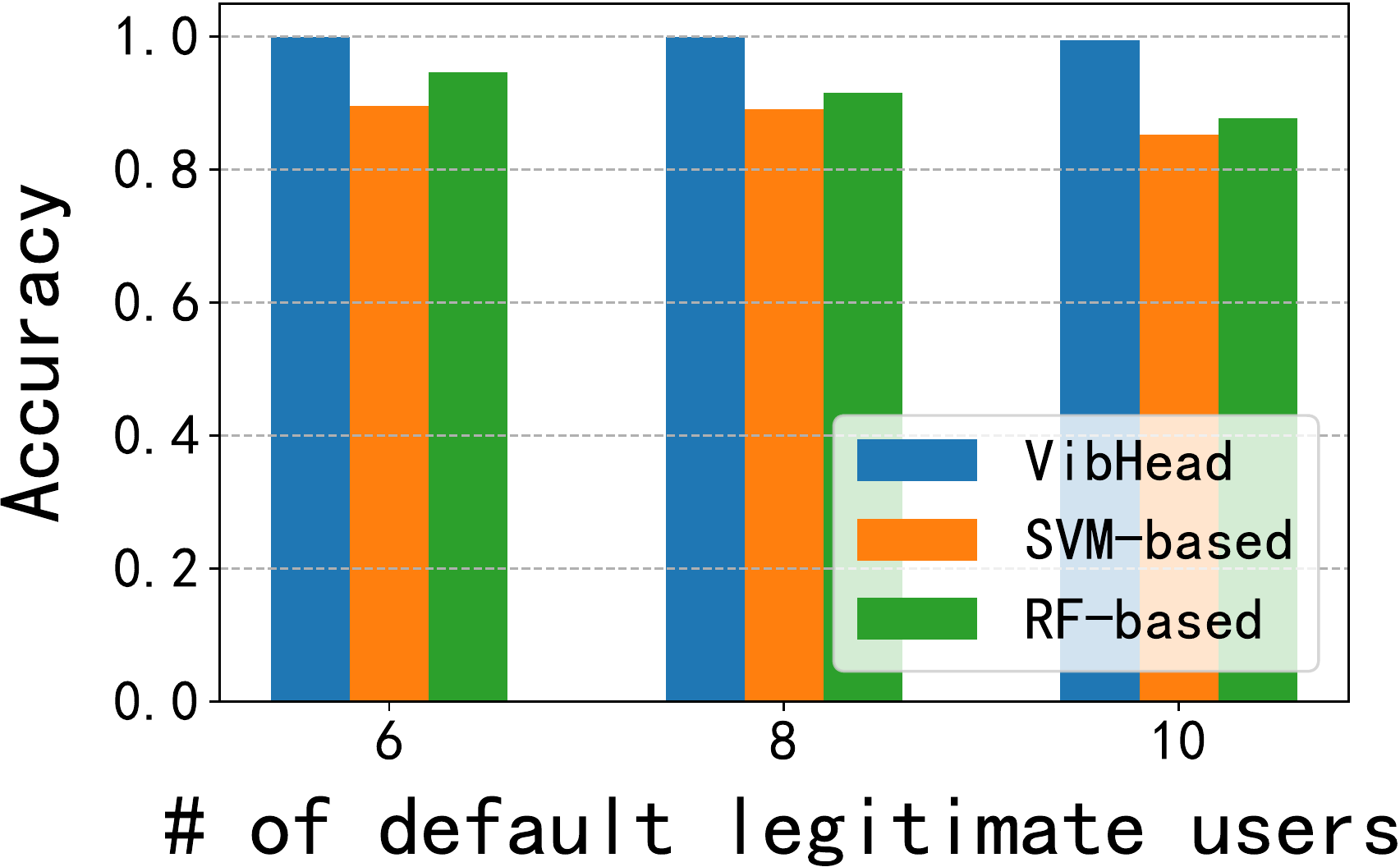}}
      \parbox{.45\textwidth}{\center\includegraphics[width=.4\textwidth]{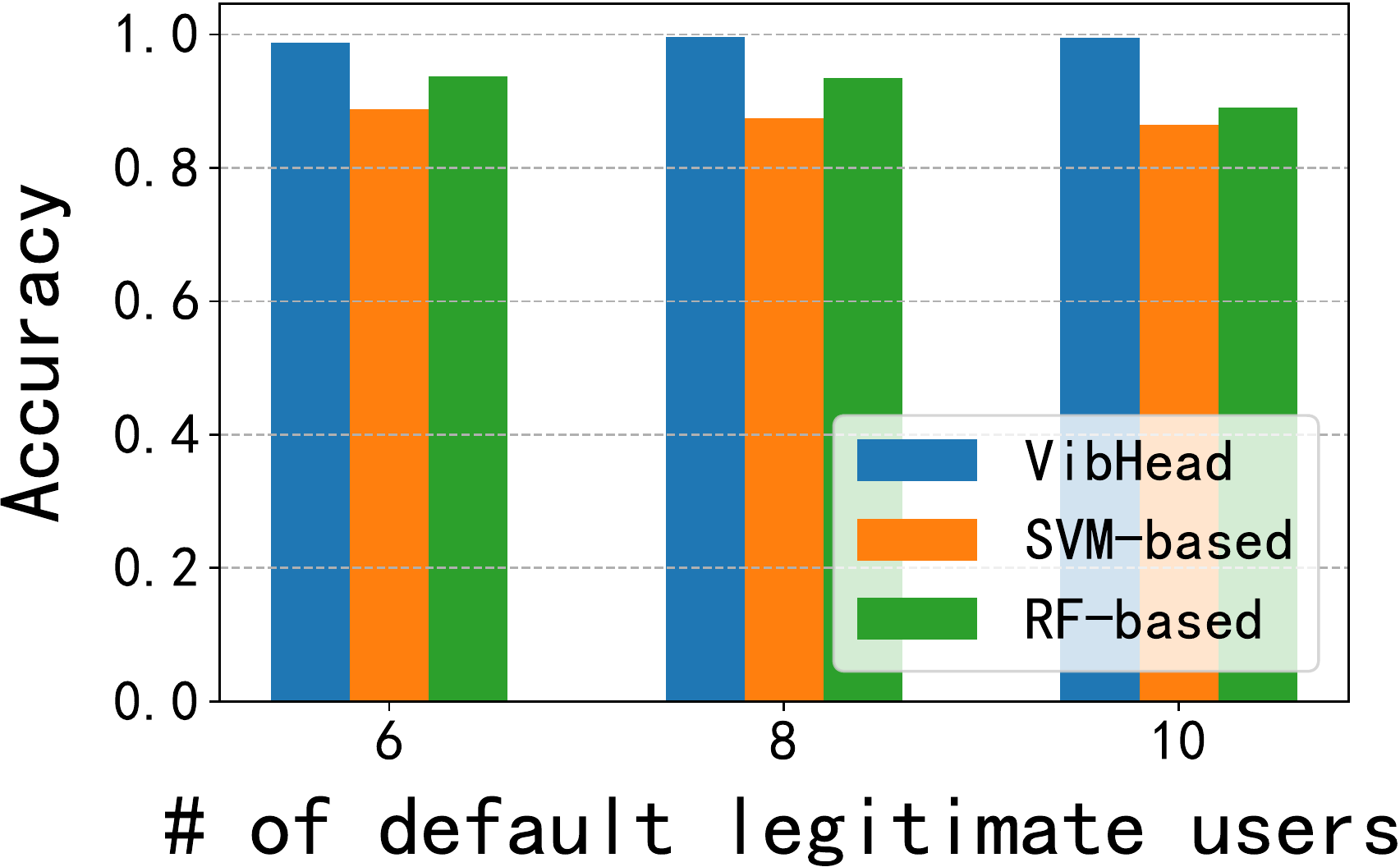}}
      \parbox{.45\columnwidth}{\center\scriptsize(c) $T=800ms$}
      \parbox{.45\columnwidth}{\center\scriptsize(d) $T=1s$}
    \caption{Classification accuracy of different models with different numbers of default legitimate users.}
    \label{fig:classaccdiffuser}
    \end{center}
    \end{figure}

    We also demonstrate how the number of default legitimate users impacts the classification accuracy. We vary the number of the default legitimate users from $6$ to $10$ and the time length of the data samples from $400$ms to $1$s. The comparison results between our VibHead and the other two reference methods are presented in Fig.~\ref{fig:classaccdiffuser}. It is revealed that, the classification accuracy of our VibHead is impacted very slightly by the number of default legitimate users. The accuracy is very close to $1$ even when there are $10$ default legitimate users. Both the SVM model and the RF model have lower accuracy, across the different numbers of default legitimate users. When there are more default legitimate users, the two reference models have their accuracy decreased noticeably.

  \subsection{Authentication Performance} \label{ssec:authperf}
    \subsubsection{FAR and FRR}
\begin{figure*}[htb!]
    \begin{center}
      \parbox{.32\textwidth}{\center\includegraphics[width=.32\textwidth]{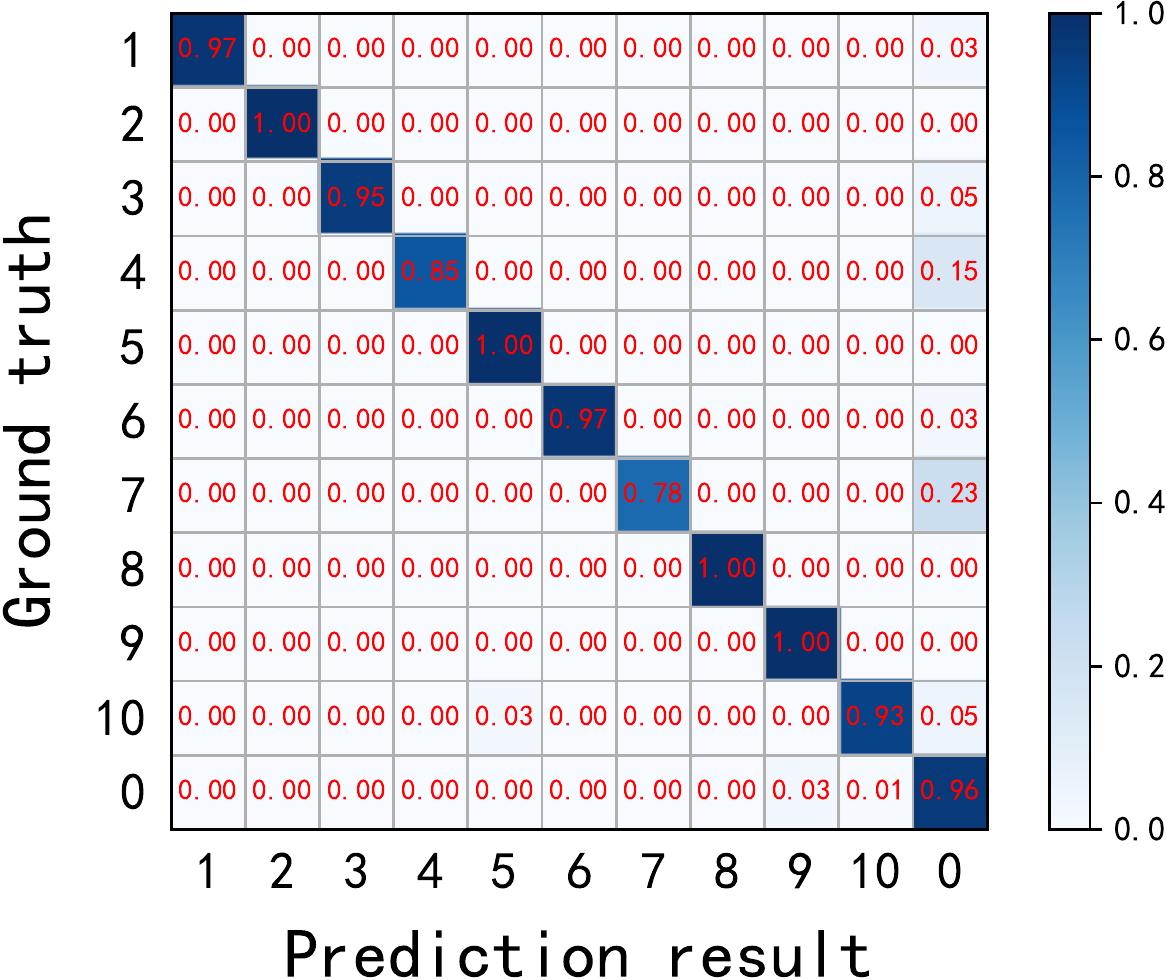}}
      \parbox{.32\textwidth}{\center\includegraphics[width=.32\textwidth]{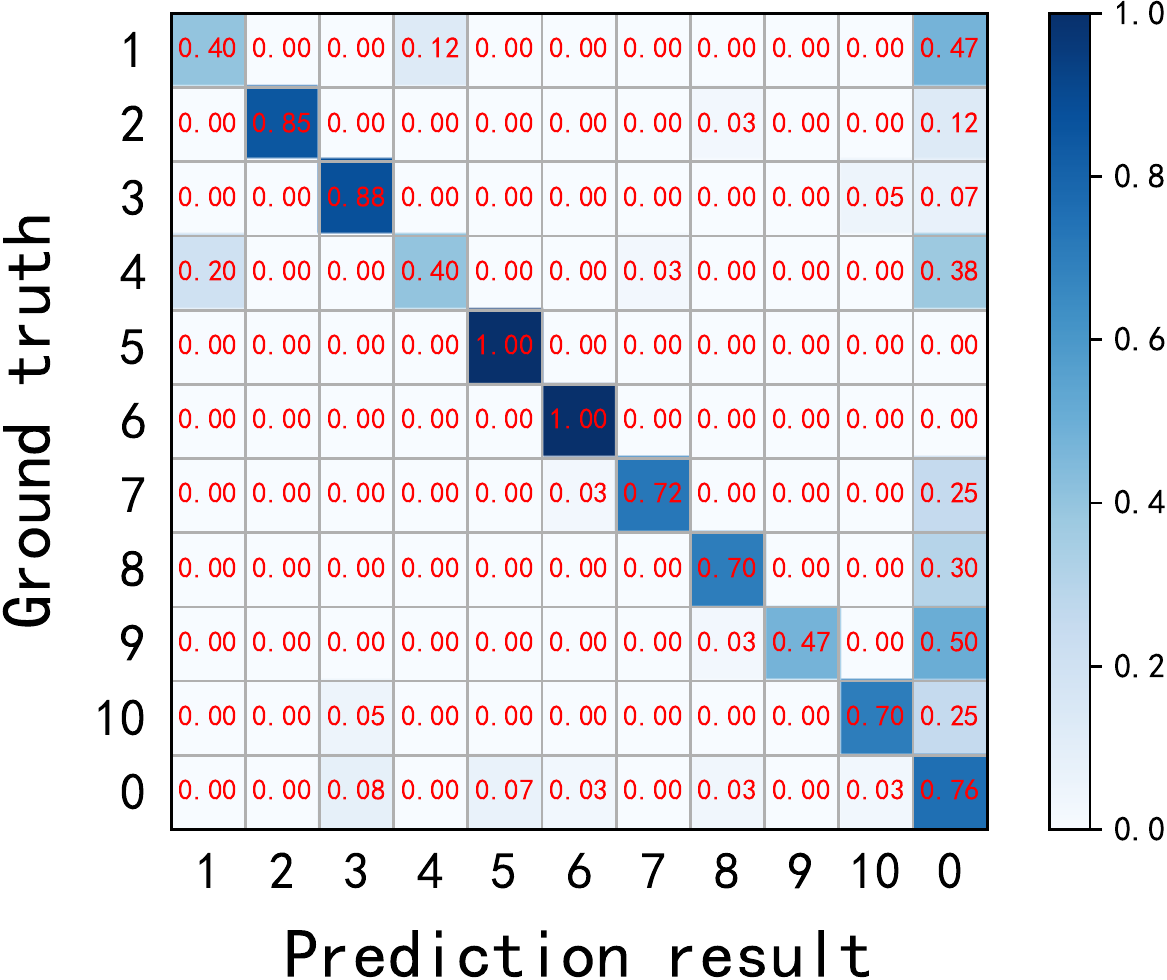}}
      \parbox{.32\textwidth}{\center\includegraphics[width=.32\textwidth]{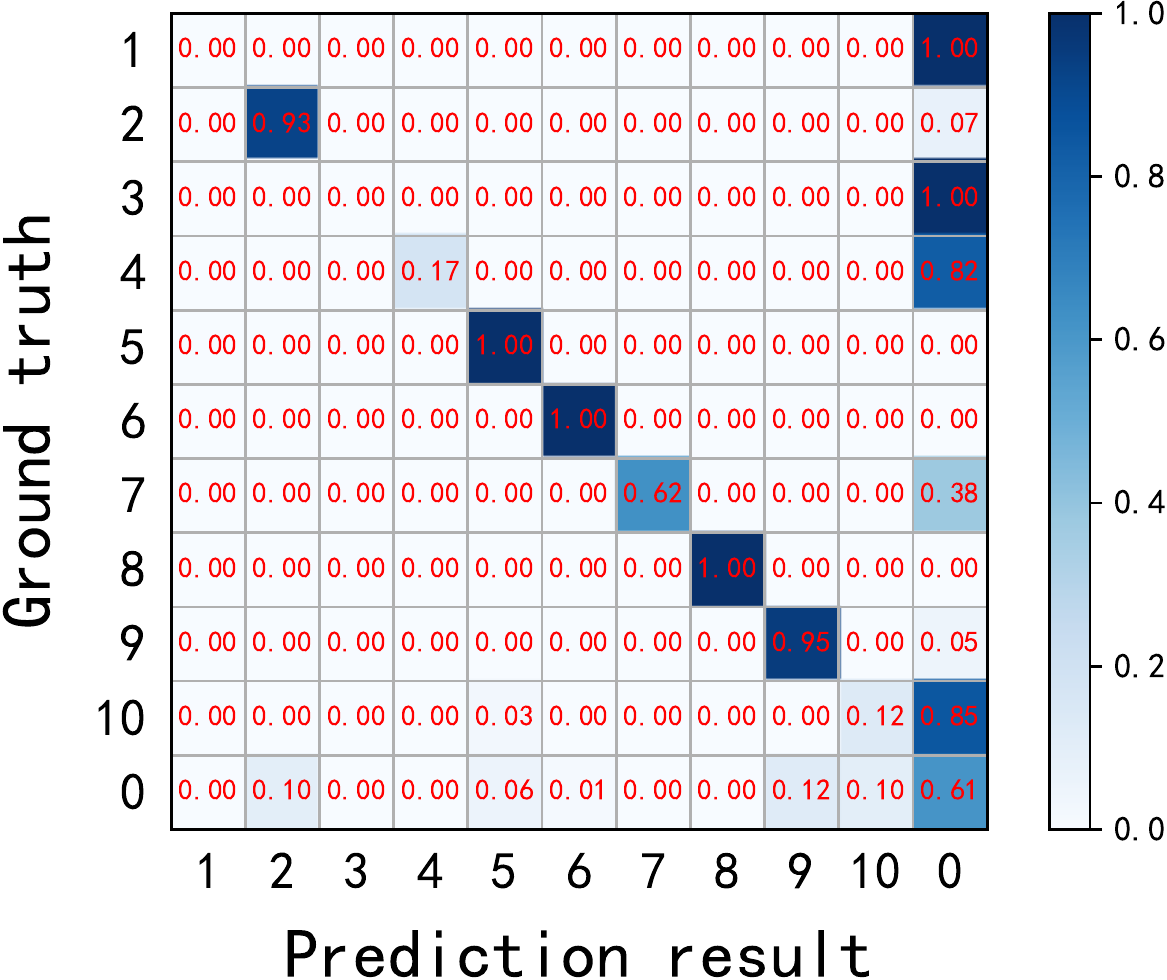}}
      \parbox{.32\columnwidth}{\center\scriptsize(a1) VibHead with $T=400ms$}
      \parbox{.32\columnwidth}{\center\scriptsize(a2) SVM with $T=400ms$}
      \parbox{.32\columnwidth}{\center\scriptsize(a3) RF with $T=400ms$}
      \parbox{.32\textwidth}{\center\includegraphics[width=.32\textwidth]{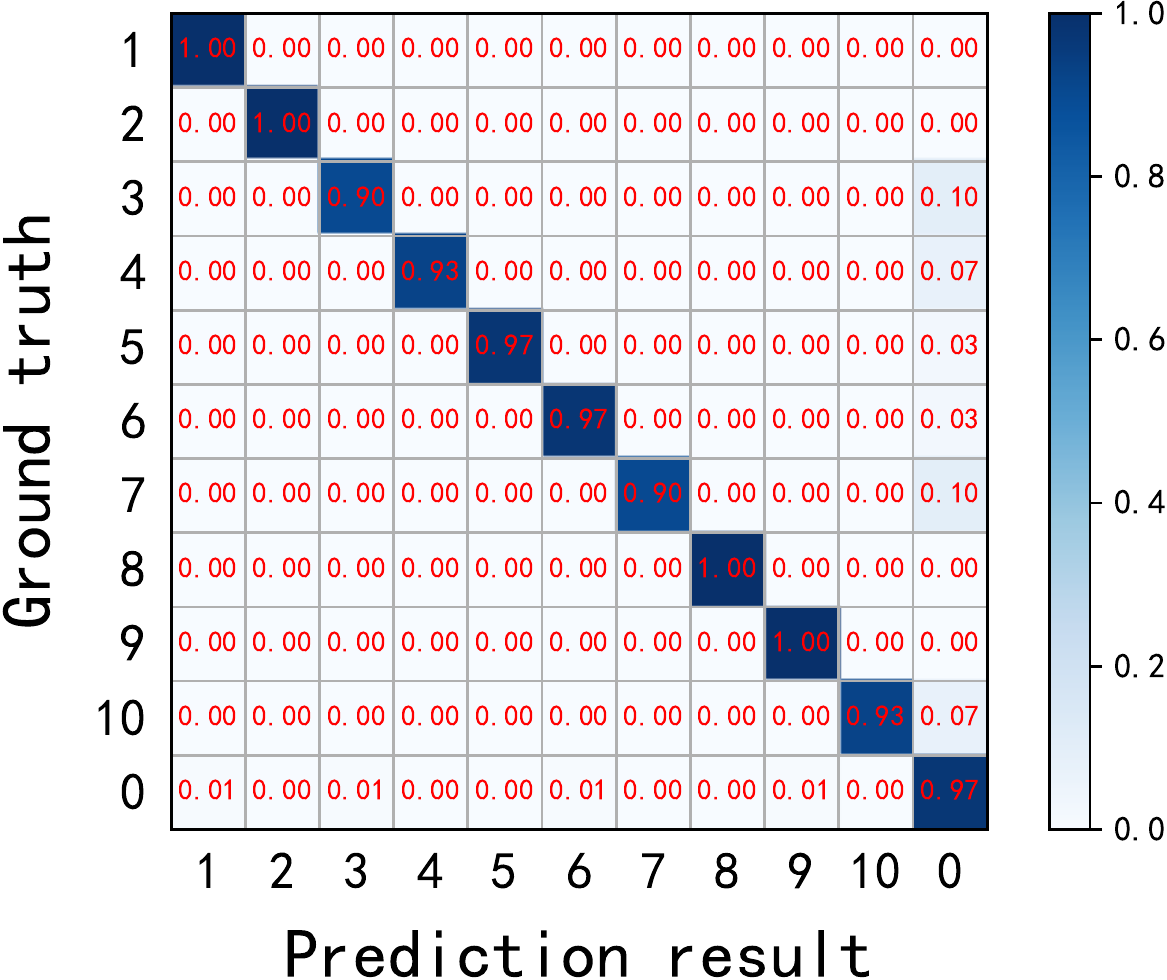}}
      \parbox{.32\textwidth}{\center\includegraphics[width=.32\textwidth]{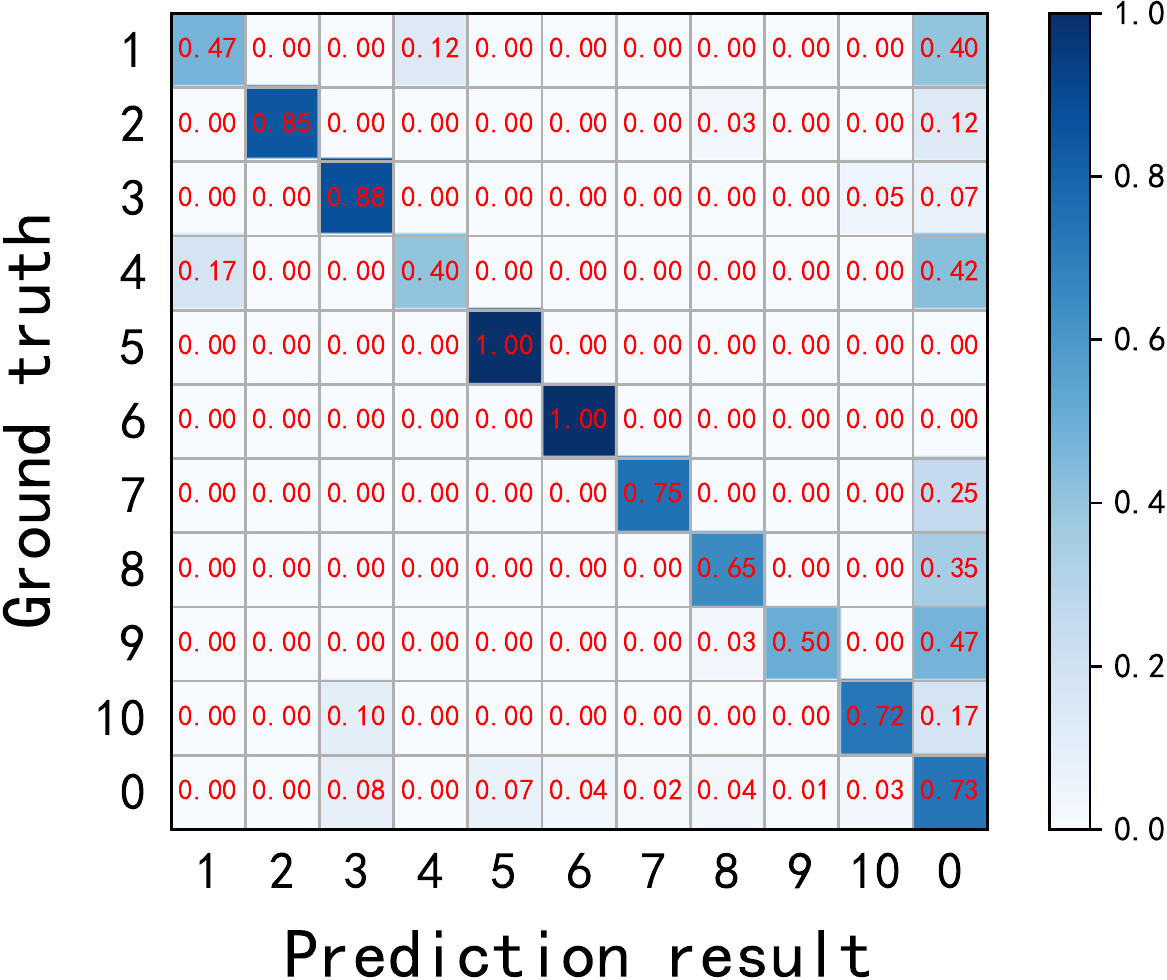}}
      \parbox{.32\textwidth}{\center\includegraphics[width=.32\textwidth]{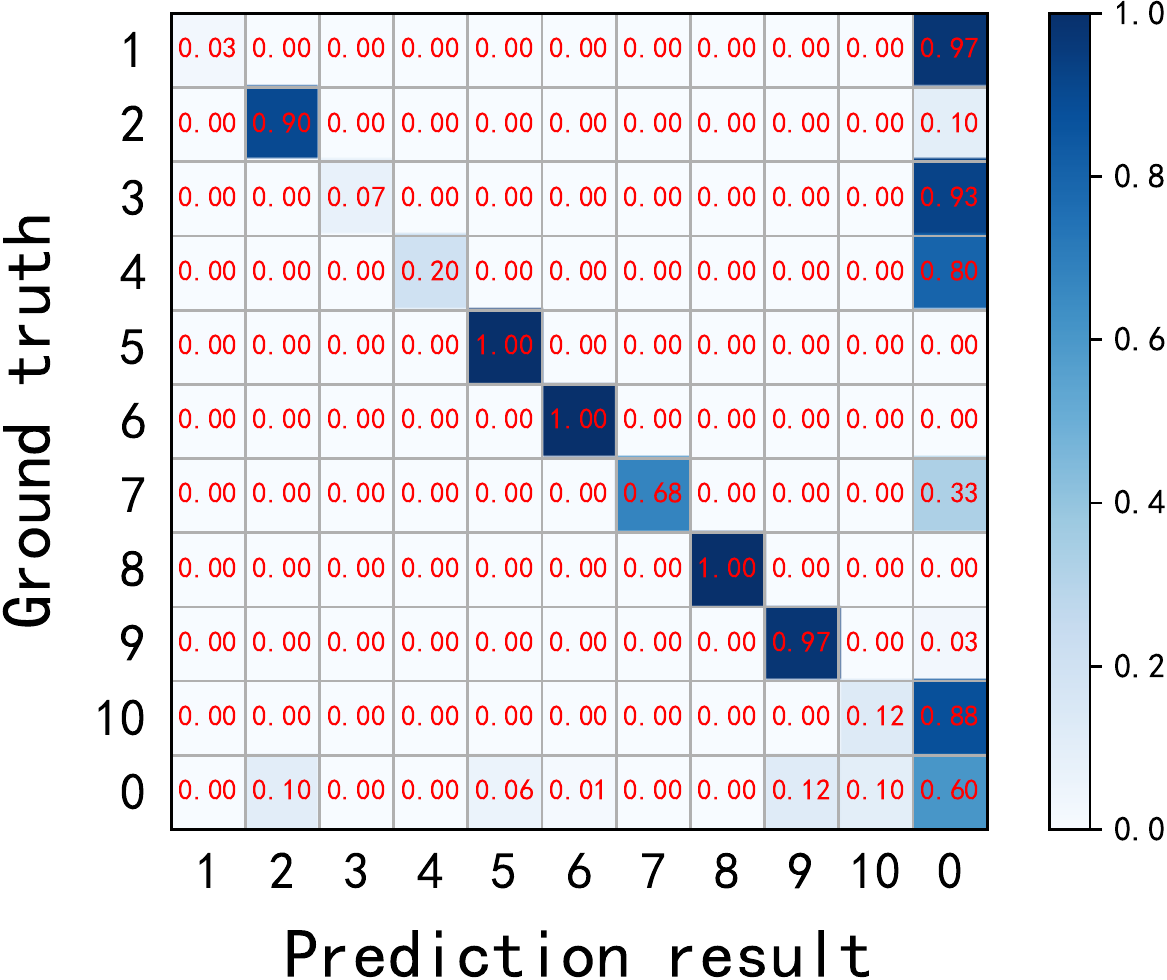}}
      \parbox{.32\columnwidth}{\center\scriptsize(a1) VibHead with $T=600ms$}
      \parbox{.32\columnwidth}{\center\scriptsize(a2) SVM with $T=600ms$}
      \parbox{.32\columnwidth}{\center\scriptsize(a3) RF with $T=600ms$}
      \parbox{.32\textwidth}{\center\includegraphics[width=.32\textwidth]{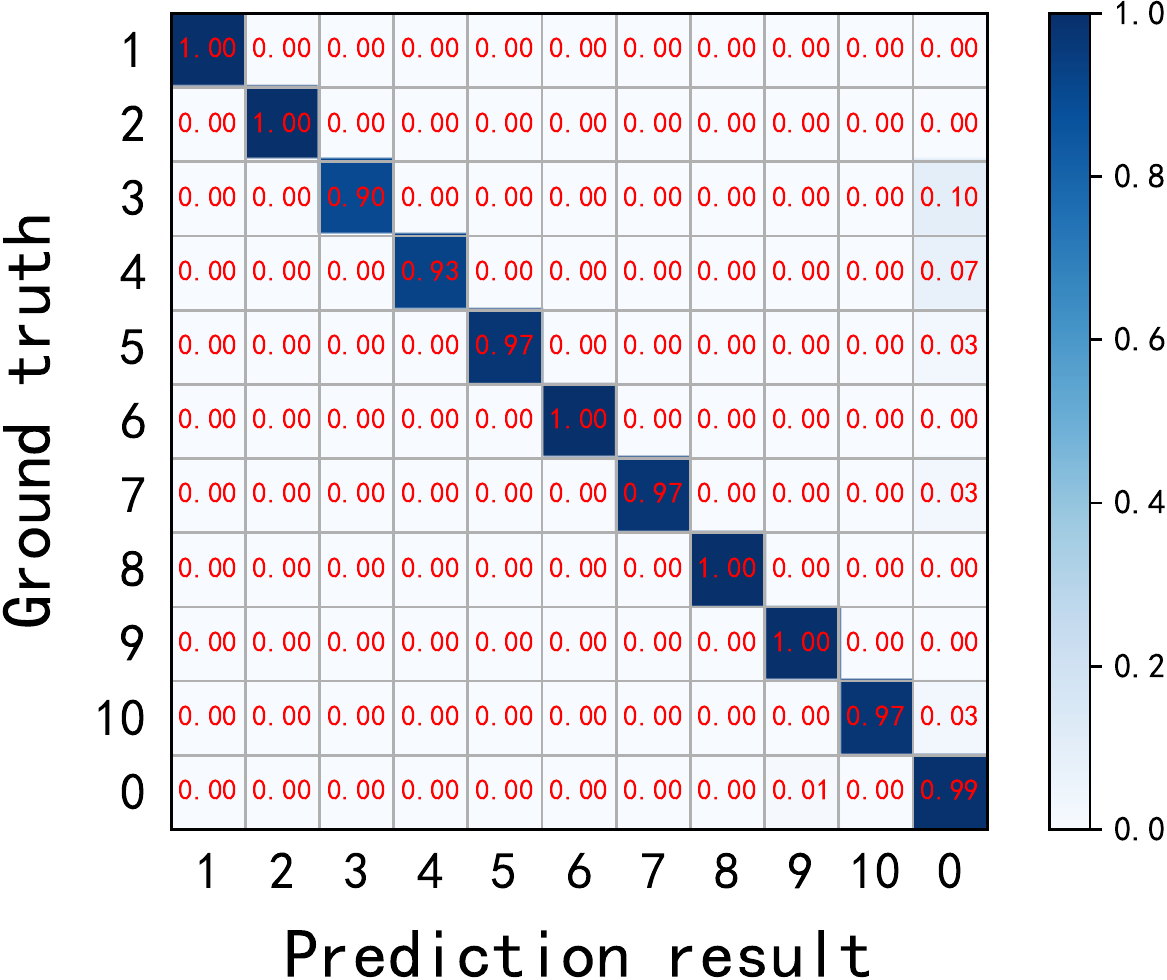}}
      \parbox{.32\textwidth}{\center\includegraphics[width=.32\textwidth]{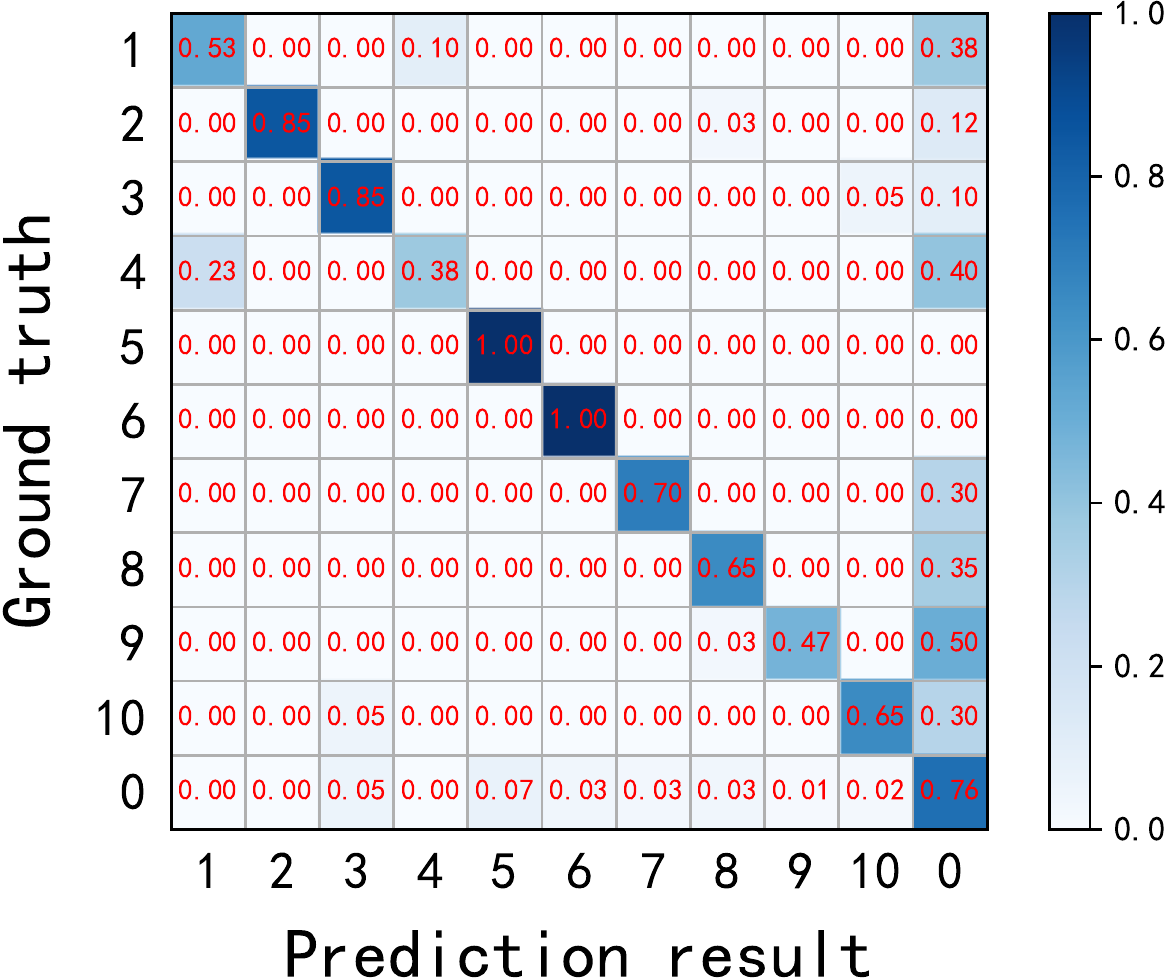}}
      \parbox{.32\textwidth}{\center\includegraphics[width=.32\textwidth]{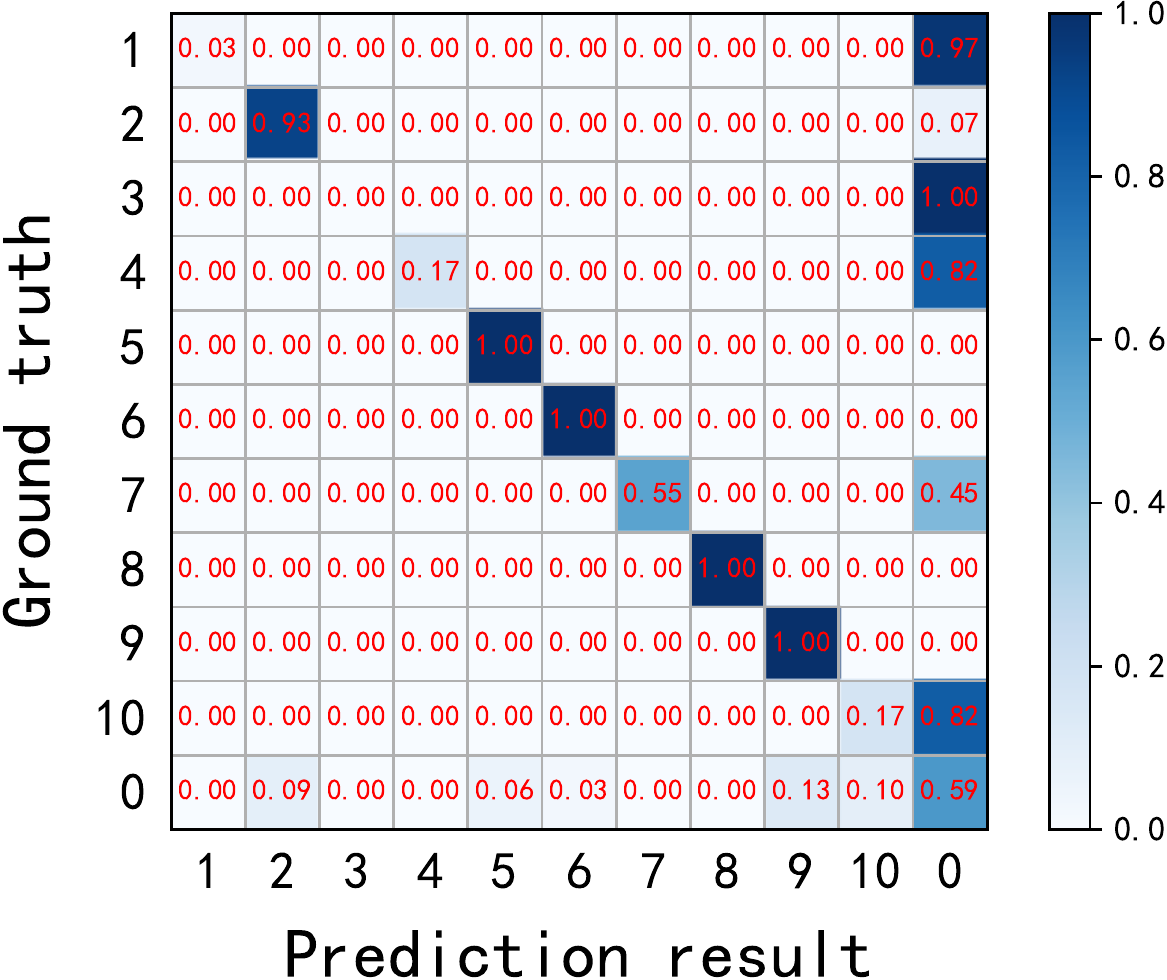}}
      \parbox{.32\columnwidth}{\center\scriptsize(a1) VibHead with $T=800ms$}
      \parbox{.32\columnwidth}{\center\scriptsize(a2) SVM with $T=800ms$}
      \parbox{.32\columnwidth}{\center\scriptsize(a3) RF with $T=800ms$}
      \parbox{.32\textwidth}{\center\includegraphics[width=.32\textwidth]{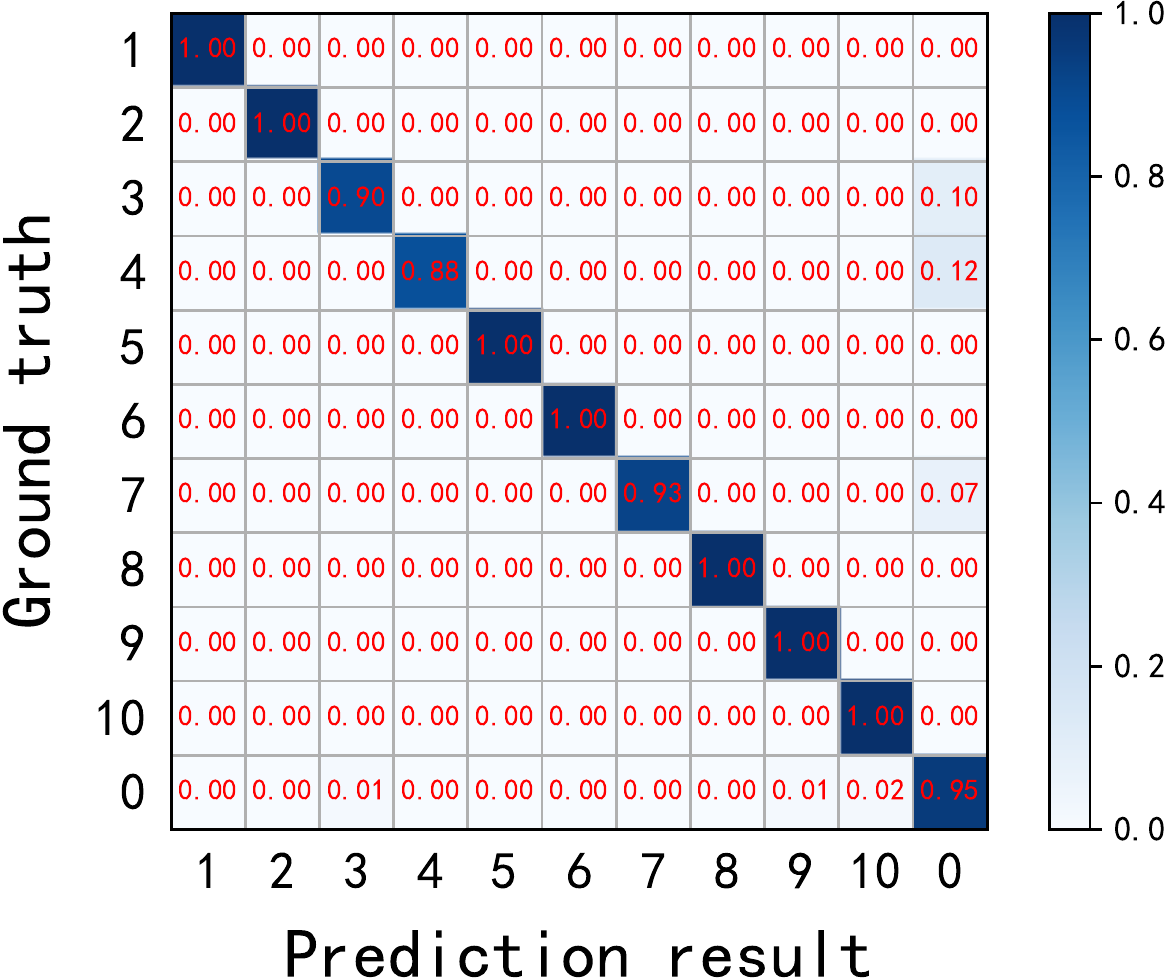}}
      \parbox{.32\textwidth}{\center\includegraphics[width=.32\textwidth]{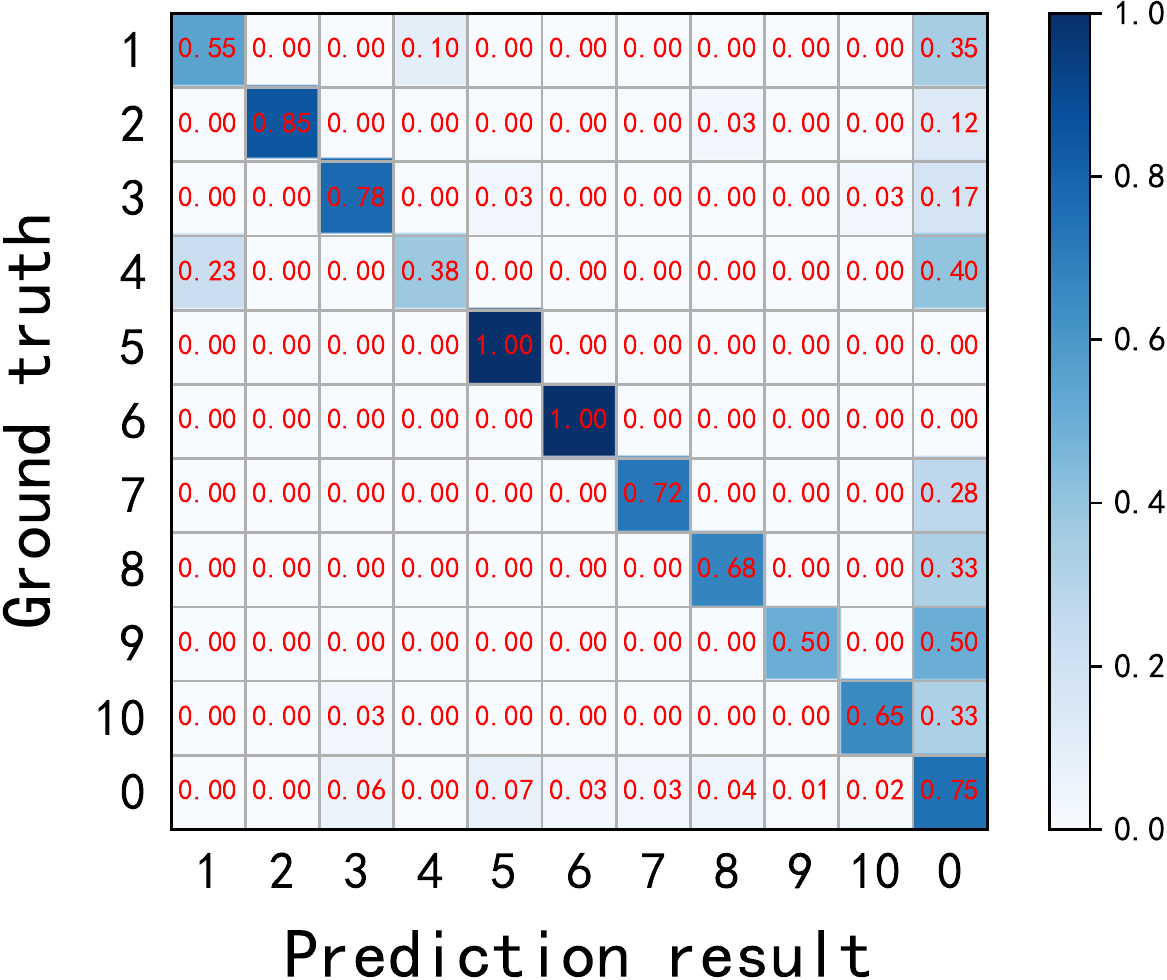}}
      \parbox{.32\textwidth}{\center\includegraphics[width=.32\textwidth]{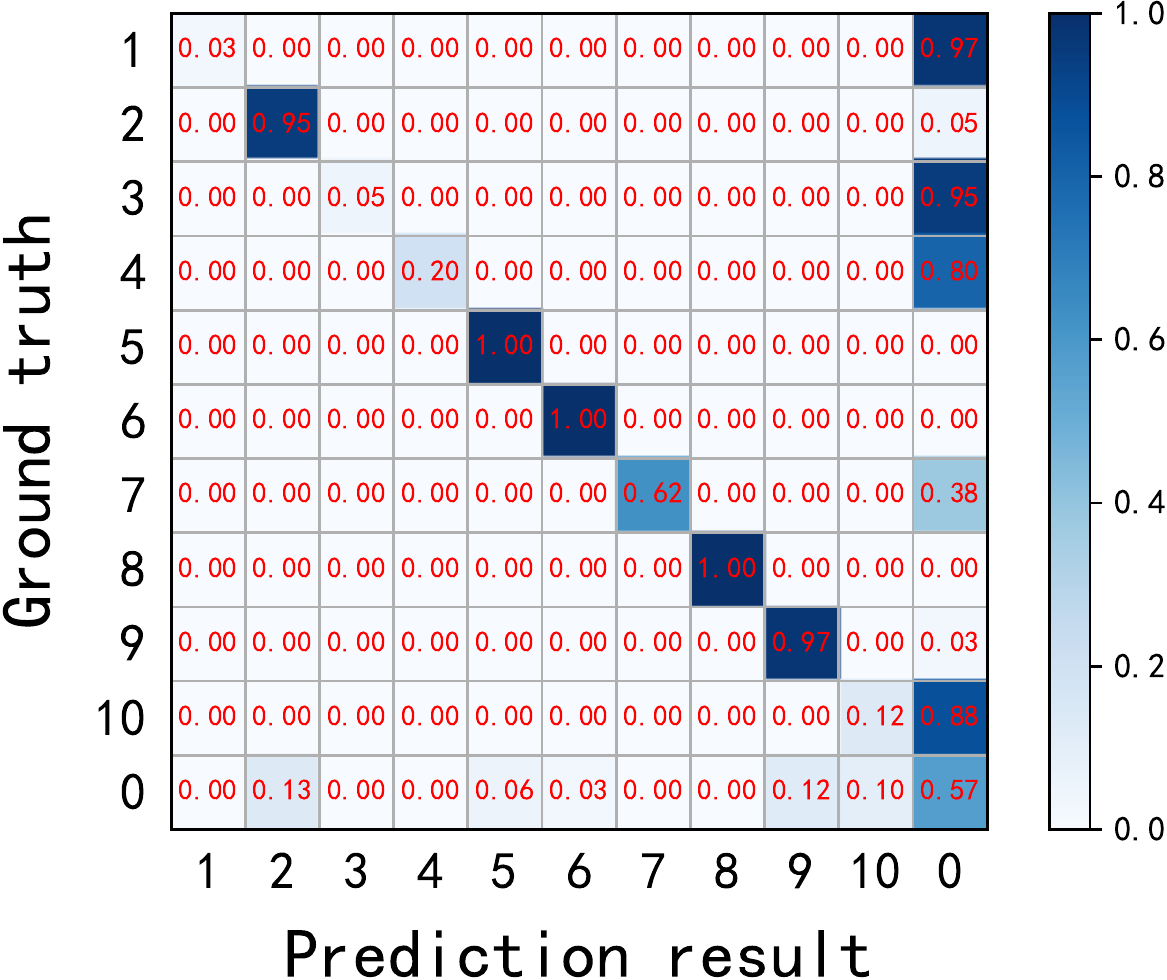}}
      \parbox{.32\columnwidth}{\center\scriptsize(a1) VibHead with $T=1000ms$}
      \parbox{.32\columnwidth}{\center\scriptsize(a2) SVM with $T=1000ms$}
      \parbox{.32\columnwidth}{\center\scriptsize(a3) RF with $T=1000ms$}
    \caption{Confusion matrices of user authentication. We use ``$0$'' to represent illegitimate users.}
    \label{fig:authconfmat}
    \end{center}
    \end{figure*}

    We hereby report the performances of the three methods in terms of user authentication. The authentication confusion matrices are illustrated in Fig~\ref{fig:authconfmat}. It is observed that, VibHead not only identifies the individual legitimate users but also recognizes the illegitimate users correctly. In contrast, the probability for the SVM-based and RF-based methods to mistakenly authenticate the login users is much higher.

\begin{figure}[htbp]
    \centering
    \begin{minipage}[t]{0.48\columnwidth}
    \centering
      \includegraphics[width=0.8\columnwidth]{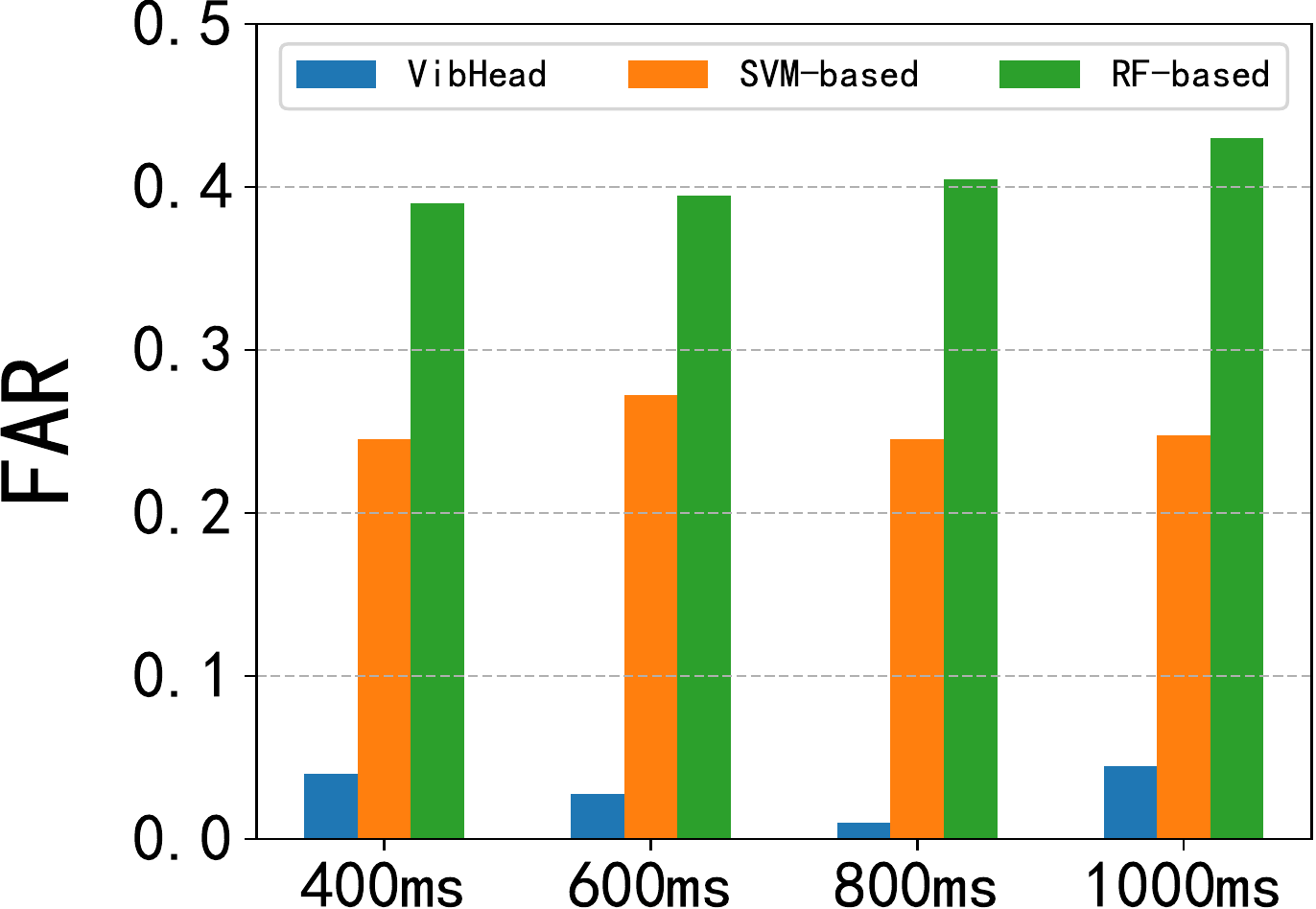}
    \caption{FARs of different authentication schemes.}
    \label{fig:far}
    \end{minipage}
    \begin{minipage}[t]{0.48\columnwidth}
    \centering
      \includegraphics[width=0.8\columnwidth]{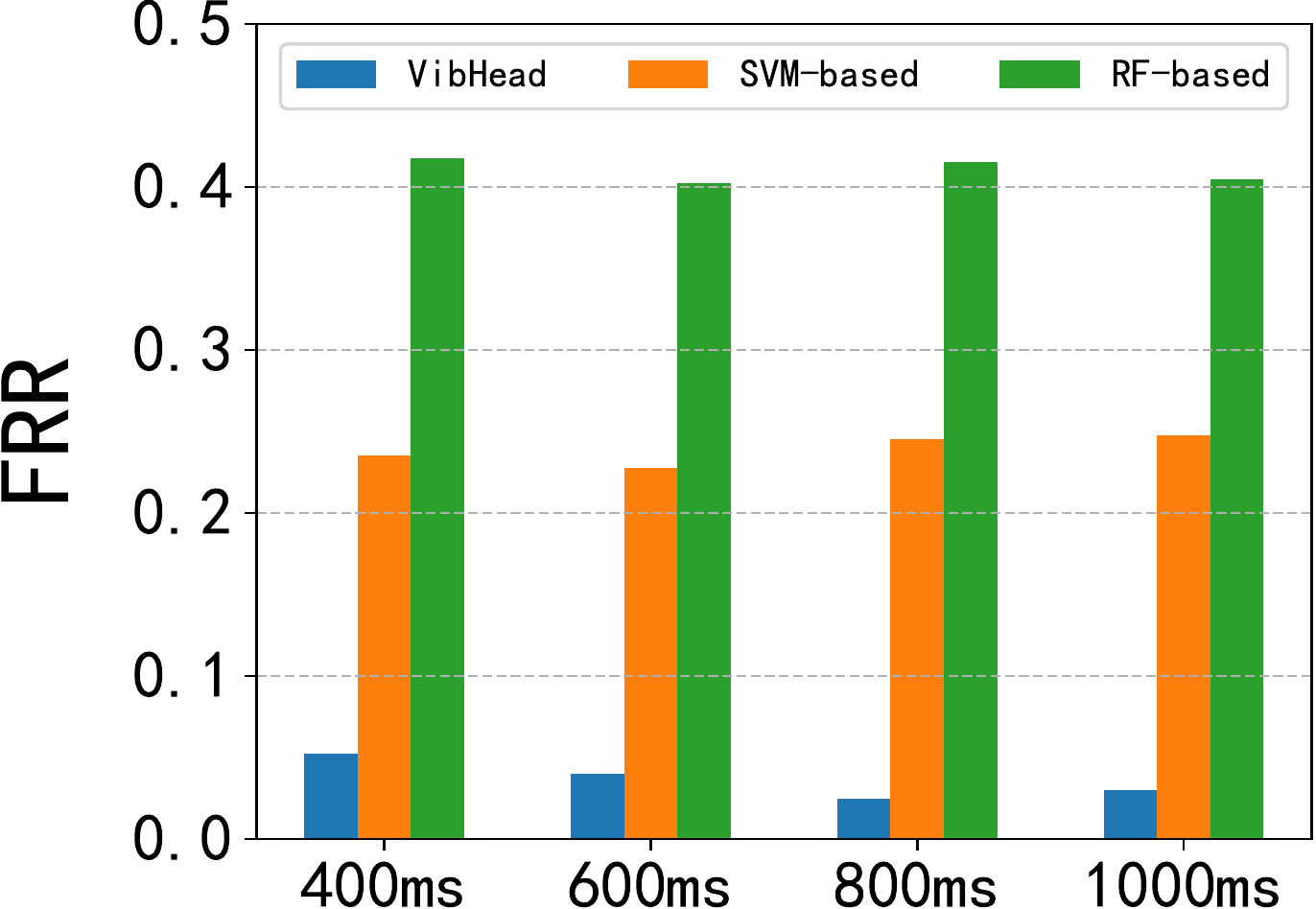}
    \caption{FRRs of different authentication schemes.}
    \label{fig:frr}
    \end{minipage}
    \end{figure}

\begin{figure*}[htb!]
    \begin{center}
      \parbox{.43\textwidth}{\center\includegraphics[width=.4\textwidth]{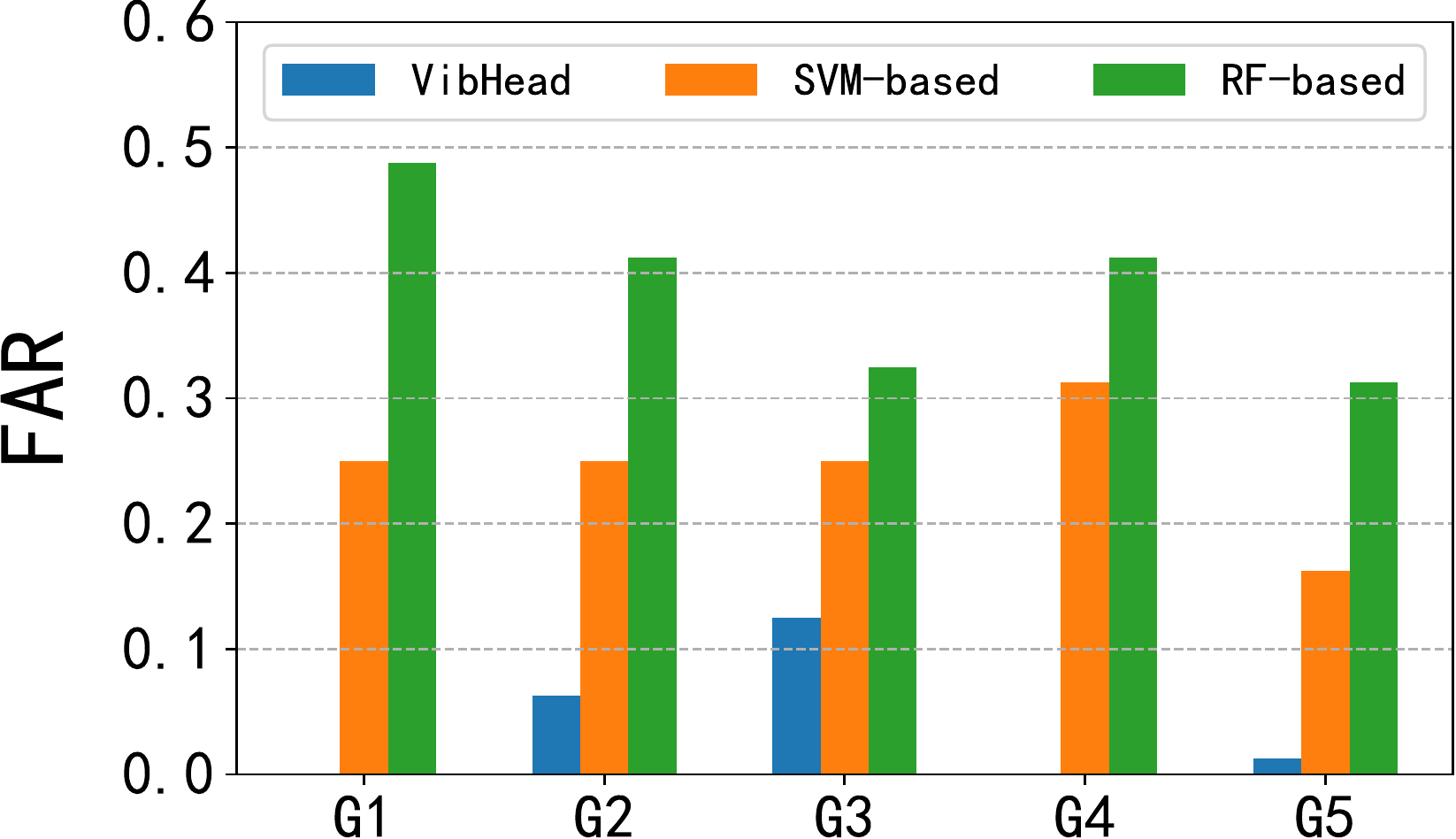}}
      \parbox{.43\textwidth}{\center\includegraphics[width=.4\textwidth]{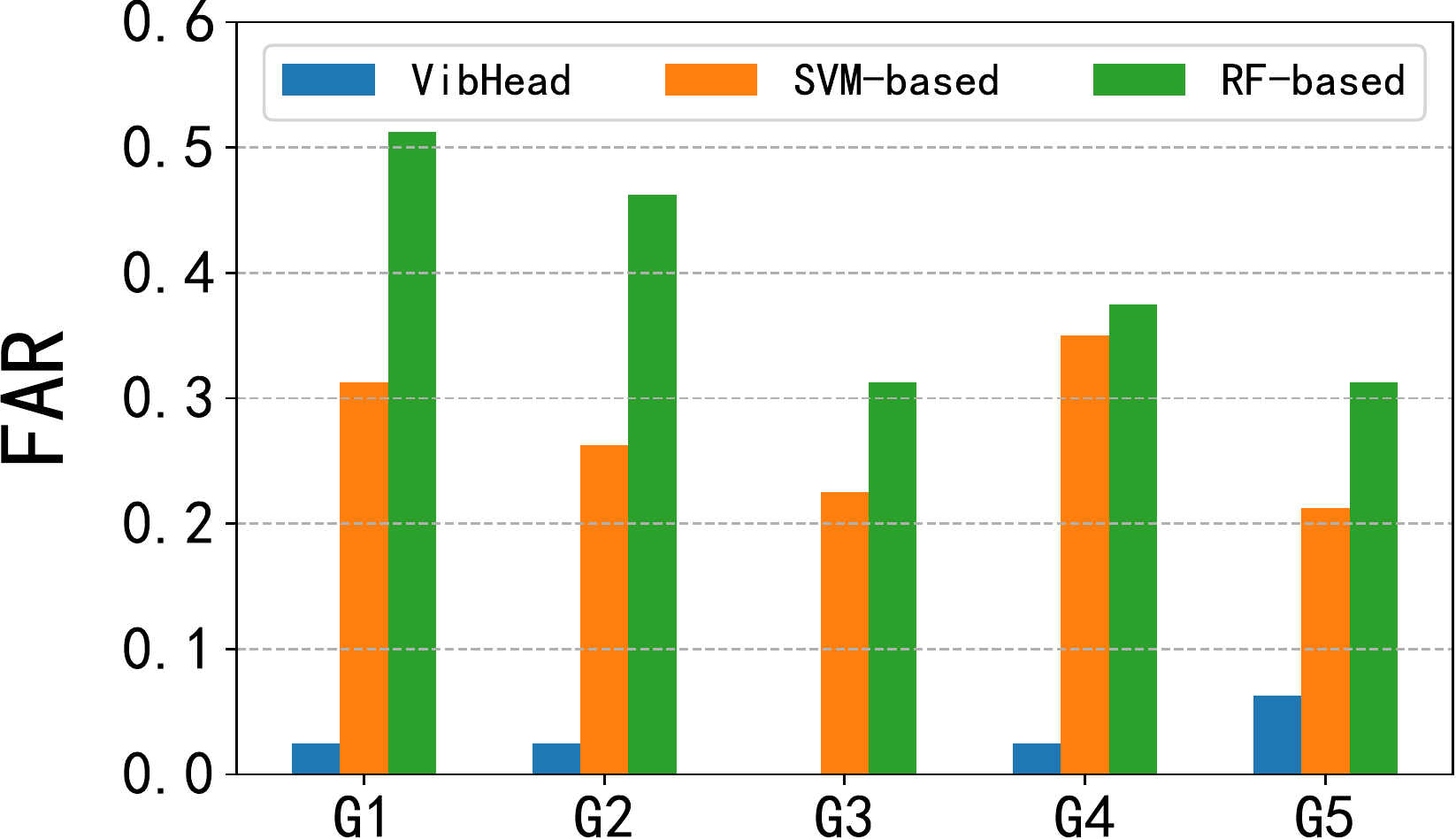}}
      \parbox{.43\columnwidth}{\center\scriptsize(a) $T=400ms$}
      \parbox{.43\columnwidth}{\center\scriptsize(b) $T=600ms$}
      \parbox{.43\textwidth}{\center\includegraphics[width=.4\textwidth]{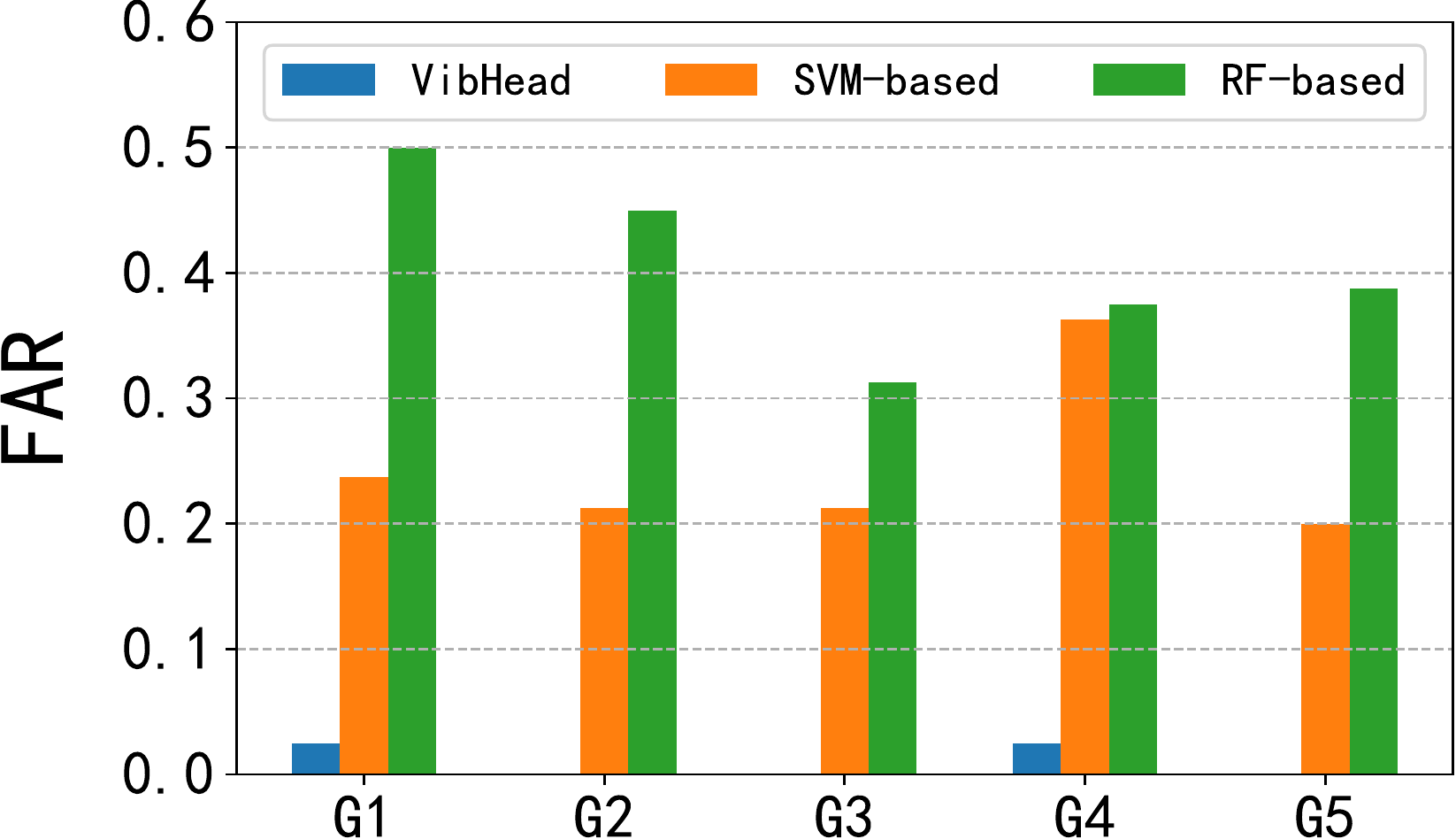}}
      \parbox{.43\textwidth}{\center\includegraphics[width=.4\textwidth]{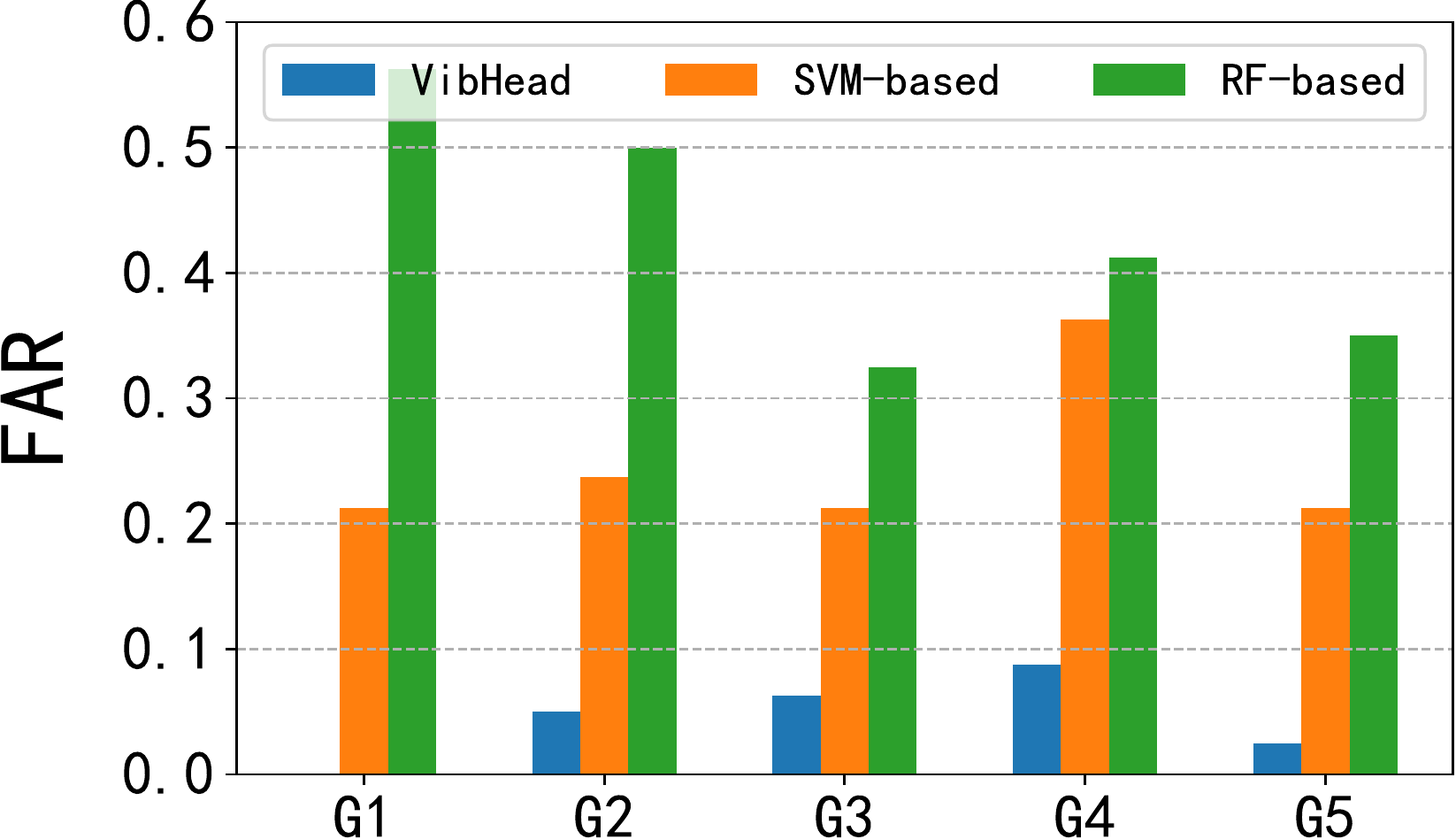}}
      \parbox{.43\columnwidth}{\center\scriptsize(c) $T=800ms$}
      \parbox{.43\columnwidth}{\center\scriptsize(d) $T=1000ms$}
    \caption{FARs of different authentication schemes for different gestures. G1, G2, G3, G4, and G5 stands for the five different gestures, i.e., standing, sitting upright, sitting-and-leaning-forward, sitting-and-leaning-backward, and walking, respectively.}
    \label{fig:farges}
    \end{center}
    \end{figure*}

\begin{figure*}[htb!]
    \begin{center}
      \parbox{.43\textwidth}{\center\includegraphics[width=.4\textwidth]{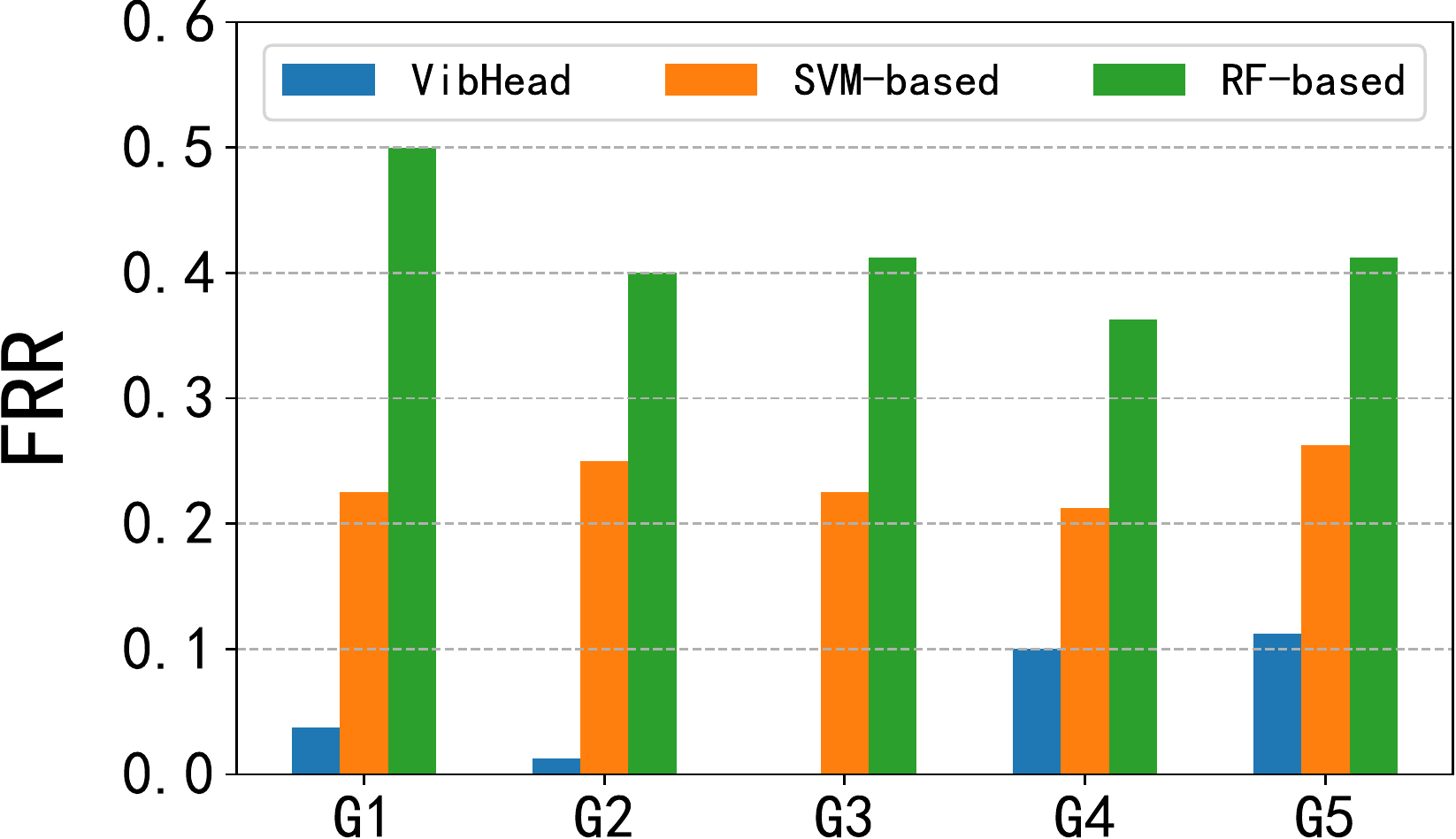}}
      \parbox{.43\textwidth}{\center\includegraphics[width=.4\textwidth]{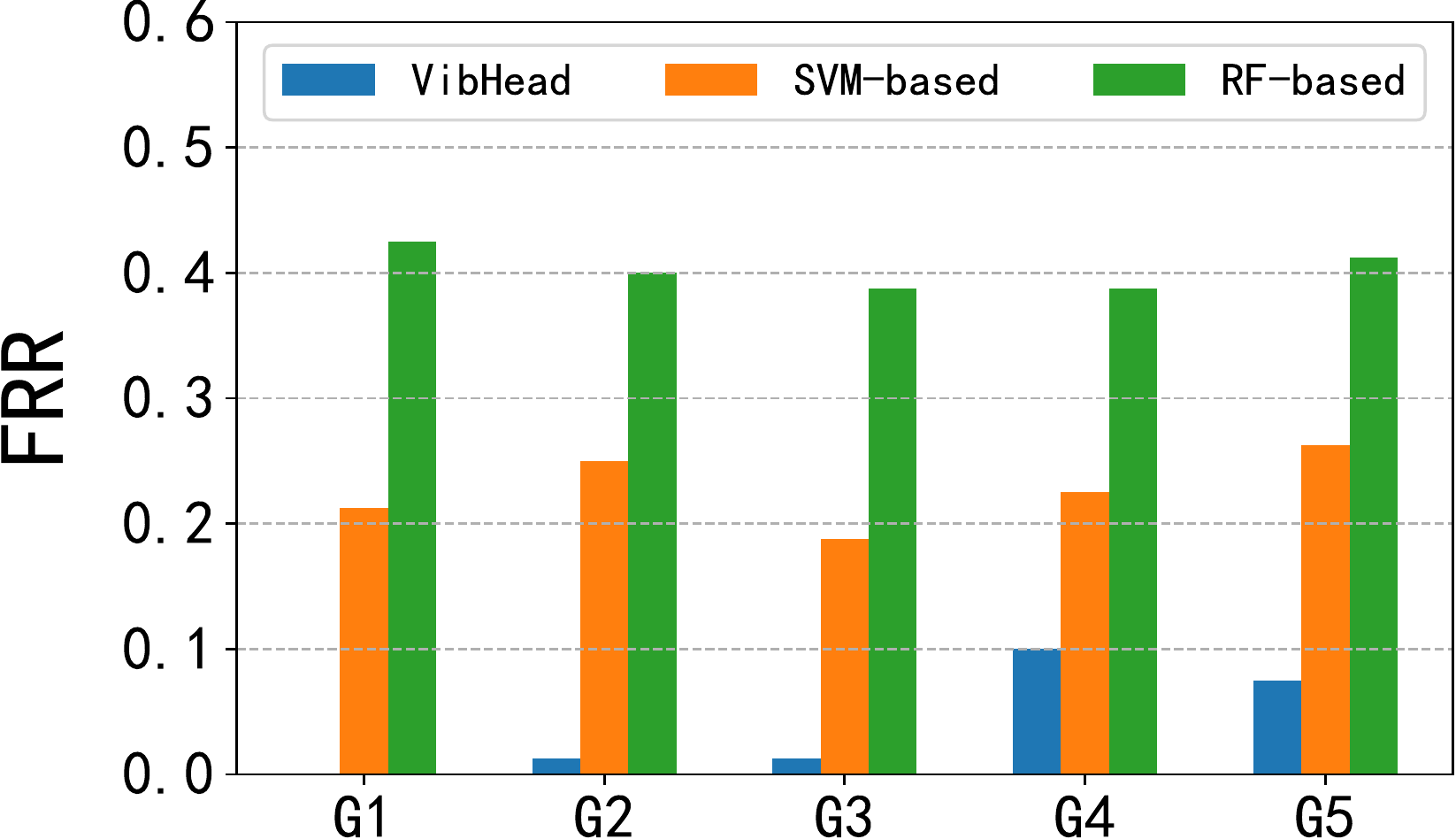}}
      \parbox{.43\columnwidth}{\center\scriptsize(a) $T=400ms$}
      \parbox{.43\columnwidth}{\center\scriptsize(b) $T=600ms$}
      \parbox{.43\textwidth}{\center\includegraphics[width=.4\textwidth]{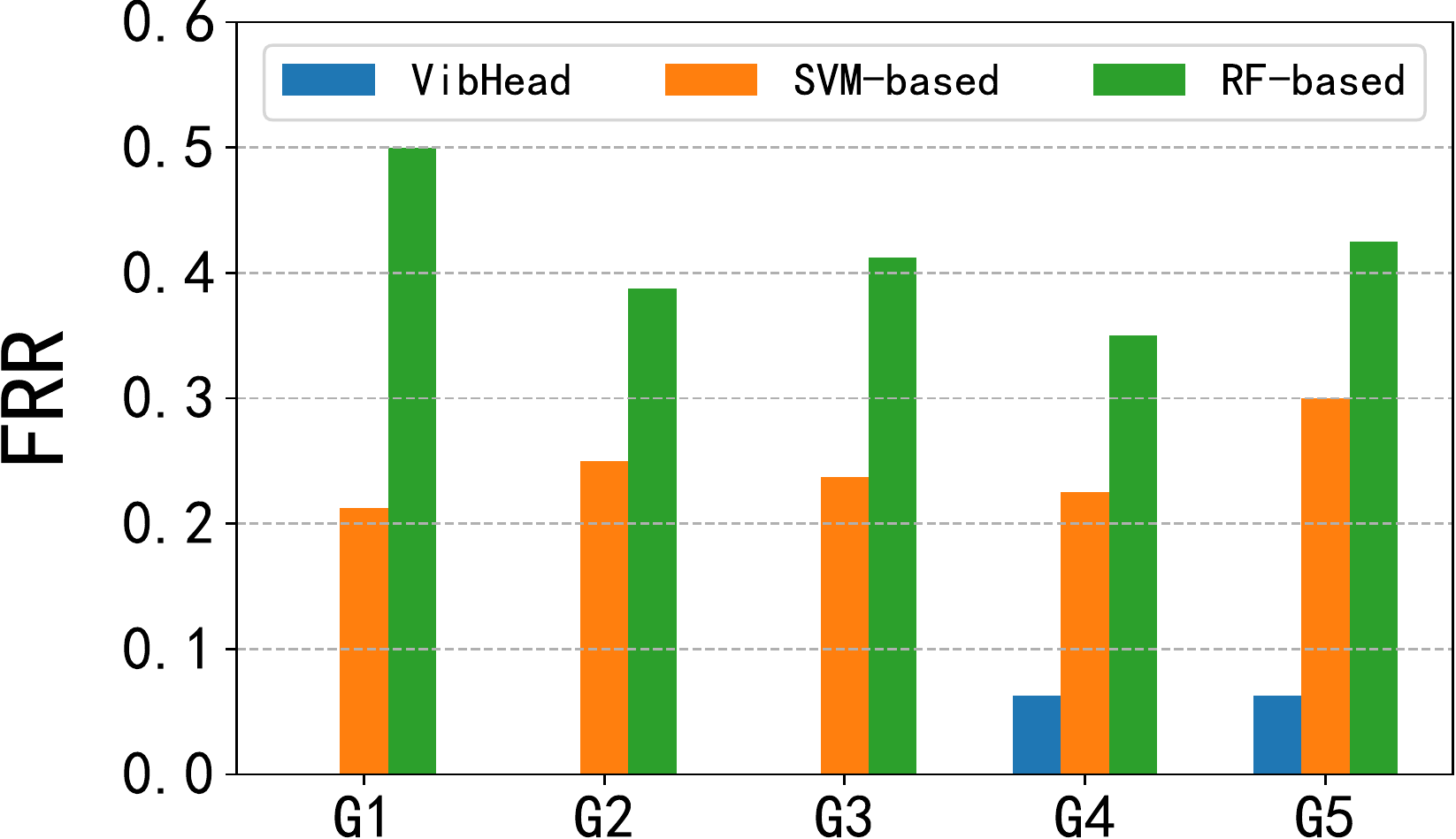}}
      \parbox{.43\textwidth}{\center\includegraphics[width=.4\textwidth]{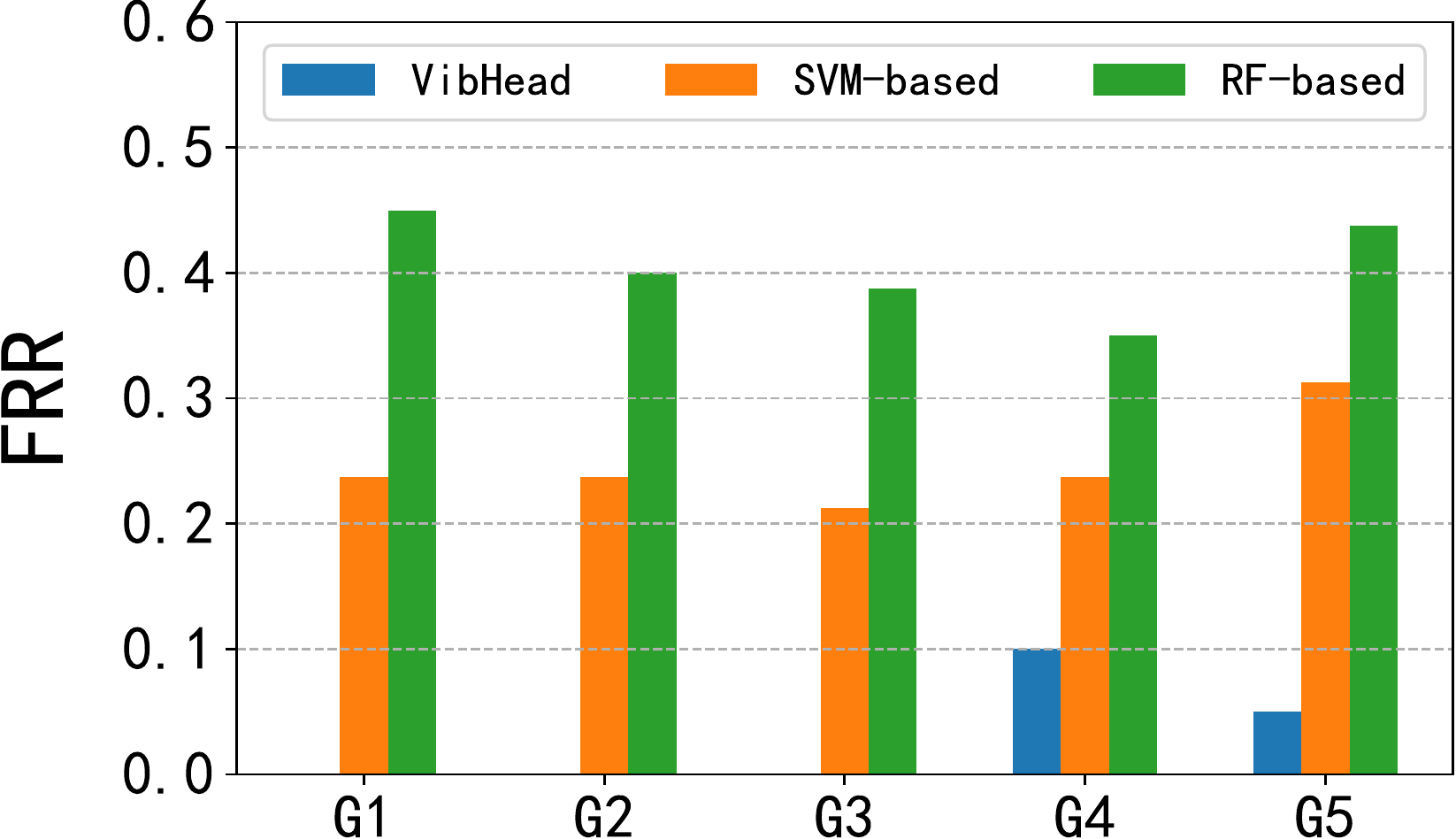}}
      \parbox{.43\columnwidth}{\center\scriptsize(c) $T=800ms$}
      \parbox{.43\columnwidth}{\center\scriptsize(d) $T=1000ms$}
    \caption{FRRs of different authentication schemes for different gestures. G1, G2, G3, G4, and G5 stands for the five different gestures, i.e., standing, sitting upright, sitting-and-leaning-forward, sitting-and-leaning-backward, and walking, respectively.}
    \label{fig:frrges}
    \end{center}
    \end{figure*}

    Furthermore, we summarize the FARs and FRRs of different authentication schemes in Fig.~\ref{fig:far} and Fig.~\ref{fig:frr}, respectively. As illustrated in Fig.~\ref{fig:far}, the FARs of our VibHead system across the different duration of data samples are much smaller than the ones of the SVM-based authentication scheme and the ones of the RF-based authentication scheme. The FARs of our VibHead system are always low, i.e., $0.05$ for $T=400 \sim 1000$ms, while the SVM-based scheme may have its FAR reaching only $0.28$, almost six times higher than that of our VibHead system. The FAR of the RF-based scheme is much worse; it is around $0.4$ for all different duration of data samples. It is shown in Fig.~\ref{fig:frr} that, VibHead also has a very low FRR. For any during of the data samples, the FRR of VibHead is below $0.05$, five times smaller than that of the SVM-based scheme, and eight times smaller than that of the RF-based scheme. Overall, when $T=800$ms, we get the best FAR and FRR trade-off with our VibHead system. We believe that, when $T$ is around $800$ms, the data samples are sufficiently informative (compare with the ones with $T=400$ and $T=200$ms) and contain less noise (than the ones with $T=800$ms), based on which we can distinguish the legitimate and illegitimate users more accurately.
    We also present the performance of different authentication schemes with different gestures in Fig.~\ref{fig:farges} and Fig.~\ref{fig:frrges}. It is demonstrated that, in term of both FAR and FRR, VibHead has oblivious advantages over the SVM-based and RF-based schemes across all the five  different gestures.
    %
    %

\begin{figure*}[htb!]
    \begin{center}
      \parbox{.43\textwidth}{\center\includegraphics[width=.4\textwidth]{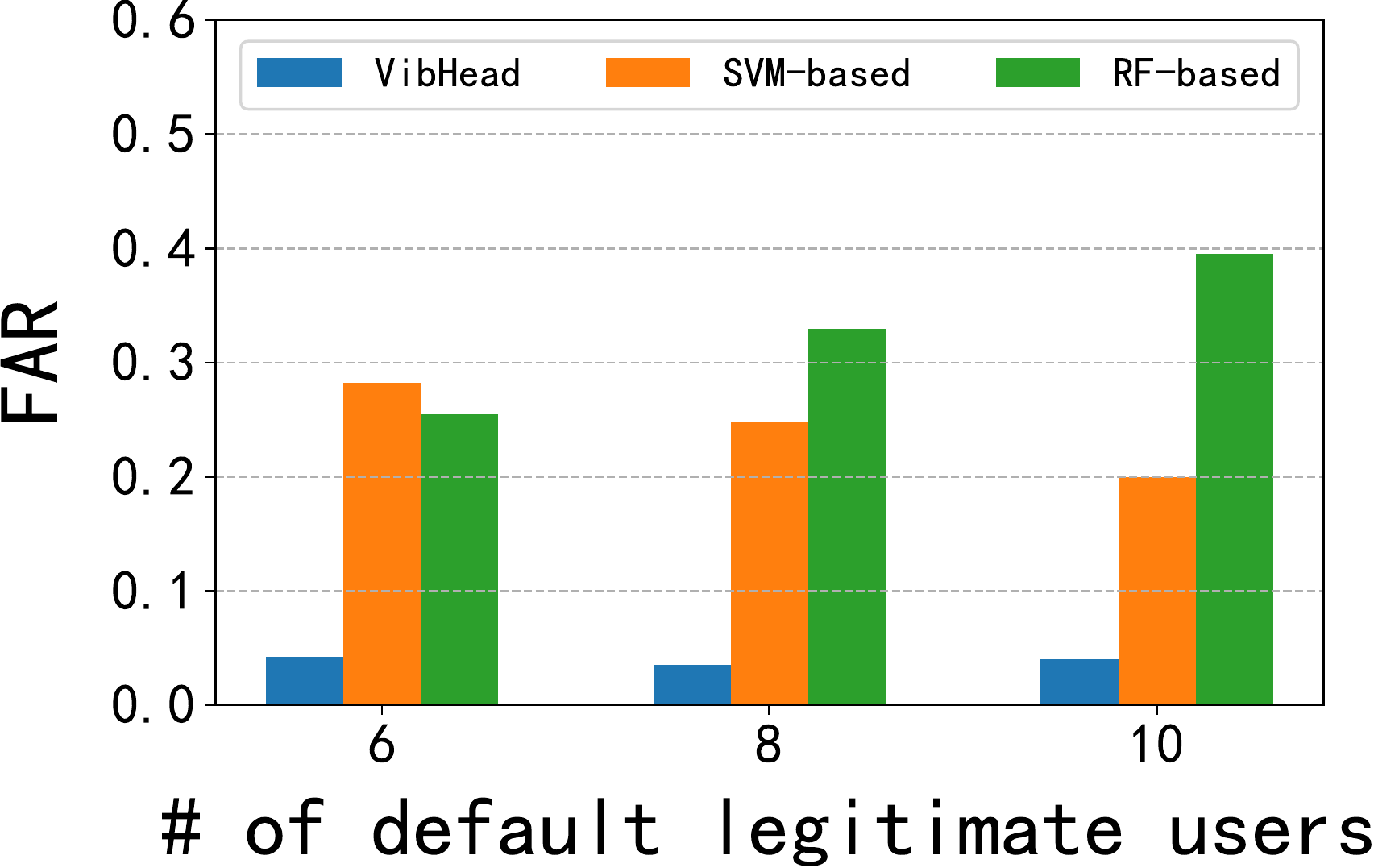}}
      \parbox{.43\textwidth}{\center\includegraphics[width=.4\textwidth]{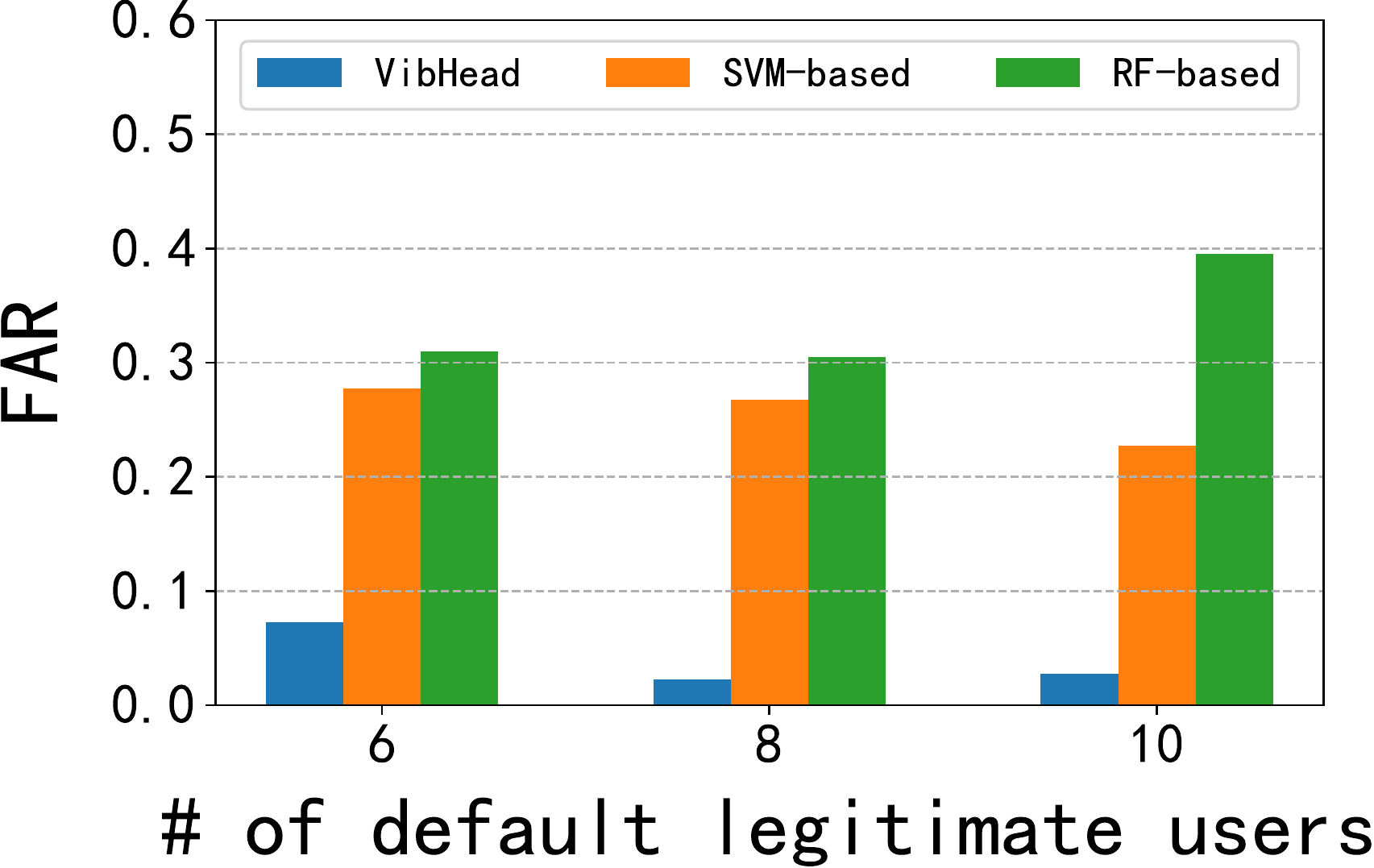}}
      \parbox{.43\columnwidth}{\center\scriptsize(a) $T=400ms$}
      \parbox{.43\columnwidth}{\center\scriptsize(b) $T=600ms$}
      \parbox{.43\textwidth}{\center\includegraphics[width=.4\textwidth]{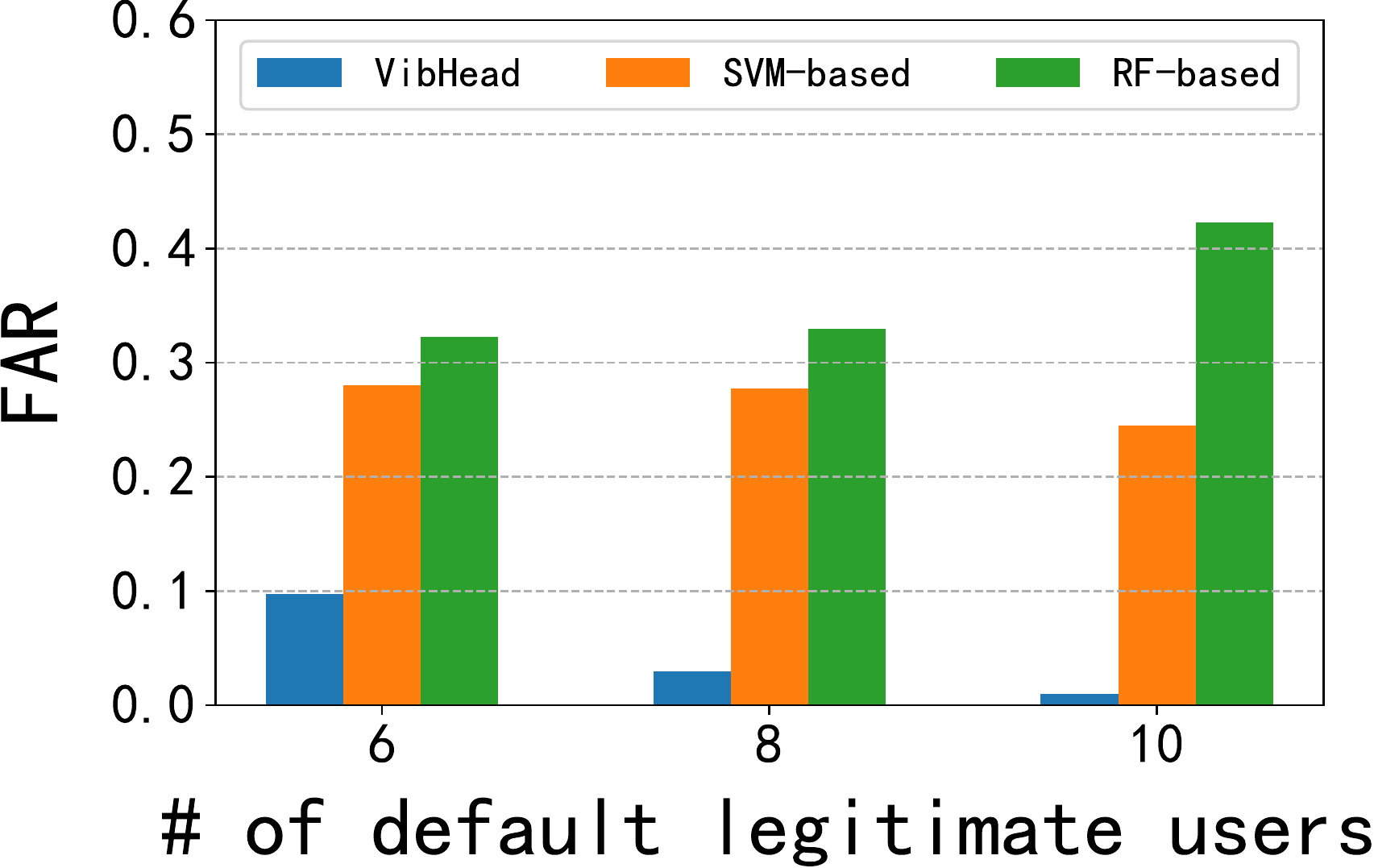}}
      \parbox{.43\textwidth}{\center\includegraphics[width=.4\textwidth]{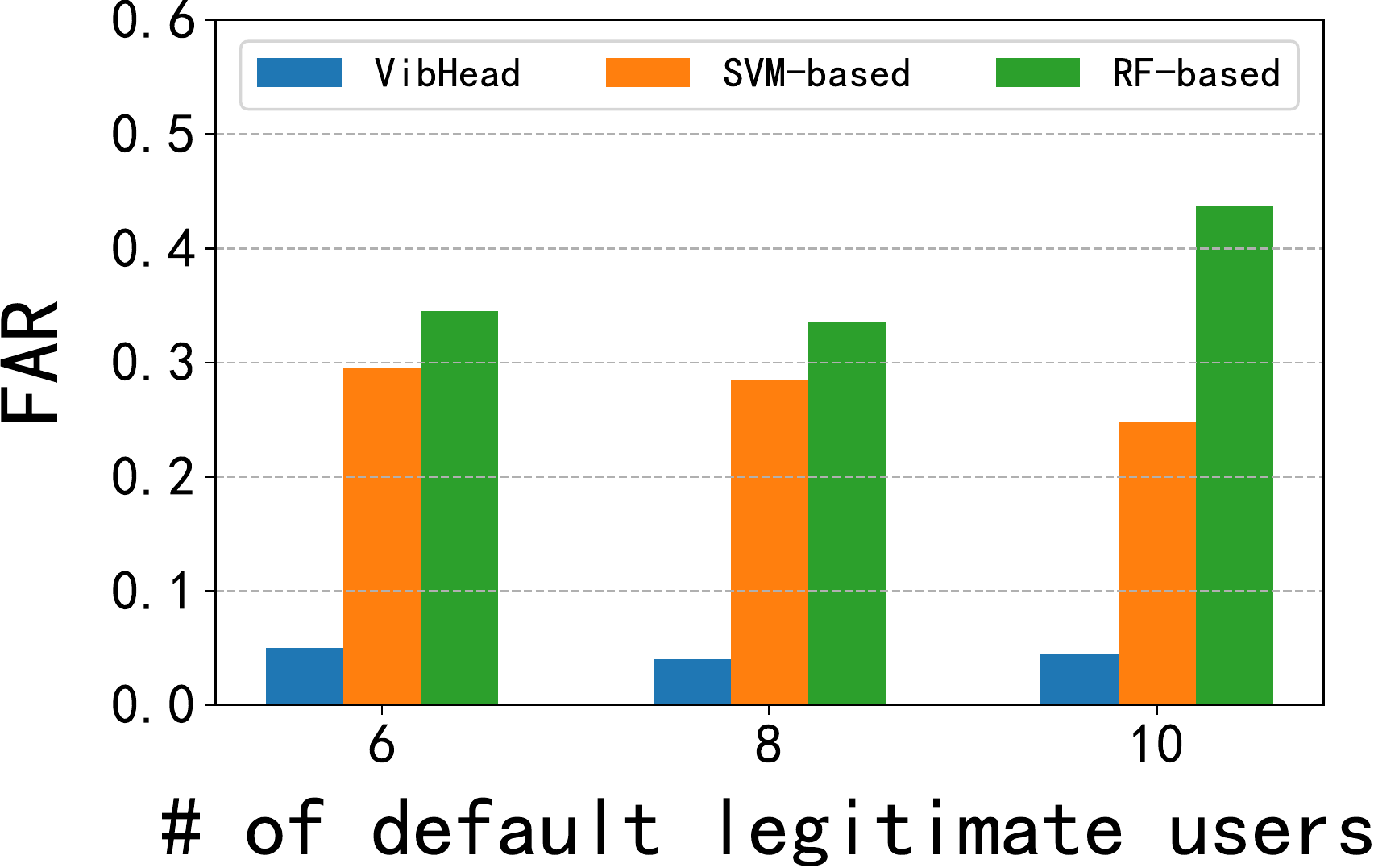}}
      \parbox{.43\columnwidth}{\center\scriptsize(a) $T=800ms$}
      \parbox{.43\columnwidth}{\center\scriptsize(b) $T=1000ms$}
    \caption{FAR with different numbers of default legitimate users.}
    \label{fig:fardiffuser}
    \end{center}
    \end{figure*}

\begin{figure*}[htb!]
    \begin{center}
      \parbox{.43\textwidth}{\center\includegraphics[width=.4\textwidth]{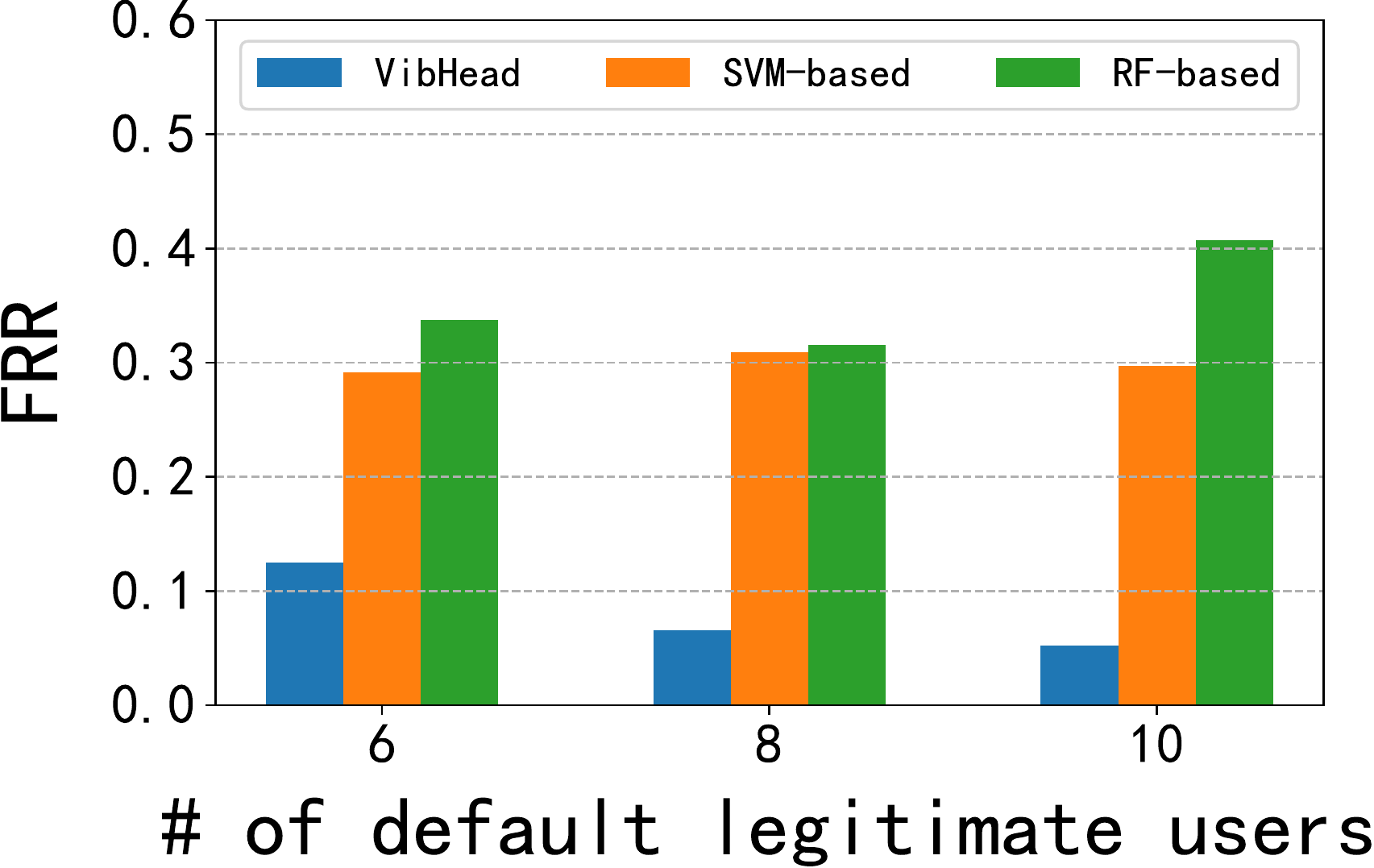}}
      \parbox{.43\textwidth}{\center\includegraphics[width=.4\textwidth]{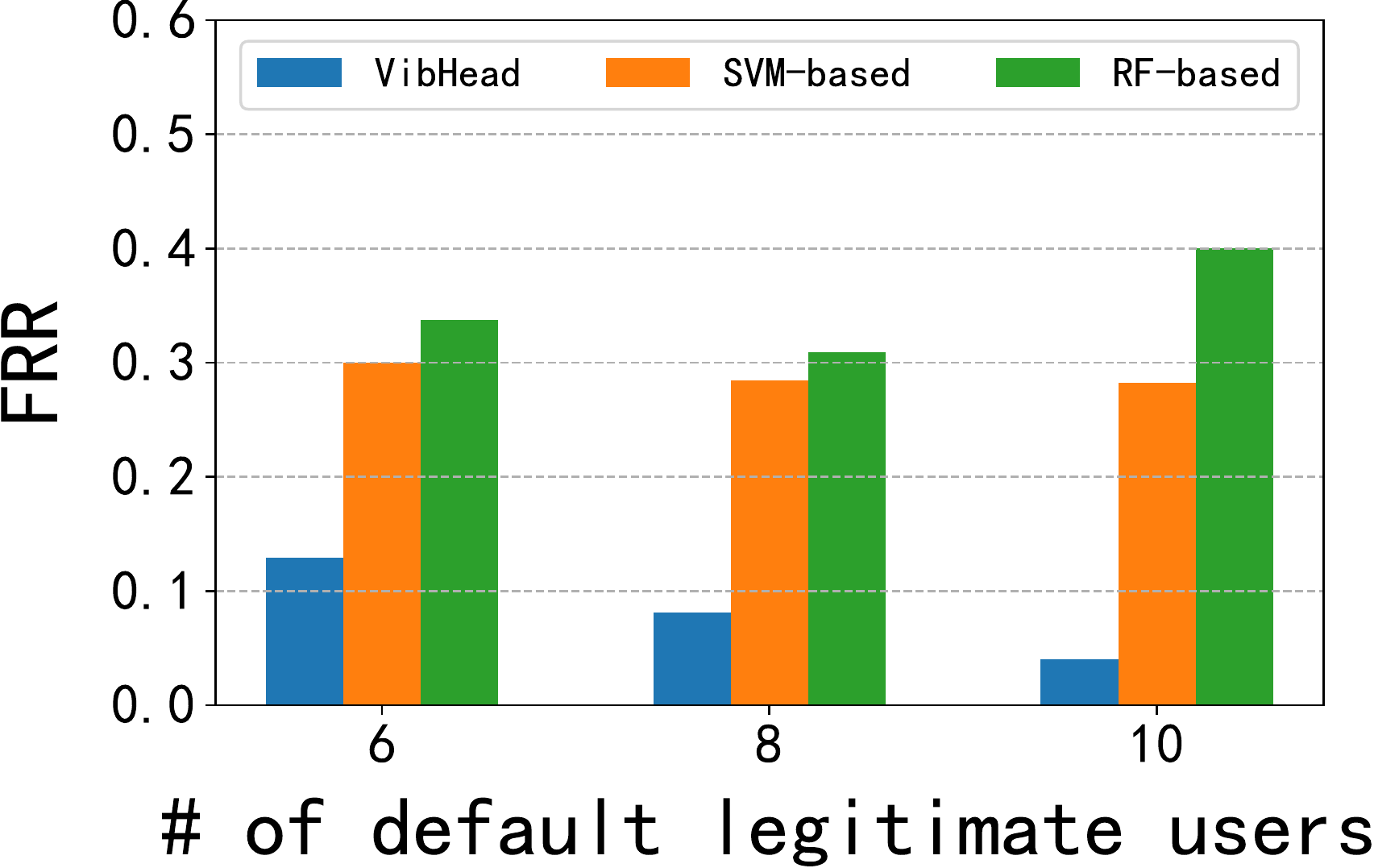}}
      \parbox{.43\columnwidth}{\center\scriptsize(a) $T=400ms$}
      \parbox{.43\columnwidth}{\center\scriptsize(b) $T=600ms$}
      \parbox{.43\textwidth}{\center\includegraphics[width=.4\textwidth]{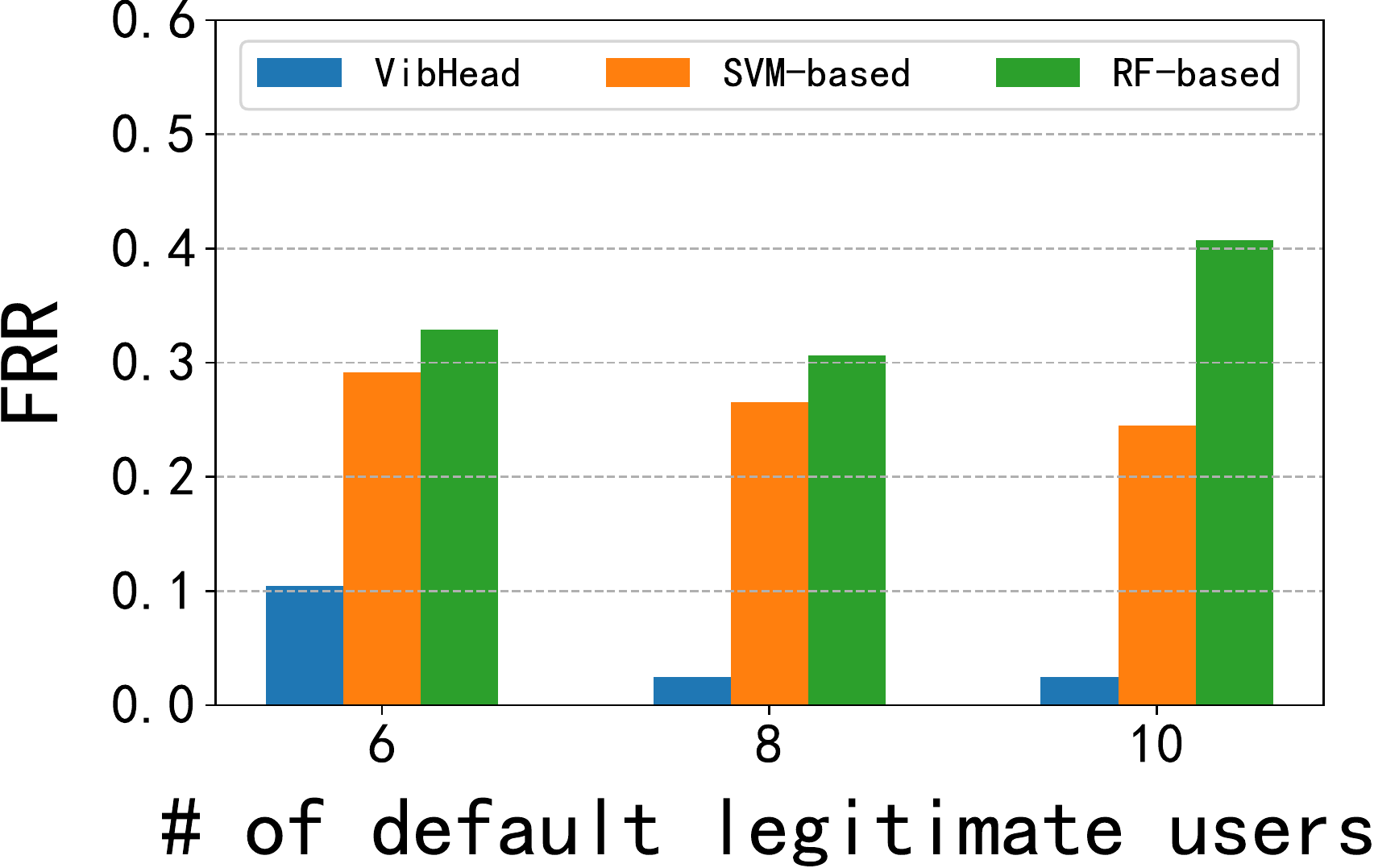}}
      \parbox{.43\textwidth}{\center\includegraphics[width=.4\textwidth]{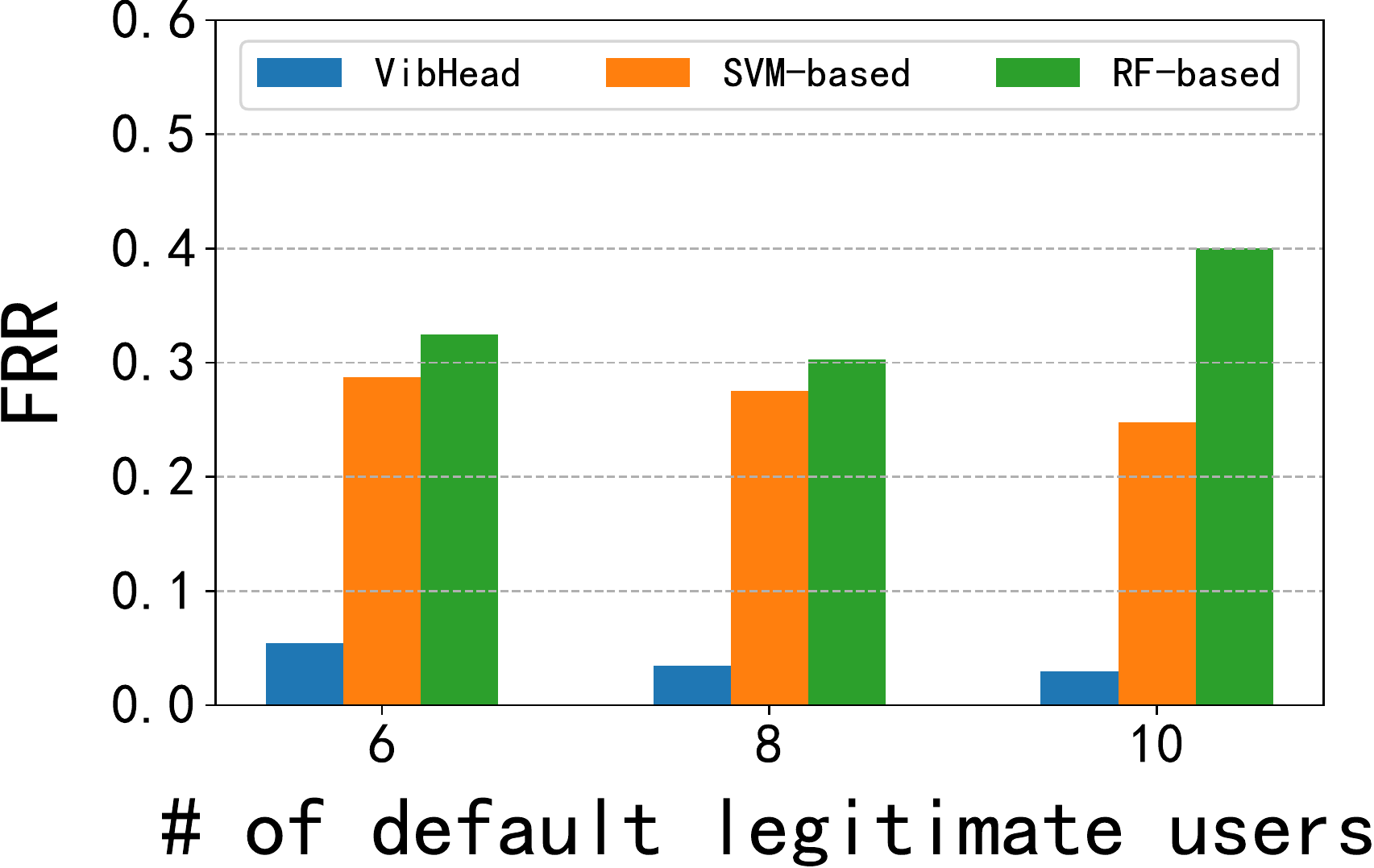}}
      \parbox{.43\columnwidth}{\center\scriptsize(a) $T=800ms$}
      \parbox{.43\columnwidth}{\center\scriptsize(b) $T=1000ms$}
    \caption{FRRs with different numbers of default legitimate users.}
    \label{fig:frrdiffuser}
    \end{center}
    \end{figure*}

    We also reveal how the different numbers of default legitimate users impact the performance of our authentication scheme. The experiment results are illustrated in Fig.~\ref{fig:fardiffuser} and Fig.~\ref{fig:frrdiffuser}. As demonstrated in Fig.~\ref{fig:fardiffuser}, our VibHead has much smaller FAR than the other two reference authentication scheme, across the different number of default legitimate users. Moreover, when there are more legitimate users, our VibHead may have smaller FAR. In terms of FRR, our VibHead also has oblivious advantage over the other two reference scheme, as illustrated in Fig.~\ref{fig:frrdiffuser}. Likewise, we can obtain smaller FRR by introducing more default legitimate users in our VibHead systems.
    

    \subsubsection{Inference Latency}
\begin{figure*}[htb!]
      \begin{center}
        \parbox{.43\textwidth}{\center\includegraphics[width=.4\textwidth]{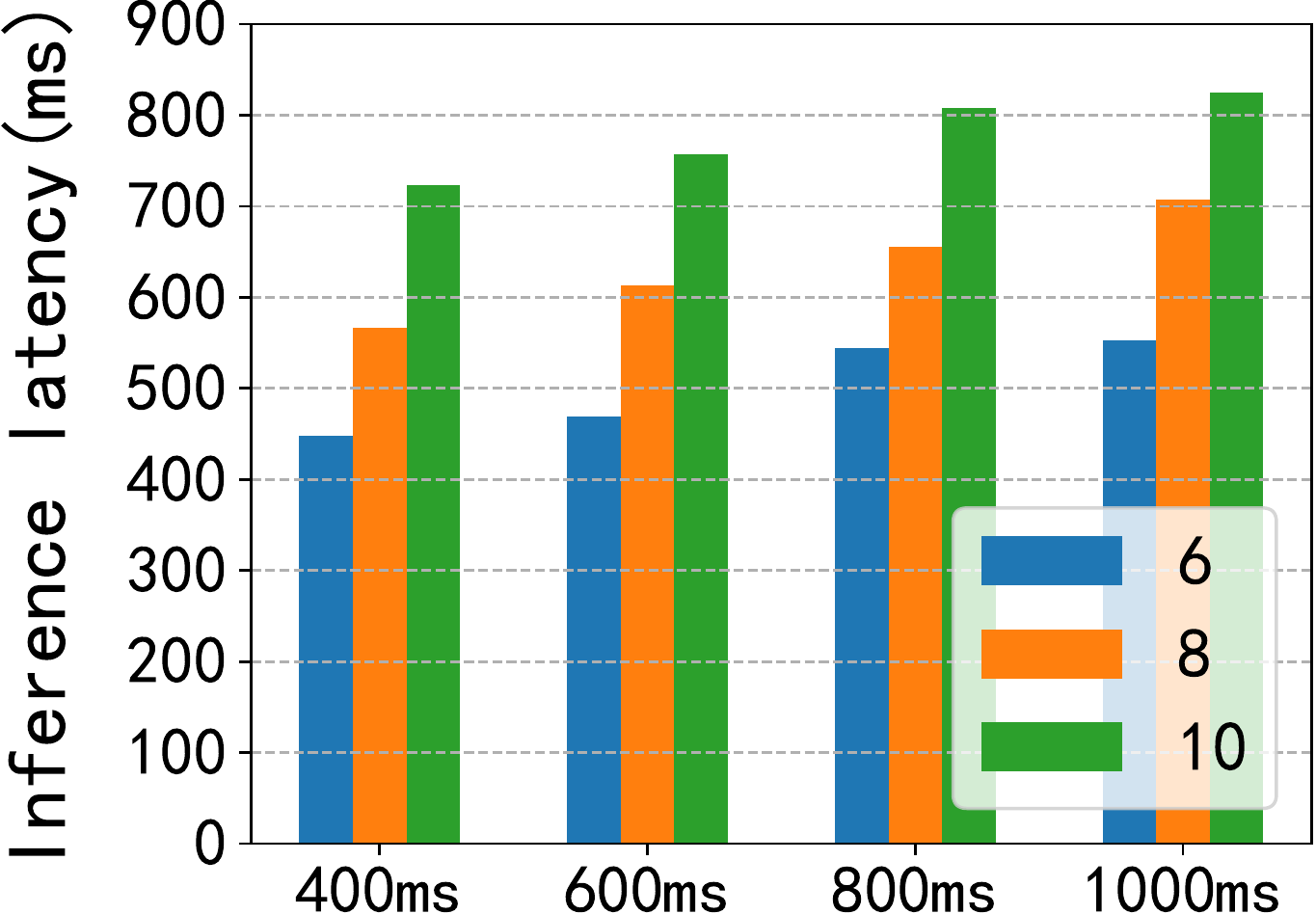}}
        \parbox{.43\textwidth}{\center\includegraphics[width=.4\textwidth]{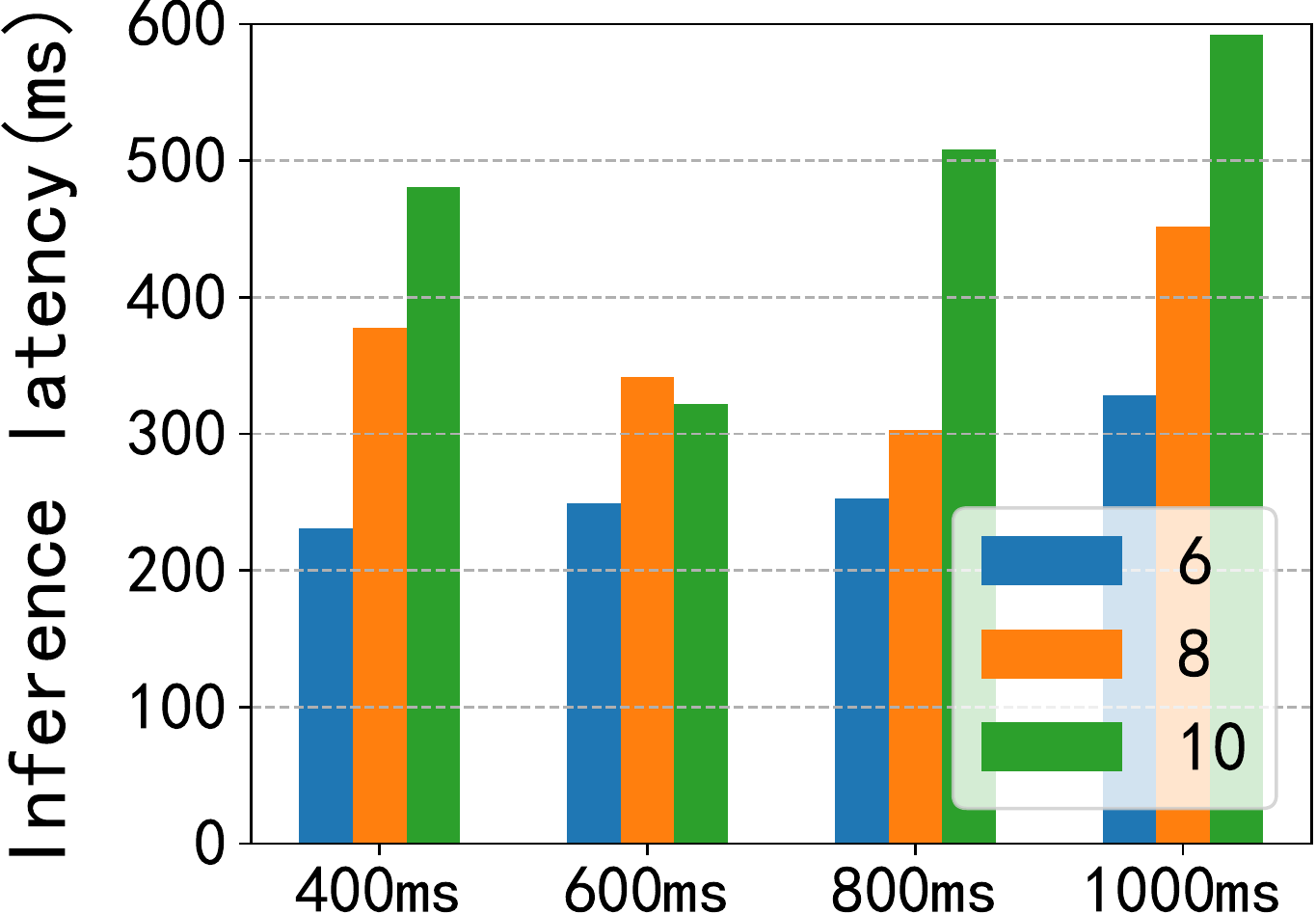}}
        \parbox{.43\columnwidth}{\center\scriptsize(a) Legitimate users}
        \parbox{.43\columnwidth}{\center\scriptsize(b) Illegitimate users}
      \caption{Inference latency.}
      \label{fig:latency}
      \end{center}
      \end{figure*}

      At last, we evaluate our VibHead system in terms of inference latency (i.e., the latency of implementing our two-step authentication scheme) when authenticating different numbers of legitimate users. Since the Microsoft HoloLens does not provide APIs to develop our third-party authentication application, the evaluation is performance on another similar device, i.e., PICO Neo3 where the deployment of our classification models is allowed. The device is equipped with Snapdragon XR2 processor and 8GB RAM. The experiment results, i.e., the average inference latency for the legitimate users and the illegitimate users, are presented in Fig.~\ref{fig:latency}. Since legitimate users should be authenticated through all classifiers $F_0$ and $\mathcal{F}_{i^*_0}$ while the authentication for illegitimate users can be done either through only $F_0$ in the first step of our authentication scheme or a handful of classifiers in $\mathcal{F}_{i^*_0}$ in the second step of our authentication scheme, the latency of authenticating illegitimate users is less than the one of authenticating legitimate users. For example, when there are six default legitimate users, the inference latency for the legitimate users is between $450$ms and $550$ms, whereas that of the illegitimate users is only $220 \sim 320$ms. Furthermore, when there are more default legitimate users, the latency is longer, since we have to adopt more classifiers in the second step to make authentication, especially for the legitimate users, as shown in Sec.~\ref{sec:authen}. Additionally, when we increase the duration of the data samples, the inference latency may be longer; nevertheless, the fluctuation is only $100 \sim 200$ms and is thus tolerable, especially considering data sampling could take hundreds of milliseconds.


\section{Conclusion}  \label{sec:conclusion}
  In this paper, we propose a vibration-based user authentication system on smart headsets, VibHead, which leverages active vibration signals to extract  physiological characters of human heads for identifying different login users. VibHead first extracts primitive data features and MFCC-based features from active vibration signals to train a set of CNN-based classifiers for user classification. We then design a two-step authentication scheme by utilizing the user classifiers. Our extensive experiments reveal that VibHead can accurately authenticate login users.


\bibliographystyle{ACM-Reference-Format}
\bibliography{vibauth}


\begin{thebibliography}{37}


\ifx \showCODEN    \undefined \def \showCODEN     #1{\unskip}     \fi
\ifx \showDOI      \undefined \def \showDOI       #1{#1}\fi
\ifx \showISBNx    \undefined \def \showISBNx     #1{\unskip}     \fi
\ifx \showISBNxiii \undefined \def \showISBNxiii  #1{\unskip}     \fi
\ifx \showISSN     \undefined \def \showISSN      #1{\unskip}     \fi
\ifx \showLCCN     \undefined \def \showLCCN      #1{\unskip}     \fi
\ifx \shownote     \undefined \def \shownote      #1{#1}          \fi
\ifx \showarticletitle \undefined \def \showarticletitle #1{#1}   \fi
\ifx \showURL      \undefined \def \showURL       {\relax}        \fi
\providecommand\bibfield[2]{#2}
\providecommand\bibinfo[2]{#2}
\providecommand\natexlab[1]{#1}
\providecommand\showeprint[2][]{arXiv:#2}

\bibitem[Babich(2019)]%
        {Babich2019}
\bibfield{author}{\bibinfo{person}{N. Babich}.}
  \bibinfo{year}{2019}\natexlab{}.
\newblock \bibinfo{title}{{How VR In Education Will Change How We Learn And
  Teach}}.
\newblock
  \bibinfo{howpublished}{\url{https://xd.adobe.com/ideas/principles/emerging-technology/virtual-reality-will-change-learn-teach/}}.
\newblock


\bibitem[Cao et~al\mbox{.}(2021)]%
        {CaoJLX-ICDCS21}
\bibfield{author}{\bibinfo{person}{H. Cao}, \bibinfo{person}{H. Jiang},
  \bibinfo{person}{D. Liu}, {and} \bibinfo{person}{J. Xiong}.}
  \bibinfo{year}{2021}\natexlab{}.
\newblock \showarticletitle{{Evidence in Hand: Passive Vibration Response-based
  Continuous User Authentication}}. In \bibinfo{booktitle}{\emph{Proc. of the
  41st IEEE ICDCS}}. \bibinfo{pages}{1020--1030}.
\newblock


\bibitem[Changer(2020)]%
        {SCGC2020}
\bibfield{author}{\bibinfo{person}{Supply Chain~Game Changer}.}
  \bibinfo{year}{2020}\natexlab{}.
\newblock \bibinfo{title}{{Virtual reality (VR) is enhancing e-commerce
  shopping!}}
\newblock
  \bibinfo{howpublished}{\url{https://supplychaingamechanger.com/how-virtual-reality-vr-is-drastically-enhancing-the-e-commerce-shopping-experience-infographic/}}.
\newblock


\bibitem[Chen et~al\mbox{.}(2018)]%
        {ChenGHWRHW-SECON18}
\bibfield{author}{\bibinfo{person}{W. Chen}, \bibinfo{person}{M. Guan},
  \bibinfo{person}{Y. Huang}, \bibinfo{person}{L. Wang}, \bibinfo{person}{R.
  Ruby}, \bibinfo{person}{W. Hu}, {and} \bibinfo{person}{K. Wu}.}
  \bibinfo{year}{2018}\natexlab{}.
\newblock \showarticletitle{{ViType: A Cost Efficient On-Body Typing System
  through Vibration}}. In \bibinfo{booktitle}{\emph{Proc. of the 15th IEEE
  SECON}}. \bibinfo{pages}{19--27}.
\newblock


\bibitem[Flavián et~al\mbox{.}(2019)]%
        {FlavianIO-JBR19}
\bibfield{author}{\bibinfo{person}{C. Flavián}, \bibinfo{person}{S.
  Ibáñez-Sánchez}, {and} \bibinfo{person}{C. Orús}.}
  \bibinfo{year}{2019}\natexlab{}.
\newblock \showarticletitle{{The impact of virtual, augmented and mixed reality
  technologies on the customer experience}}.
\newblock \bibinfo{journal}{\emph{Journal of Business Research}}
  \bibinfo{volume}{100}, \bibinfo{number}{4} (\bibinfo{year}{2019}),
  \bibinfo{pages}{547--560}.
\newblock


\bibitem[Fu et~al\mbox{.}(2022)]%
        {FuHS-TCH22}
\bibfield{author}{\bibinfo{person}{Y. Fu}, \bibinfo{person}{Y. Hu}, {and}
  \bibinfo{person}{V. Sundstedt}.} \bibinfo{year}{2022}\natexlab{}.
\newblock \showarticletitle{{A Systematic Literature Review of Virtual,
  Augmented, and Mixed Reality Game Applications in Healthcare}}.
\newblock \bibinfo{journal}{\emph{ACM Trans. on Computing for
  HealthcareVolume}} \bibinfo{volume}{3}, \bibinfo{number}{2}
  (\bibinfo{year}{2022}), \bibinfo{pages}{22:1--22:27}.
\newblock


\bibitem[Funk et~al\mbox{.}(2019)]%
        {FunkMMKMM-CHI19}
\bibfield{author}{\bibinfo{person}{M. Funk}, \bibinfo{person}{K. Marky},
  \bibinfo{person}{I. Mizutani}, \bibinfo{person}{M. Kritzler},
  \bibinfo{person}{S. Mayer}, {and} \bibinfo{person}{F. Michahelles}.}
  \bibinfo{year}{2019}\natexlab{}.
\newblock \showarticletitle{{LookUnlock: Using Spatial-Targets for
  User-Authentication on HMDs}}. In \bibinfo{booktitle}{\emph{Extended
  Abstracts of the 2019 CHI Conference on Human Factors in Computing Systems}}.
\newblock


\bibitem[Husa and Tourani(2021)]%
        {HusaT-CNS21}
\bibfield{author}{\bibinfo{person}{E. Husa} {and} \bibinfo{person}{R.
  Tourani}.} \bibinfo{year}{2021}\natexlab{}.
\newblock \showarticletitle{{Vibe: An Implicit Two-Factor Authentication using
  Vibration Signals}}. In \bibinfo{booktitle}{\emph{Proc. of the 9th CNS}}.
  \bibinfo{pages}{236--244}.
\newblock


\bibitem[(IDC)(2022)]%
        {IDC2022}
\bibfield{author}{\bibinfo{person}{International Data~Corporation (IDC)}.}
  \bibinfo{year}{2022}\natexlab{}.
\newblock \bibinfo{title}{{Worldwide Augmented and Virtual Reality Spending
  Guide}}.
\newblock
  \bibinfo{howpublished}{\url{https://www.idc.com/getfile.dyn?containerId=IDC_P34919&attachmentId=47456852}}.
\newblock


\bibitem[Karnik et~al\mbox{.}(2022)]%
        {KarnikBBKD-JIII22}
\bibfield{author}{\bibinfo{person}{N. Karnik}, \bibinfo{person}{U. Bora},
  \bibinfo{person}{K. Bhadri}, \bibinfo{person}{P. Kadambi}, {and}
  \bibinfo{person}{P. Dhatrak}.} \bibinfo{year}{2022}\natexlab{}.
\newblock \showarticletitle{{A comprehensive study on current and future trends
  towards the characteristics and enablers of industry 4.0}}.
\newblock \bibinfo{journal}{\emph{Journal of Industrial Information
  Integration}}  \bibinfo{volume}{27} (\bibinfo{year}{2022}),
  \bibinfo{pages}{100294}.
\newblock


\bibitem[Kumar et~al\mbox{.}(2022)]%
        {KumarLCSHPTH-MM22}
\bibfield{author}{\bibinfo{person}{A. Kumar}, \bibinfo{person}{L. Lee},
  \bibinfo{person}{J. Chauhan}, \bibinfo{person}{X. Su}, \bibinfo{person}{M.
  Hoque}, \bibinfo{person}{S. Pirttikangas}, \bibinfo{person}{S. Tarkoma},
  {and} \bibinfo{person}{P. Hui}.} \bibinfo{year}{2022}\natexlab{}.
\newblock \showarticletitle{{PassWalk: Spatial Authentication Leveraging
  Lateral Shift and Gaze on Mobile Headsets}}. In
  \bibinfo{booktitle}{\emph{Proc. of the 30th ACM Multimedia}}.
  \bibinfo{pages}{952--960}.
\newblock


\bibitem[Laput et~al\mbox{.}(2016)]%
        {LaputXH-UIST16}
\bibfield{author}{\bibinfo{person}{G. Laput}, \bibinfo{person}{R. Xiao}, {and}
  \bibinfo{person}{C. Harrison}.} \bibinfo{year}{2016}\natexlab{}.
\newblock \showarticletitle{{ViBand: High-Fidelity Bio-Acoustic Sensing Using
  Commodity Smartwatch Accelerometers}}. In \bibinfo{booktitle}{\emph{Proc. of
  the 29th UIST}}. \bibinfo{pages}{321--333}.
\newblock


\bibitem[Lee et~al\mbox{.}(2021)]%
        {LeeCL-CCS21}
\bibfield{author}{\bibinfo{person}{S. Lee}, \bibinfo{person}{W. Choi}, {and}
  \bibinfo{person}{D. Lee}.} \bibinfo{year}{2021}\natexlab{}.
\newblock \showarticletitle{{Usable User Authentication on a Smartwatch using
  Vibration}}. In \bibinfo{booktitle}{\emph{Proc. of the 27th ACM CCS}}.
  \bibinfo{pages}{304--319}.
\newblock


\bibitem[Li et~al\mbox{.}(2022)]%
        {LiZLPL-RCIM22}
\bibfield{author}{\bibinfo{person}{C. Li}, \bibinfo{person}{P. Zheng},
  \bibinfo{person}{S. Li}, \bibinfo{person}{Y. Pang}, {and} \bibinfo{person}{C.
  Lee}.} \bibinfo{year}{2022}\natexlab{}.
\newblock \showarticletitle{{AR-assisted digital twin-enabled robot
  collaborative manufacturing system with human-in-the-loop}}.
\newblock \bibinfo{journal}{\emph{Robotics and Computer-Integrated
  Manufacturing}}  \bibinfo{volume}{76} (\bibinfo{year}{2022}),
  \bibinfo{pages}{102321}.
\newblock


\bibitem[Li et~al\mbox{.}(2019)]%
        {LiFK-CCS19}
\bibfield{author}{\bibinfo{person}{J. Li}, \bibinfo{person}{K. Fawaz}, {and}
  \bibinfo{person}{Y. Kim}.} \bibinfo{year}{2019}\natexlab{}.
\newblock \showarticletitle{{Velody: Nonlinear Vibration Challenge-Response for
  Resilient User Authentication}}. In \bibinfo{booktitle}{\emph{Proc. of the
  25th ACM CCS}}. \bibinfo{pages}{1201--1213}.
\newblock


\bibitem[Li et~al\mbox{.}(2016)]%
        {LiAZXLG-PerCom16}
\bibfield{author}{\bibinfo{person}{S. Li}, \bibinfo{person}{A. Ashok},
  \bibinfo{person}{Y. Zhang}, \bibinfo{person}{C. Xu}, \bibinfo{person}{J.
  Lindqvist}, {and} \bibinfo{person}{M. Gruteser}.}
  \bibinfo{year}{2016}\natexlab{}.
\newblock \showarticletitle{{Whose move is it anyway? Authenticating smart
  wearable devices using unique head movement patterns}}. In
  \bibinfo{booktitle}{\emph{Proc. of the 14th IEEE PerCom}}.
  \bibinfo{pages}{1--9}.
\newblock


\bibitem[Liebers et~al\mbox{.}(2021)]%
        {LiebersHBGS-VRST21}
\bibfield{author}{\bibinfo{person}{J. Liebers}, \bibinfo{person}{P. Horn},
  \bibinfo{person}{C. Burschik}, \bibinfo{person}{U. Gruenefeld}, {and}
  \bibinfo{person}{S. Schneegass}.} \bibinfo{year}{2021}\natexlab{}.
\newblock \showarticletitle{{Using Gaze Behavior and Head Orientation for
  Implicit Identification in Virtual Reality}}. In
  \bibinfo{booktitle}{\emph{Proc. of the 27th VRST}}.
  \bibinfo{pages}{22:1--22:9}.
\newblock


\bibitem[Liu et~al\mbox{.}(2017a)]%
        {LiuCGW-SECON17}
\bibfield{author}{\bibinfo{person}{J. Liu}, \bibinfo{person}{Y. Chen},
  \bibinfo{person}{M. Gruteser}, {and} \bibinfo{person}{Y. Wang}.}
  \bibinfo{year}{2017}\natexlab{a}.
\newblock \showarticletitle{{VibSense: Sensing Touches on Ubiquitous Surfaces
  through Vibration}}. In \bibinfo{booktitle}{\emph{Proc. of the 14th IEEE
  SECON}}. \bibinfo{pages}{1--9}.
\newblock


\bibitem[Liu et~al\mbox{.}(2017b)]%
        {LiuWCS-CCS17}
\bibfield{author}{\bibinfo{person}{J. Liu}, \bibinfo{person}{C. Wang},
  \bibinfo{person}{Y. Chen}, {and} \bibinfo{person}{N. Saxena}.}
  \bibinfo{year}{2017}\natexlab{b}.
\newblock \showarticletitle{{VibWrite: Towards Finger-input Authentication on
  Ubiquitous Surfaces via Physical Vibration}}. In
  \bibinfo{booktitle}{\emph{Proc. of the 24th ACM CCS}}.
  \bibinfo{pages}{73--87}.
\newblock


\bibitem[Mathis et~al\mbox{.}(2020)]%
        {MathisWVK-CHI20}
\bibfield{author}{\bibinfo{person}{F. Mathis}, \bibinfo{person}{J. Williamson},
  \bibinfo{person}{K. Vaniea}, {and} \bibinfo{person}{M. Khamis}.}
  \bibinfo{year}{2020}\natexlab{}.
\newblock \showarticletitle{{RubikAuth: Fast and Secure Authentication in
  Virtual Reality}}. In \bibinfo{booktitle}{\emph{Extended Abstracts of the
  2020 CHI Conference on Human Factors in Computing Systems}}.
  \bibinfo{pages}{1--9}.
\newblock


\bibitem[Murty and Yegnanarayana(2006)]%
        {MurtyY-SPL06}
\bibfield{author}{\bibinfo{person}{K. Murty} {and} \bibinfo{person}{B.
  Yegnanarayana}.} \bibinfo{year}{2006}\natexlab{}.
\newblock \showarticletitle{{Combining Evidence from Residual Phase and MFCC
  Features for Speaker Recognition}}.
\newblock \bibinfo{journal}{\emph{IEEE Signal Processing Letters}}
  \bibinfo{volume}{13}, \bibinfo{number}{1} (\bibinfo{year}{2006}),
  \bibinfo{pages}{52--55}.
\newblock


\bibitem[Mustafa et~al\mbox{.}(2018)]%
        {MustafaMSM-IWSPA18}
\bibfield{author}{\bibinfo{person}{T. Mustafa}, \bibinfo{person}{R. Matovu},
  \bibinfo{person}{A. Serwadda}, {and} \bibinfo{person}{N. Muirhead}.}
  \bibinfo{year}{2018}\natexlab{}.
\newblock \showarticletitle{{Unsure How to Authenticate on Your VR Headset?:
  Come on, Use Your Head!}}. In \bibinfo{booktitle}{\emph{Proc. of the 4th ACM
  IWSPA}}. \bibinfo{pages}{23--30}.
\newblock


\bibitem[Pedregosa et~al\mbox{.}(2011)]%
        {Pedregosa-JMLR11}
\bibfield{author}{\bibinfo{person}{F. Pedregosa}, \bibinfo{person}{G.
  Varoquaux}, \bibinfo{person}{A. Gramfort}, \bibinfo{person}{V. Michel},
  \bibinfo{person}{B. Thirion}, \bibinfo{person}{O. Grisel},
  \bibinfo{person}{M. Blondel}, \bibinfo{person}{P. Prettenhofer},
  \bibinfo{person}{R. Weiss}, \bibinfo{person}{V. Dubourg}, \bibinfo{person}{J.
  VanderPlas}, \bibinfo{person}{A. Passos}, \bibinfo{person}{D. Cournapeau},
  \bibinfo{person}{M. Brucher}, \bibinfo{person}{M. Perrot}, {and}
  \bibinfo{person}{E. Duchesnay:}.} \bibinfo{year}{2011}\natexlab{}.
\newblock \showarticletitle{{Scikit-learn: Machine Learning in Python}}.
\newblock \bibinfo{journal}{\emph{Journal of Machine Learning Research}}
  \bibinfo{volume}{12} (\bibinfo{year}{2011}), \bibinfo{pages}{2825--2830}.
\newblock


\bibitem[Pfeuffer et~al\mbox{.}(2019)]%
        {PfeufferGPMBA-CHI19}
\bibfield{author}{\bibinfo{person}{K. Pfeuffer}, \bibinfo{person}{M. Geiger},
  \bibinfo{person}{S. Prange}, \bibinfo{person}{L. Mecke}, \bibinfo{person}{D.
  Buschek}, {and} \bibinfo{person}{F. Alt}.} \bibinfo{year}{2019}\natexlab{}.
\newblock \showarticletitle{{Behavioural Biometrics in VR: Identifying People
  from Body Motion and Relations in Virtual Reality}}. In
  \bibinfo{booktitle}{\emph{Proc. of ACM CHI}}. \bibinfo{pages}{1–12}.
\newblock


\bibitem[PlayStation.Blog(2022)]%
        {SONY2022}
\bibfield{author}{\bibinfo{person}{PlayStation.Blog}.}
  \bibinfo{year}{2022}\natexlab{}.
\newblock \bibinfo{title}{{PlayStation VR2 and PlayStation VR2 Sense
  controller: the next generation of VR gaming on PS5}}.
\newblock
  \bibinfo{howpublished}{\url{https://blog.playstation.com/2022/01/04/playstation-vr2-and-playstation-vr2-sense-controller-the-next-generation-of-vr-gaming-on-ps5/}}.
\newblock


\bibitem[Roy and Choudhury(2016)]%
        {RoyC-NSDI16}
\bibfield{author}{\bibinfo{person}{N. Roy} {and} \bibinfo{person}{R.
  Choudhury}.} \bibinfo{year}{2016}\natexlab{}.
\newblock \showarticletitle{{Ripple II: Faster Communication through Physical
  Vibration}}. In \bibinfo{booktitle}{\emph{Proc. of the 13th USENIX NSDI}}.
  \bibinfo{pages}{671--684}.
\newblock


\bibitem[Roy et~al\mbox{.}(2015)]%
        {RoyGC-NSDI15}
\bibfield{author}{\bibinfo{person}{N. Roy}, \bibinfo{person}{M. Gowda}, {and}
  \bibinfo{person}{R. Choudhury}.} \bibinfo{year}{2015}\natexlab{}.
\newblock \showarticletitle{{Ripple: Communicating through Physical
  Vibration}}. In \bibinfo{booktitle}{\emph{Proc. of the 12th USENIX NSDI}}.
  \bibinfo{pages}{265--278}.
\newblock


\bibitem[Shen et~al\mbox{.}(2019)]%
        {ShenWLXZHR-TDSC19}
\bibfield{author}{\bibinfo{person}{Y. Shen}, \bibinfo{person}{H. Wen},
  \bibinfo{person}{C. Luo}, \bibinfo{person}{W. Xu}, \bibinfo{person}{T.
  Zhang}, \bibinfo{person}{W. Hu}, {and} \bibinfo{person}{D. Rus}.}
  \bibinfo{year}{2019}\natexlab{}.
\newblock \showarticletitle{{GaitLock: Protect Virtual and Augmented Reality
  Headsets Using Gait}}.
\newblock \bibinfo{journal}{\emph{IEEE Trans. on Dependable and Secure
  Computing}} \bibinfo{volume}{16}, \bibinfo{number}{3} (\bibinfo{year}{2019}),
  \bibinfo{pages}{484--497}.
\newblock


\bibitem[Sluganovic et~al\mbox{.}(2016)]%
        {SluganovicRRM-CCS16}
\bibfield{author}{\bibinfo{person}{I. Sluganovic}, \bibinfo{person}{M.
  Roeschlin}, \bibinfo{person}{K. Rasmussen}, {and} \bibinfo{person}{I.
  Martinovic}.} \bibinfo{year}{2016}\natexlab{}.
\newblock \showarticletitle{{Using Reflexive Eye Movements for Fast
  Challenge-Response Authentication}}. In \bibinfo{booktitle}{\emph{Proc. of
  the 23rd ACM CCS}}. \bibinfo{pages}{1056--1067}.
\newblock


\bibitem[Tanprasert and Yoon(2022)]%
        {TanprasertY-CHI22}
\bibfield{author}{\bibinfo{person}{T. Tanprasert} {and} \bibinfo{person}{D.
  Yoon}.} \bibinfo{year}{2022}\natexlab{}.
\newblock \showarticletitle{{AR Music Visualizers: Application Space and Design
  Guidelines}}. In \bibinfo{booktitle}{\emph{Extended Abstracts of the 2022 CHI
  Conference on Human Factors in Computing Systems}}.
  \bibinfo{pages}{298:1--298:6}.
\newblock


\bibitem[Wang et~al\mbox{.}(2022)]%
        {WangBJM-NMI22}
\bibfield{author}{\bibinfo{person}{G. Wang}, \bibinfo{person}{A. Badal},
  \bibinfo{person}{X. Jia}, \bibinfo{person}{J. Maltz}, \bibinfo{person}{K.
  Mueller}, \bibinfo{person}{K. Myers}, \bibinfo{person}{C. Niu},
  \bibinfo{person}{M. Vannier}, \bibinfo{person}{P. Yan}, \bibinfo{person}{Z.
  Yu}, {and} \bibinfo{person}{R. Zeng}.} \bibinfo{year}{2022}\natexlab{}.
\newblock \showarticletitle{{Development of metaverse for intelligent
  healthcare}}.
\newblock \bibinfo{journal}{\emph{Nature Machine Intelligence}}
  \bibinfo{volume}{4}, \bibinfo{number}{11} (\bibinfo{year}{2022}),
  \bibinfo{pages}{922--929}.
\newblock


\bibitem[Wang and Zhang(2021)]%
        {WangZ-CHI21}
\bibfield{author}{\bibinfo{person}{X. Wang} {and} \bibinfo{person}{Y. Zhang}.}
  \bibinfo{year}{2021}\natexlab{}.
\newblock \showarticletitle{{Nod to Auth: Fluent AR/VR Authentication with User
  Head-Neck Modeling}}. In \bibinfo{booktitle}{\emph{Extended Abstracts of the
  2021 CHI Conference on Human Factors in Computing Systems}}.
  \bibinfo{pages}{452:1--452:7}.
\newblock


\bibitem[Wen et~al\mbox{.}(2016)]%
        {WenRD-CHI16}
\bibfield{author}{\bibinfo{person}{H. Wen}, \bibinfo{person}{J. Rojas}, {and}
  \bibinfo{person}{A. Dey}.} \bibinfo{year}{2016}\natexlab{}.
\newblock \showarticletitle{{Serendipity: Finger Gesture Recognition using an
  Off-the-Shelf Smartwatch}}. In \bibinfo{booktitle}{\emph{Proc. of ACM CHI}}.
  \bibinfo{pages}{3847--3851}.
\newblock


\bibitem[Xu et~al\mbox{.}(2020)]%
        {XuYCHZCL-MobiCom20}
\bibfield{author}{\bibinfo{person}{X. Xu}, \bibinfo{person}{J. Yu},
  \bibinfo{person}{Y. Chen}, \bibinfo{person}{Q. Hua}, \bibinfo{person}{Y.
  Zhu}, \bibinfo{person}{Y. Chen}, {and} \bibinfo{person}{M. Li}.}
  \bibinfo{year}{2020}\natexlab{}.
\newblock \showarticletitle{{TouchPass: towards behavior-irrelevant on-touch
  user authentication on smartphones leveraging vibrations}}. In
  \bibinfo{booktitle}{\emph{Proc. of the 26th ACM MobiCom}}.
  \bibinfo{pages}{24:1--24:13}.
\newblock


\bibitem[Yu et~al\mbox{.}(2019)]%
        {YuLFM-APCCAS19}
\bibfield{author}{\bibinfo{person}{Z. Yu}, \bibinfo{person}{H. Liang},
  \bibinfo{person}{C. Fleming}, {and} \bibinfo{person}{K. Man}.}
  \bibinfo{year}{2019}\natexlab{}.
\newblock \showarticletitle{{An Exploration of Usable Authentication Mechanisms
  for Virtual Reality Systems}}. In \bibinfo{booktitle}{\emph{Proc. of IEEE
  APCCAS}}. \bibinfo{pages}{458--460}.
\newblock


\bibitem[Zhu et~al\mbox{.}(2020)]%
        {ZhuJXML-IMWUT20}
\bibfield{author}{\bibinfo{person}{H. Zhu}, \bibinfo{person}{W. Jin},
  \bibinfo{person}{M. Xiao}, \bibinfo{person}{S. Murali}, {and}
  \bibinfo{person}{M. Li}.} \bibinfo{year}{2020}\natexlab{}.
\newblock \showarticletitle{{BlinKey: A Two-Factor User Authentication Method
  for Virtual Reality Devices}}.
\newblock \bibinfo{journal}{\emph{Proc. of the ACM IMWUT}} \bibinfo{volume}{4},
  \bibinfo{number}{4} (\bibinfo{year}{2020}), \bibinfo{pages}{164:1--164:29}.
\newblock


\bibitem[Álvarez Marín and Velázquez-Iturbide:(2021)]%
        {MarinV-TLT21}
\bibfield{author}{\bibinfo{person}{A. Álvarez Marín} {and}
  \bibinfo{person}{J. Velázquez-Iturbide:}.} \bibinfo{year}{2021}\natexlab{}.
\newblock \showarticletitle{{Augmented Reality and Engineering Education: A
  Systematic Review}}.
\newblock \bibinfo{journal}{\emph{IEEE Trans. on Learning Technologies,}}
  \bibinfo{volume}{14}, \bibinfo{number}{6} (\bibinfo{year}{2021}),
  \bibinfo{pages}{817--831}.
\newblock


\end{thebibliography}


\end{document}